**Graphene Schottky diodes: an experimental review of the rectifying graphene/semiconductor heterojunction**

Antonio Di Bartolomeo
Dipartimento di Fisica "E. R. Caianiello" and Centro Interdipartimentale Nano-Mates
Università degli Studi di Salerno
Via Giovanni Paolo II, 132
I-84084 Fisciano (SA) – Italy

E-mail: antonio.dibartolomeo@fisica.unisa.it

**Abstract**
In the past decade graphene has been one of the most studied material for several unique and excellent properties. Due to its two dimensional nature, physical and chemical properties and ease of manipulation, graphene offers the possibility of integration with the exiting semiconductor technology for next-generation electronic and sensing devices. In this context, the understanding of the graphene/semiconductor interface is of great importance since it can constitute a versatile standalone device as well as the building-block of more advanced electronic systems. Since graphene was brought to the attention of the scientific community in 2004, the device research has been focused on the more complex graphene transistors, while the graphene/semiconductor junction, despite its importance, has started to be the subject of systematic investigation only recently. As a result, a thorough understanding of the physics and the potentialities of this device is still missing. The studies of the past few years have demonstrated that graphene can form junctions with 3D or 2D semiconducting materials which have rectifying characteristics and behave as excellent Schottky diodes. The main novelty of these devices is the tunable Schottky barrier height, a feature which makes the graphene/semiconductor junction a great platform for the study of interface transport mechanisms as well as for applications in photo-detection, high-speed communications, solar cells, chemical and biological sensing, etc. In this paper, we review the state-of-the art of the research on graphene/semiconductor junctions, the attempts towards a modeling and the most promising applications.

**Table of Contents**

## 1. Introduction

The extraordinary properties of graphene, the two-dimensional layer of $sp^2$ carbon atoms, have made it the most studied material of the past decade and are at the origin of a myriad of theoretical and experimental studies [1]-[12] and of a huge variety of applications [13]-[21]. Graphene is an excellent material for electronic devices for its high electron mobility ($\sim 1 \div 1.5 \times 10^{+4}\, cm^2V^{-1}s^{-1}$ for graphene on $SiO_2$/Si substrate and up to $2 \cdot 10^{+5}\, cm^2V^{-1}s^{-1}$ for suspended graphene [22]), great electric current carrying capacity ($\sim 10^{+8}\, A/cm^2$ on $SiO_2$/Si substrate [23]), high thermal conductivity ($\sim 4.84 - 5.3 \times 10^{+3}\, Wm^{-1}K^{-1}$ for suspended graphene [24] record mechanical strength (Young's modulus is ~1 *TPa* [25]), resilience to high temperatures (melting temperature estimated as $4510\,K$ [26]) and humidity [27]-[28], resistance to molecule diffusion and chemical stability. Being practically an all-surface material, graphene is also an ideal material for sensing applications: compared to any other material, it offers the largest detecting area, which favors interaction with the ambient. Further, due to its important advantage of being naturally compatible with thin film processing, graphene is easy to integrate into existing semiconductor device technologies. It is readily scalable, has low contact resistance with most common metals as Ti, Cr, Ni, Pa (a contact resistance as low as $\sim 100\,\Omega\mu m$ has been reported with Ni and Pa [29]-[32]), and can form rectifying junctions with several semiconductor materials. The graphene-semiconductor (G/S) junction (henceforth referred to as G/S junction or GSJ for brevity) is one of the simplest conceivable devices in a hybrid graphene-semiconductor technology. The understanding of its properties and the mastering of its fabrication process are important prerequisites towards a graphene integrated electronics. The challenge in the fabrication is the ability to establish an intimate G/S contact by avoiding chemical-structural modifications to the semiconductor while simultaneously preserving the superior properties of graphene.

The GSJ offers great opportunity to study the physics occurring at the interface between a 2D and a 3D material, as well as between a zero and a definite bandgap system, and can be a convenient platform to investigate electronic properties and transport mechanisms. Surprisingly, it has become the subject of systematic investigation only in the last five years. Graphene transistors have for long dominated the research scenario, with the consequence that the understanding of the physics and the manufacturing techniques of a GSJ are still in their early stage. Nonetheless, the GSJ has already been demonstrated as a rectifying or a barrier-variable device, a photovoltaic cell, a bias-tunable photodetector, a chemical sensor and as a building-block of more complex graphene-based electronic systems, such as Schottky-barrier based field-effect transistors (FETs) or high-electron-mobility transistors (HEMTs).

While the current–voltage (*I-V*) behavior of a GSJ can be roughly described by the well-known ideal diode equation, the details of the measured characteristics often require modifications of the standard thermionic theory and the consideration of additional effects. As for conventional metal/semiconductor diodes, the quality of the interface dramatically affects the junction properties. Impurities and defects may significantly alter the *I-V* curve. Additionally, there are important effects, which origin from the peculiar band structure and density of states of graphene and from its two-dimensional nature. The low density of states close to the neutrality point makes the graphene Fermi level extremely sensitive to the amount of carriers injected into or from the semiconductor. The position of the Fermi level affects the Schottky barrier height (SBH), which in turn controls the current-voltage relationship. These features make the current of the G/S junction tunable in several ways and can be conveniently exploited in practical applications.

In this paper, we review the unique physics occurring at the G/S interface and summarize the progress on fabrication and applications. The paper is organized as follows. First, we start with a reminder of the metal/semiconductor (MS) Schottky junction (SJ) and of some relevant properties of graphene to set the language and the formalism that will be used in the rest of the paper. Second, we focus on the GSJ by reviewing the most significant experimental results and the theoretical models proposed to date. Third, we present the most promising applications, which include photodetectors, solar cells and chemical sensors. Finally, we introduce the important class of van der Waals heterojunctions formed by graphene with 2D layered semiconductors, whose properties and applications are currently the subject of a very active research field.

## 2. The Schottky junction

The intimate contact between a metal and a semiconductor can result in two ideal devices: the ohmic junction or the rectifying (also called Schottky) junction. Ohmic junctions are usually formed with highly doped semiconductors. In an ideal ohmic junction, the current $I$ varies linearly with the applied voltage $V$ and the ratio $V/I$ is the combination of the contact ($R_c$) and the series ($R_s$) resistance:

$$\frac{V}{I} = R_c + R_s \qquad (1)$$

($R_s$ is the lumped resistance of the two materials of the junction). The lower the contact resistance the better the quality of the ohmic contact. Ohmic junctions with low $R_c$ are critical for the performance of high-speed semiconductor devices. In contrast, ideal Schottky junctions act as a perfect diode with high current and very low contact resistance in one direction (the so-called forward direction which correspond to the "on" state of the diode) and negligible current or infinite resistance in the opposite direction (reverse direction or "off" state). Schottky junctions are usually formed with lightly doped semiconductors.

In practice, metal/semiconductor (MS) junctions are fabricated by the evaporation or sputtering of a metal onto the cleaned surface of a semiconductor wafer. Real M/S junctions are neither perfectly ohmic nor perfectly rectifying. A good approximation of one of the two ideal cases can be achieved only with an appropriate choice of the metal and semiconductor materials and with due care in the fabrication process.

In this section we review the metal/semiconductor Schottky junction (SJ) with the purpose of introducing the physical concepts and the formalism which will be later adapted to the GSJ. Exhaustive treatments of M/S contacts can be found in any semiconductor device textbook [33]-[37] or in extensive review articles as Ref. [38]-[39].

Schottky diodes are important electronic components, used in many applications such as solar cells, photodetectors, clamped transistors, MESFETs, HEMTs, microwave mixers, RF attenuators, rectifiers, varactors, Zener diodes and several integrated circuits.

*(a) The Schottky barrier.* Fig. 1 introduces the Schottky model of the M/S junction and some relevant quantities. We assume here that an intimate contact between the metal and the semiconductor with no interfacial layer is established. We typically take a n-type substrate as example; for p-type semiconductors everything works in a similar manner, with electrons replaced by holes and reversed energy axis.

With reference to Fig. 1, the vacuum level or the free-electron energy, $E_0$, which is a convenient reference level for energy, is the energy state of electrons with zero kinetic energy outside the material, either metal or semiconductor. $E_0$, as all other energies in this context, is expressed in *eV*. The difference between $E_0$ and the Fermi level $E_F$ in any material is called the workfunction $\Phi$:

$$\Phi = E_0 - E_F \,. \qquad (2)$$

The Fermi energy $E_F$ represents the highest occupied electron energy state at $T = 0K$ in a metal; in a non-degenerate semiconductor it lies in the gap between the valence and conduction band (Fig. 1(b)), and, as in metals, it separates the occupied from the unoccupied states at $T = 0K$. $E_F$ is the parameter that appears in the Fermi-Dirac distribution function

$$f(E) = \frac{1}{1 + e^{(E-E_F)/kT}} \,, \qquad (3)$$

which expresses the probability that an electron occupies a state with energy $E$ at the absolute temperature $T$ ($k = 8.62 \cdot 10^{-5} eV/K$ is the Boltzmann constant). According to eq. (3), for $T > 0K$, electrons can occupy levels above the Fermi level with a rapidly decreasing probability as the energy moves away from $E_F$. For metals, the workfunction $\Phi_M = E_0 - E_{FM}$ is the energy required to remove an electron at the Fermi level $E_{FM}$ to the vacuum and has a value which depends only on the type of metal (metal workfunctions are comprised between ~2 and 6 *eV*). In a given semiconductor, the position of $E_{FS}$ depends on the doping: $E_{FS}$ is closer to $E_c$ (the lowest allowed energy level of the conduction band, $\approx 4.05 eV$ for silicon (Si)) for a n-type semiconductor or closer to $E_v$ (the highest allowed energy level of the valence band, $\approx 5.17 eV$ for Si) for p-type doping, as shown in Fig. 1(d) and (e). The electron and hole densities in a semiconductor, denoted by $n$ and $p$ respectively, are in fact related to the Fermi energy by the relations:

$$n = N_c e^{-(Ec - E_{FS})/kT} \qquad (4)$$

and

$$p = N_v e^{-(E_{FS} - Ev)/kT} \qquad (5)$$

where $N_c = 2\left(2\pi m_e^* kT/h^2\right)^{3/2}$ and $N_v = 2\left(2\pi m_p^* kT/h^2\right)^{3/2}$ are the effective densities of states in the conduction and valence band, and $m_e^*$ and $m_p^*$ are the effective masses of electrons and holes, respectively ($h = 4.136 \times 10^{-15} eVs$ is the Planck constant). Since the Fermi level is not fixed, the workfunction, $\Phi_S = E_0 - E_{FS}$, in a semiconductor varies according to the doping and is not a good parameter to consider. A constant quantity that, alike $\Phi_M$, can be used to characterize a semiconductor material is rather the difference between the vacuum level and the conduction band edge, known as the electron affinity and is denoted as $X$:

$$X = E_0 - E_c \qquad (6)$$

(examples of electron affinity are: $4.05 eV$ for Si, $4.07 eV$ for gallium arsenide (GaAs), $4.0 eV$ for germanium (Ge)).

When a contact between a metal with workfunction $\Phi_M$ is established with a semiconductor having a different workfunction $\Phi_S$ ($\Phi_M > \Phi_S$ in the example of Fig. 1(a)-(d)), charge transfer occurs until the respective Fermi levels align at equilibrium. In the example of Fig. 1(a) and (b), the different position of the Fermi levels implies that electrons in the n-type semiconductor have an average total energy higher than that in the metal; when the contact is established, the disparity in average energy causes the transfer of electrons from the semiconductor to the metal. The transfer of charge results in the formation of a layer at the semiconductor interface depleted of free charge carriers, called depletion layer. The removal of electrons (and similarly of holes for p-type substrates) leaves behind the space immobile charge of the uncompensated dopant ions (for this reason, the depletion layer is also called the space charge region). In Fig. 1(c) such layer of immobile positive ions is shown over a distance $w$ from the junction and corresponds in the band diagram of Fig. 1(d) to the region with bands bent upwards. In the depletion layer with up-bent bands of Fig. 1(d), the Fermi level is close to neither $E_c$ nor $E_v$ as should be according to eq. (4) and eq. (5) in the region with reduced values of *n* and *p*. The formation of a depletion region in the semiconductor is a necessary condition for the achievement of a Schottky rectifying junction.

The contact between a metal and a n-type semiconductor with $\Phi_M < \Phi_S$ would result in electron injection from the metal to the semiconductor. No depletion layer will be formed in this case, since the metal can be considered an infinite electron reservoir, and the junction would be ohmic (unless some pinning of the Fermi level, that we will discuss later, takes place). When a depletion layer is formed in the semiconductor, the space charge is mirrored by a very thin layer of opposite-sign charge at the metal surface (Fig. 1(c)). These two layers of opposite charge, which make the metal/semiconductor junction (MSJ) somehow resembling to a parallel plate capacitor, give rise to an electric field $\vec{E}_i$ and to a potential $\phi_i = \Phi_i / e$ ($e$ denotes the absolute value of the electron charge) at the junction (Fig. 1(d)), which prevent further net charge diffusion between the semiconductor and the metal. $\phi_i$ (expressed in $V$) is called built-in potential; the corresponding energy, $\Phi_i = e\phi_i$ (in $eV$), which is the energy barrier against the diffusion of electrons from the semiconductor to the metal, is obtained as

$$\Phi_i = \Phi_M - \Phi_S . \qquad (7)$$

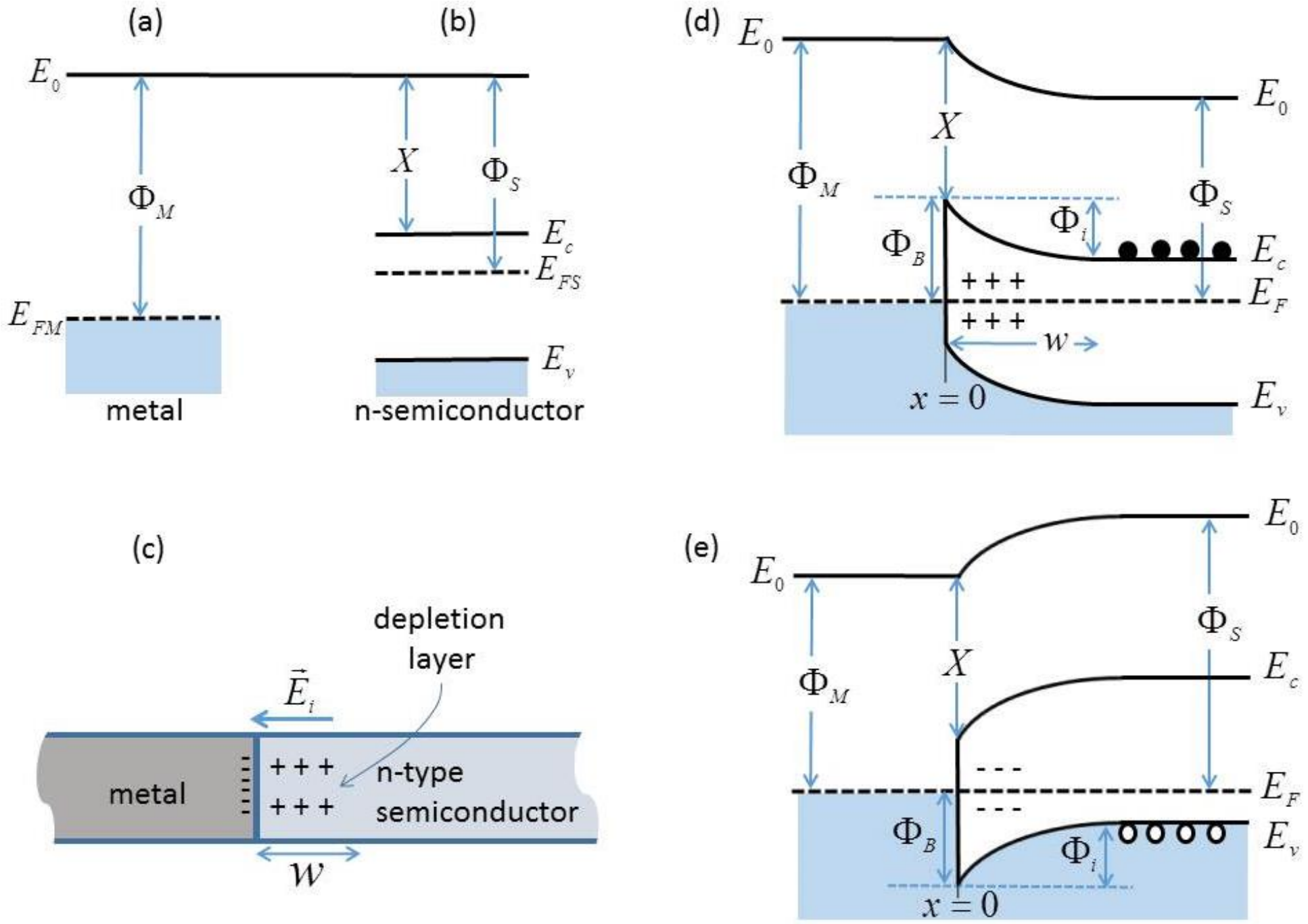

Fig. 1 - A Schottky barrier formed by a metal of high workfunction contacting a n-type semiconductor. (a) Metal workfunction $\Phi_M$ and Fermi energy $E_{FM}$. (b) Semiconductor workfunction $\Phi_S$, electron affinity $X$ and band structure with a bandgap between $E_c$ and $E_v$ and Fermi energy $E_{FS}$. (c) Charge at the metal/semiconductor (MS) junction: the negative charge at the metal surface is approximately a delta function, while the positive charge consisting entirely of immobile ionized donors extends over a distance $w$ inside the semiconductor creating the space charge (or depletion) region. (d) Idealized equilibrium band diagram for the M/S junction. The physical junction is set at $x = 0$. $\Phi_i$ is the energy barrier to the flow of electrons (black dots) from the semiconductor to the metal, while $\Phi_B$ is the Schottky barrier height (SBH) for the electron flow in the opposite direction. $w$ is the extension of the depletion layer and corresponds to the bent part of the energy bands. (e) Idealized equilibrium band diagram of a metal with a p-type semiconductor under the assumption that $\Phi_M < \Phi_S$ (empty circles represent holes).

The electric field $\vec{E}_i = -\vec{\nabla}\phi_i$, which in the example of Fig. 1 is oriented from the positive charge of the depletion layer of the n-type semiconductor towards the negative charge at the metal surface, is the built-in electric field and opposes to the motion of electrons from the semiconductor to the metal. It has the maximum value at the physical M/S interface ($x = 0$ in Fig. 1) and decreases with distance until it vanishes at the edge of the depletion layer (*i.e.* at $x = w$). The most important feature of the M/S energy diagram at the equilibrium, as shown in Fig. 1(d) for the n-type semiconductor, is the appearance of a discontinuity of the allowed energy states, which results in the formation of an energy barrier at the metal/semiconductor interface, $\Phi_B$, known as the Schottky barrier (SB). $\Phi_B$ is the barrier against electron's flow from the metal to the n-type semiconductor and plays a similar role as $\Phi_i$ with the important difference that while $\Phi_i$ can be modified by the application of an external voltage bias (as we will see next), $\Phi_B$ is unaffected by the voltage bias (apart for second order effects). $\Phi_B$ is called the Schottky barrier height (SBH) and can be related to the metal workfunction and to the semiconductor electron affinity, as shown in Fig. 1(d):

$$\Phi_B = |\Phi_M - \mathrm{X}| . \qquad (8)$$

Eq. (8) is known as the Schottky–Mott relation. $\Phi_M$ and $\mathrm{X}$ are both properties of the crystal lattice and cannot be modified by normal level doping o by a voltage bias and, in principle, so is the SBH.
A similar barrier, for the flow of holes from the metal to the semiconductor is formed at the metal/p-type semiconductor junction when $\Phi_M < \Phi_S$, as shown in Fig. 1(e). The Schottky/Mott relation of a MSJ on a p-type substrate, can be written as

$$\Phi_B = E_g - |\Phi_M - \mathrm{X}| , \qquad (9)$$

where

$$E_g = E_c - E_v \qquad (10)$$

is the bandgap of the semiconductor. According to eq. (8) and (9), the sum of the SBHs for electrons and holes of a given metal on a n- and p-doped semiconductor are expected to be equal to the bandgap:

$$\Phi_{B,n} + \Phi_{B,p} = E_g . \qquad (11)$$

The Schottky barrier is the most important feature of a M/S rectifying junction and the SBH, according to Schottky-Mott relations (8) and (9), is controllable by the choice of materials and is independent of the semiconductor doping level. A larger SBH, which as we will show results in better rectifying characteristics, is usually achieved with elevated workfunction metals on n-type semiconductors or with low workfunction metals on p-type semiconductors. However, experimental measurements show that the relation (8) or (9) are only qualitatively valid, with SBH often nearly independent of metal workfunction [40]-[44].
The quantitative discrepancy is due to the fact that the simple Schottky model neglects certain effects which arise at a junction between two dissimilar materials (heterojunction). A M/S junction includes the termination of the semiconductor crystal. The semiconductor surface contains surface (or interface) states, with energies within the semiconductor bandgap, due to incomplete covalent bonds (called Shockley-Tamm states) and other effects (as foreign atoms bonded at the surface and crystal defects) which can lead to charges at the M/S interface. Some of these energy states are acceptor like and may be neutral (when empty) or negative (when occupied by an electron); other surface states are donor like and may be neutral (when occupied by an electron) or positive (when unoccupied). J. Bardeen [45] pointed out the importance of surface states in determining the SBH. Furthermore, the contact is rarely an atomically sharp discontinuity between the metal and the semiconductor crystal, which is what we call an "intimate" contact where possible chemical bonding can take place. Typically, there is a thin interfacial layer which is neither semiconductor nor metal. Silicon crystals for example are usually covered by a thin (10-20 *Å*) oxide layer even after etching or cleavage in atmospheric conditions. Therefore, deposition of a metal on silicon surface leaves a glassy interfacial layer at the junction. Although electrons can tunnel through it, this thin layer can affect the properties of the junction and more importantly make the SBH sensitive to the voltage bias, as we will explain in the following. Considering this intermediate layer, some surface states are physically closer to the bulk semiconductor and remain in thermal equilibrium with the bulk states even when a voltage bias is changed rapidly (> 1 *kHz*). These states are called fast surface states because the electrons occupying them come into equilibrium quickly. In contrast, states situated more remote from the bulk semiconductor within the intermediate layer are called slow surface states, since they take relatively longer times to reach thermal equilibrium with the bulk states.
The effect of acceptor like surface states on a n-type semiconductor is shown in Fig. 2(a). The occupation of some of the surface states by electrons induces a positive depletion layer in the silicon near the surface, causing the energy bands to bend upwards near the surface, even if the semiconductor is not in contact with a metal. When a junction with a metal with $\Phi_M > \Phi_S$ is formed (Fig. 2(b)), the transfer of electrons from the semiconductor to the metal further bends the conduction band away from the Fermi level. This bending also removes negative charge trapped in the surface states and lifts these states above $E_F$. Electrons trapped in the acceptor like interface states contribute to the overall charge transfer from the semiconductor to the metal needed to establish the thermal equilibrium. The larger the density of surface states, the more the trapped charge is removed for each incremental increase of $E_c - E_F$ (*i.e.* of the band bending) near the contact. However, if the density of surface states is very large, a negligible movement of the Fermi level at the semiconductor surface transfers sufficient charge to establish thermal equilibrium. In this case the Fermi level is said to be pinned by the high density of states. The pinning of the Fermi level happens any time a large amount of electronic states is clustered near the Fermi level, because slight changes in the Fermi energy position at the semiconductor surface result in very sizeable charge transfer.
We have already mentioned that any real junction has a thin glassy gap, of a few atomic layers, between the metal and the semiconductor. Simply, this layer is usually omitted in the drawing of the band diagrams of the metal/semiconductor Schottky junction; however this layer is necessary for the continuity of the vacuum level at the interface ($E_0$ must be a continuous, single-value function in space, otherwise energy conservation will be violated [40]). Therefore, the proper treatment of MSJ requires the consideration of a physical gap at the interface. Such interfacial layer of atomic dimensions allows electrical connection because electrons can tunnel through it. Moreover, this layer can contain impurities and added interface states and, more important, can sustain a voltage drop. Otherwise stated, there is an electric field in the extra-thin gap between the metal and the semiconductor (that many authors take into account in the simplified band diagram by considering a charge dipole at the interface). A realistic picture of the band structure of a metal/n-type semiconductor junction, including the mentioned thin layer and acceptor-like surface states, is shown in Fig. 2(b). The electric field in the gap layer in this case is supported by a negative charge on the metal surface that together with the negative charge of the occupied acceptor surface states mirrors the positive charge of the depletion layer in the semiconductor. Fig. 2(b) shows that the gap layer and the interface states originate a lower SBH than the Schottky–Mott one:

$$\Phi_B = \left|\Phi_M - X - \Delta\right|. \qquad (12)$$

When both acceptor and donor type surface states are present, the net charge at the semiconductor surface is zero only when the Fermi level coincides with the so called neutral level, $\Phi_0$ (Fig. 2(c)). The neutral level is the level above $E_v$ that separates the donor from the acceptor-like states: above $\Phi_0$ the states are acceptor type and below it the states are donor type. Consequently when the Fermi level at the surface coincides with the neutral level, the net interface charge is zero. In any other condition, there is a localized charge at the silicon surface which can affect the SBH, as explained above. The neutral energy level also tends to pin the semiconductor Fermi level at the surface before the metal contact is formed. When the Fermi level is moved away from the neutral point a charge appears at the interface (this charge is negative in the example of Fig. 2(c) where the Fermi level has moved above the neutral level, filling some acceptor-like states). Knowing the density of the interface states, $D_{it}$ ($states/\left(cm^2 eV\right)$), under some simplifying assumptions, it is possible to evaluate $\Delta$ in eq. (12) and express the SBH as:

$$\Phi_B = c_2\left(\Phi_m - X\right) + \left(1 - c_2\right)\left(E_g - \Phi_0\right) = c_2\Phi_m + c_3, \qquad (13)$$

where $c_2 = \dfrac{\varepsilon_i}{\varepsilon_i + e^2\delta D_{it}}$, $c_3 = (1-c_2)(E_g - \Phi_0)$, and $\varepsilon_i$ and $\delta$ are the dielectric constant and the thickness of the thin interfacial layer [46]. According to eq. (13), if the density of the interface states is very high, $D_{it} \to \infty$ and $c_2 \to 0$, then

$$\Phi_B = E_g - \Phi_0, \qquad (14)$$

meaning that the Fermi level at the interface is pinned by the surface states at the value $\Phi_0$ above $E_v$ and that the SBH is independent of the metal work function, being completely determined by the surface properties of the semiconductor. In contrast, when $D_{it} \to 0$, eq. (13) is reduced to the Schottky–Mott eq. (8).

In eq. (13), for a given semiconductor, $c_2$ and $c_3$ can be obtained experimentally by fabricating junctions with metals of different $\Phi_m$, and measuring the respective $\Phi_B$; then the $\Phi_0$ and $D_{it}$ which characterize the interfacial properties of the semiconductor are given by

$$\Phi_0 = E_g - e\frac{c_2 \mathrm{X} + c_3}{1-c_2} \qquad (15)$$

and

$$D_{it} = \frac{(1-c_2)\varepsilon_i}{c_2 \delta e^2}. \qquad (16)$$

As an example, Ref. [40] reports $\Phi_B = 0.27\,\Phi_m - 0.52$, for n-Si, which according to eq. (15) and eq. (16) corresponds to $\Phi_0 = 0.33eV$ and $D_{it} = 4\times10^{13}\ states/(cm^2 eV)$.

Another effect which lowers the SBH with respect to the prediction of Schottky–Mott relation is the so called Schottky effect. We refer the reader to standard textbooks for an extensive treatment of it. We just mention here that when an electron is at a distance *x* close to a metal, a positive charge is induced at the metal surface. This charge generates a force which is equivalent to that obtained with an equal positive charge (image charge) at distance –*x* inside the metal. The interaction between the electron and its image charge corresponds to a potential energy $-e^2/(16\pi\varepsilon_0 x^2)$. This energy, combined with the electron potential energy in an external electric field $\vec{E}$ at the metal surface tending to pull electrons from the metal, $-e|\vec{E}|x$, results in an effective lowering of the barrier height as shown in Fig. 2(d). When the metal is in contact with a semiconductor, the appropriate interface electric field has to be considered and the free-space dielectric constant has to be replaced by the dielectric constant of the semiconductor $\varepsilon_s = k_s\varepsilon_0$ ($k_s$ is the relative permittivity, ~11.7 for Si) . In this case, the SBH lowering can be expressed as

$$\Delta\Phi_B = \sqrt{\frac{e^3|\vec{E}_{\max}|}{4\pi\varepsilon_s}}, \qquad (17)$$

where $\vec{E}_{\max}$ is the maximum electric field at the M/S junction [46]. If a voltage bias $V$ is applied to the junction, since $\vec{E}_{\max}$ is related to the bias $V$ by the relation

$$|\vec{E}_{\max}| = \sqrt{\frac{2eN}{\varepsilon_s}(\phi_i - V - kT/e)}, \qquad (18)$$

where $N$ is the doping density per volume, the Schottky effect of eq. (17) introduces a dependence of the SBH on the fourth root of the applied voltage. We note also that eq. (17) and eq. (18) make the SBH (and the junction current, as we will see) to depend also on the substrate doping level. This dependence enables a fine tuning of $\Delta\Phi_B$ and a SBH adjustment though the control of $N$, which is practically performed by ion implantation in a thin layer (~10 *nm* or less) over the semiconductor surface. However, we should notice that the Schottky effect in a M/S structure is less important than in a metal-vacuum system, because of the larger value of $\varepsilon_s$.

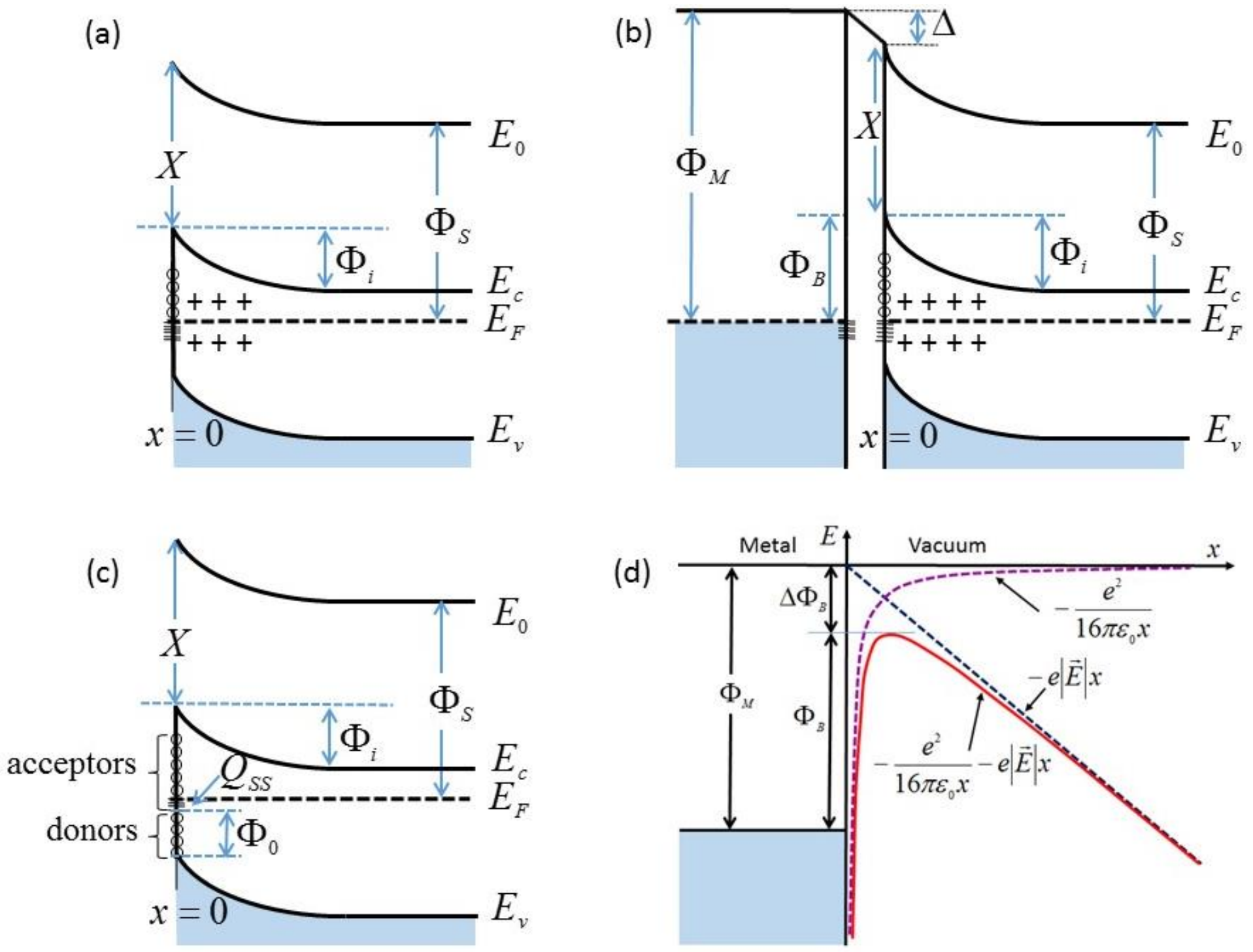

Fig. 2 - (a) Band diagram of a n-type semiconductor with acceptor-like surface states. Some of them are filled with electrons (–). The uncompensated positive ions (+) cause the upward band bending. (b) Band structure of a real metal/semiconductor junction at equilibrium. An electrically transparent thin interfacial layer, which allows electron tunneling, guarantees the continuity of the free electron levels and sustains a voltage drop $\Delta/e$. Full (–) and empty (o) acceptor-like surface states are shown. (c) Acceptor and donor-like states at the semiconductor surface separated by the neutral level, $\Phi_0$. An interface trap charge arises when $\Phi_0$ and $E_F$ do not coincide. (d) Energy band diagram between a metal surface and vacuum, showing (red continuous line) how the application of an external electric field $\vec{E}$ lowers the effective barrier height by the amount $\Delta\Phi_B$.

*(b) Thermionic emission and I-V characteristic*. Differently from a *p-n* junction, the current transport in a M/S junction is mainly due to majority carriers, that is to electrons for n-type or holes for p-type semiconductors. With the moderately doped semiconductors which are employed to fabricate the Schottky diodes (typical dopant concentration is $N \leq 5\times10^{+17} cm^{-3}$), the emission of thermally excited electrons (or holes) from the semiconductor to the metal over the potential barrier $\Phi_i$ (thermionic emission, TE) is the dominating process contributing to the MSJ current (Fig. 3(a)). Accordingly, the *I-V* characteristics of M/S Schottky junctions (Fig. 3(c)) are quite accurately reproduced using the thermionic emission theory. Other conduction mechanisms are the thermionic field emission (TFE) or field emission (FE), which include tunneling through the barrier (Fig. 3(a)). The relative contribution of the conduction mechanism depends on the doping $N$ in the semiconductor as well as on the temperature $T$. By defining a tunneling parameter

$$E_{00} = \frac{eh}{4\pi}\sqrt{\frac{N}{\varepsilon_s m^*}} \qquad (19)$$

($m^*$ is the carrier effective mass), following Ref. [47]-[48], the conditions

$$\begin{aligned} &E_{00}/kT \le 0.2, \\ &0.2 < E_{00}/kT \le 5, \qquad (20) \\ &E_{00}/kT > 5 \end{aligned}$$

categorize the dominating mechanism as TE, TFE or FE respectively. According to eq. (20), at room temperature, n-type Si substrates with $N \le 3\times10^{+17} cm^{-3}$ fit in the first category, while TFE becomes important in the doping range $3\times10^{+17} cm^{-3} < N < 2\times10^{+20} cm^{-3}$ and FE takes over at heavier doping.
Other mechanisms that may contribute to the M/S current are generation/recombination in the space-charge region, diffusion of electrons in the depletion region or injection of holes from the metal that diffuse in the semiconductor and recombine in the neutral region. In addition, there may be edge leakage current due to high electric field at the metal-contact periphery or interface current due to traps at the M/S interface.
Most of the devices we will be dealing with studying the graphene-semiconductor junction, having $N \le 10^{+17} cm^{-3}$, fit in the first category. Hence, here we consider only thermionic emission.
Fig. 3(d) and (e) show the energy band diagrams when a positive (forward) or negative (reverse) voltage bias $V$ is applied to the metal with respect to a n-type semiconductor (Fig. 3(b)). $V$ affects the width of the depletion layer which is narrowed (widened) in forward (reverse) bias according to :

$$w = \sqrt{\frac{2\varepsilon_s}{eN}\left(\phi_i - V - kT/e\right)}, \qquad (21)$$

where $\phi_i$ is the built-in potential ($kT/e \approx 26\, mV$ at room temperature is often neglected) [46]. In the reverse (forward) bias, the Fermi energy of the bulk semiconductor $E_{FS}$ shifts down (up) with respect to that of the metal, $E_{FM}$, allowing an increase (decrease) of the of the potential barrier $\phi_i - V$, that results in rectification, as we will demonstrate.
At the interface, the work function of the metal is independent of voltage bias due to the high density of states of the metal at the Fermi level. $E_{FM}$ is practically unaffected by the bias and is pinned to a level $\Phi_M$ from the vacuum $E_0$. This can be easily understood considering that the density of electron per volume is given by

$$n = \int_0^\infty f(E)N(E)dE, \qquad (22)$$

where $N(E)$ $states/(cm^2 eV)$ is the density of allowed states per volume and energy, and $f(E)$ is Fermi-Dirac distribution of eq. (3). Using eq. (22), the variation $\delta n$ of the electron density resulting from charge exchange with the semiconductor because of the bias-induced widening/narrowing of the depletion layer corresponds to a small variation of the metal Fermi level $\delta E_{FM}$, expressed as

$$\delta n = \int_0^\infty \frac{\partial f(E)}{\partial E_{FM}} \delta E_{FM} N(E)dE. \qquad (23)$$

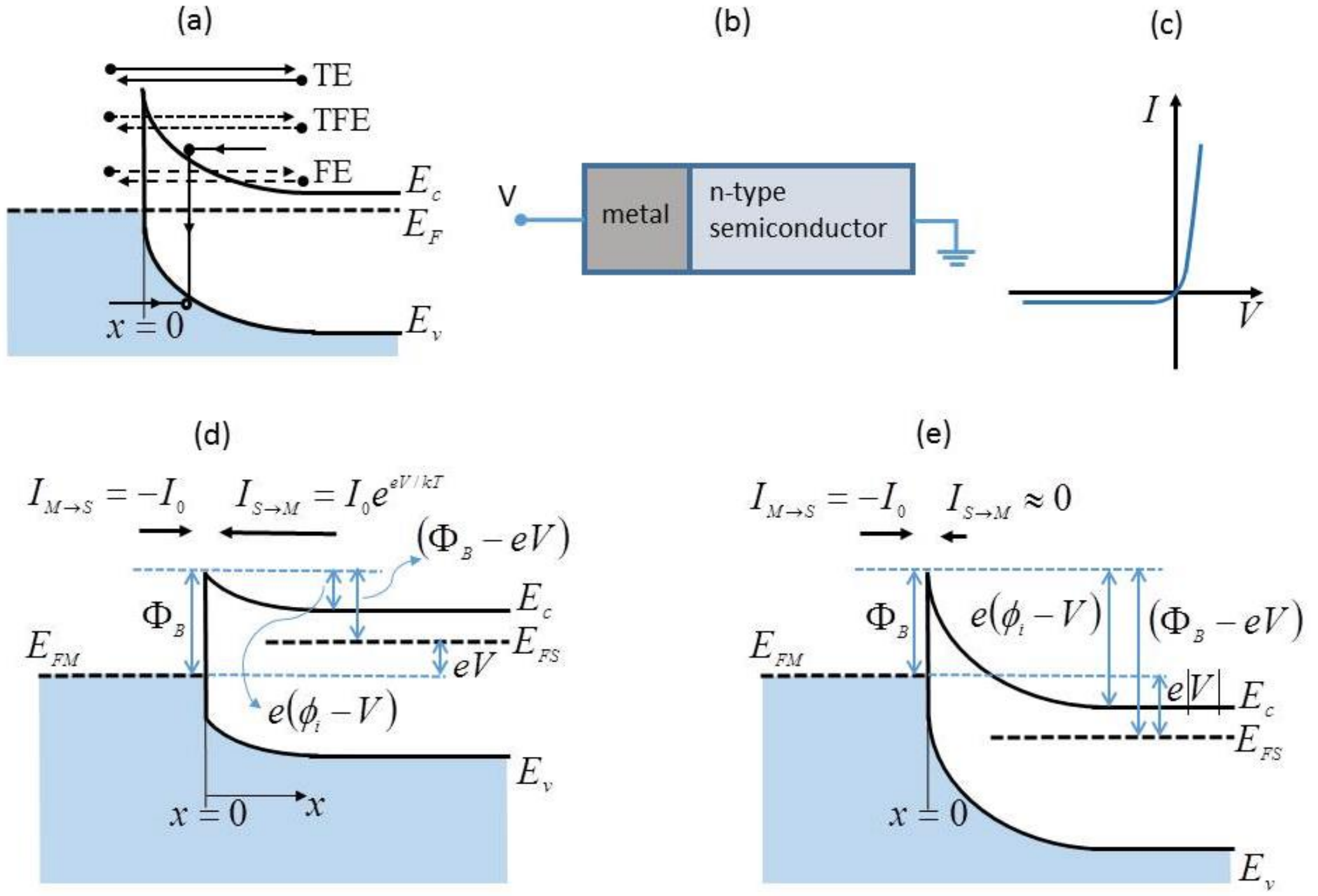

Fig. 3 - (a) Principal transport processes across a M/S Schottky junction: TE=thermionic emission, TFE=thermionic field emission, FE=field emission and electron-hole recombination. (b) Schematic of the voltage bias of the MSJ. (c) Ideal *I-V* characteristic of a rectifying Schottky junction. (d) Band diagrams at the ideal metal/n-type semiconductor Schottky junction in forward bias ($V>0$): the barrier for the transfer of electrons from the semiconductor to the metal is reduced. (e) Band diagrams in reverse bias ($V<0$): the barrier for the transfer of electrons from the semiconductor to the metal is increased. The Schottky barrier is assumed unaffected by the bias. The arrows associated with currents in (d) and (e) indicate the direction of the electron flow.

The derivative of the Fermi function, $\frac{\partial f(E)}{\partial E_{FM}}=-\frac{\partial f(E)}{\partial E}$, is a peak-shaped function, non-zero only over a narrow window around $E_{FM}$ ($\sim 4kT\approx 100\,meV$ wide at room temperature), and with area over it equal to 1, which reduces to the Dirac's delta function, $-\frac{\partial f(E)}{\partial E}=\delta(E-E_{FM})$, at $T=0\,K$. Therefore eq. (23) can be approximated as

$$\delta n\approx \delta E_{FM}N(E_{FM}) \qquad (24)$$

or, considering that $N(E_{FM})=\frac{3}{2}\frac{n}{E_{FM}}$ [49],

$$\frac{\delta n}{n}\approx\frac{3}{2}\frac{\delta E_{FM}}{E_{FM}}. \qquad (25)$$

In a metal $n\approx 10^{+22\div 23}cm^{-3}$ and the $\delta n/n$ achievable by the charge exchange with a semiconductor (where the charge density is the doping concentration $N$ which is order of magnitude lower than $n$ in the metal) is practically null and so is $\delta E_{FM}/E_{FM}$. In a semiconductor the density of states of electrons in the conduction band is

$$N(E)=\frac{4\pi}{h^3}(2m^*)^{3/2}\sqrt{E-E_c} \quad \text{for} \quad E>E_c \qquad (26)$$

and is zero for $E \le E_c$. Since at finite temperature the Fermi function goes rapidly to zero for $E > E_F$, only the $N(E)$ with $E$ closer to $E_c$ ($E \approx E_c$) contribute to eq. (22). This means that the high $N(E_{FM})$ of eq. (24) has to be replaced by the low $N(E \approx E_c)$ as given by eq. (26) and that a variation $\delta n$ of the electrons in the conduction band of the semiconductor requires a non-negligible change of the Fermi energy, $\delta E_{FS}$, with respect to $E_c$. Therefore the position of the Fermi level of the semiconductor is changed by the voltage bias. These considerations are important for the understanding of some specific features of the graphene/semiconductor junction, where differently from a metal the Fermi level of graphene will be bias dependent.

With reference to Fig. 3(d), the thermionic emission theory assumes that the Fermi energy in the semiconductor is flat all the way to $x=0$. Then, according to eq. (4), the electron density at $x=0$ can be expressed as

$$n = N_C e^{-\frac{E_C-E_F}{kT}} = N_C e^{-\frac{\Phi_B - eV}{kT}} . \qquad (27)$$

The current, $I_{S\to M}$, due to electrons flowing from the semiconductor to the metal can be obtained by multiplying the density of charge $-en$ at the interface for the area A of the junction and the average of the absolute value of the *x*-component of the thermal velocity of the electrons $\bar{v}_{th,x}$ (using the Maxwell-Boltzmann velocity distribution $\bar{v}_{th,x} = \int |v_x| \left(\frac{m^*}{2\pi kT}\right)^{3/2} e^{-\frac{m^* v^2}{2kT}} d^3v = \sqrt{\frac{2kT}{\pi m^*}}$ ):

$$I_{S\to M} = \frac{1}{2}A(-en)(-\bar{v}_{th,x}) = A\frac{4\pi e m^* k^2}{h^3}T^2 e^{-\frac{\Phi_B - eV}{kT}} \qquad (28)$$

(the factor $1/2$ has been included to consider only electrons that move in the negative *x* direction with reference to Fig. 3(a)). According to eq. (28), $I_{S\to M}$ depends on the SBH, but is independent of the shape of the barrier. However, it strongly depends on the applied voltage bias, being exponentially increased (decreased) by a positive (negative) $V$.

The total current through the junction can be obtained by adding to $I_{S\to M}$ the current $I_{M\to S}$ corresponding to the flow of electrons from the metal to the semiconductor (holes, which for the n-type semiconductor of our example are minority carriers, give a usually negligible contribution to the junction current; also we are neglecting here other contributions as the electron-hole generation/recombination in the depletion layer of the semiconductor).

At zero bias, there is no net flow of current: the electron current from the semiconductor to the metal is balanced by the current in the opposite direction:

$$I_{S\to M} + I_{M\to S} = 0 \qquad (29)$$

or

$$I_{S\to M} = -I_{M\to S} = A\frac{4\pi e m^* k^2}{h^3}T^2 e^{-\frac{\Phi_B}{kT}} = AA^* T^2 e^{-\frac{\Phi_B}{kT}} , \qquad (30)$$

where

$$A^* = 4\pi e m^* k^2 / h^3 \qquad (31)$$

is known as the Richardson constant ($\approx 112\, Acm^{-2}K^{-2}$ for n-Si and $\approx 32\, Acm^{-2}K^{-2}$ for p-Si). Although the Richardson constant should depends only on the material properties of the semiconductor, it has been shown that the properties of the metal can cause some variations in its value [50]. Variations in $A^*$ can also be caused by inhomogeneity in SBH, by interfacial layers, quantum mechanical reflections and tunnel of carriers [51]-[53].
Since the application of a bias does not change the Fermi level of the metal, $\Phi_B$, is unaffected by a bias and so is the flow of electrons from the metal to the semiconductor $I_{M\to S}$.

Finally, by adding eq. (28) and eq. (30), we obtain the total current though the junction with the applied voltage bias $V$:

$$I = I_{S\to M} + I_{M\to S} = I_0\left(e^{\frac{eV}{kT}} - 1\right). \qquad (32)$$

We have defined $I_0$, the so-called the reverse saturation current or diode leakage current, as

$$I_0 = AA^*T^2 e^{-\frac{\Phi_B}{kT}}. \qquad (33)$$

$I_0$ is an important figure of merit, the lower it is the better is the diode. It strongly depends on the temperature and the SBH: even a small change $\delta\Phi_B$ can have observable effects on it. The image force lowering $\Delta\Phi_B$ of eq. (17) can be included in eq. (33) by replacing $\Phi_B$ with $\Phi_B - \Delta\Phi_B$.
Eq. (32) is the ideal diode equation, which is valid both in forward and reverse bias, and describes the qualitative behavior of the *I-V* curve of a rectifying M/S junction (Fig. 3(c)). Accordingly, in forward bias, the current is dominated by $I_{S\to M}$ and increases exponentially with the voltage $V$, while for $V < 0$ the current is almost constant and is given by $I_{M\to S} = -I_0$ ($I_{S\to M}$ is in this case strongly suppressed being the barrier to the flow of electrons from the semiconductor to the metal $\Phi_i - eV > \Phi_B$).
Deviations from the ideal behavior are often observed in real devices. In forward bias, the rise of the current can be better reproduced by inserting a phenomenological parameter $\eta$ in the exponential of eq. (33). Also, at higher currents a series resistance $R_s$, which includes the lump resistance of the semiconductor, metal and contacts, becomes important since it lowers the effective voltage applied to the junction:

$$V_{eff} = V - R_S I. \qquad (34)$$

Eq. (32) can then be rewritten as

$$I = I_0\left(e^{\frac{e(V - R_S I)}{\eta kT}} - 1\right). \qquad (35)$$

$\eta$ is called the ideality factor and is another important metric of the M/S junction (good quality real junction have $\eta \approx 1.0 \div 1.2$). It measures the deviation from the thermionic emission, taking into account the degree to which defects and other additional non-thermionic effects mediate the transport. As already mentioned, such effects include thermionic field emission and field emission, generation/recombination, image-force-lowering of the SBH, Schottky barrier inhomogeneity, bias-dependence of the SBH, edge leakage, etc. [38]-[39].

If an interfacial layer, such as an oxide, is present at the MSJ, then the current is reduced, the effective SBH is increased and a higher ideality factor is needed to fit the *I-V* curve. A factor accounting for the suppression of current due to the tunneling probability, $e^{-\alpha_T \sqrt{\varsigma}\delta}$, where $\alpha_T$ (in $eV^{-0.5} \cdot Å^{-1}$) is a dimensional constant, $\varsigma$ (in $eV$) and $\delta$ (in *Å*) are the barrier height and the thickness of the oxide layer, have to be included in eq. (35), which is rewritten as:

$$I = I_0\, e^{-\sqrt{\varsigma}\delta}\left(e^{\frac{e(V-R_S I)}{\eta kT}} - 1\right). \qquad (36)$$

As often done, the constant $\alpha_T = 2\sqrt{2em^*}/\hbar \approx 1.01\ eV^{-0.5} \cdot Å^{-1}$, has been omitted in eq. (36) since it approaches unity when the effective mass $m^*$ in the insulator is equal to the free electron mass. The effect of the oxide layer is an increase the effective SBH, as can easily be understood by putting together eq. (33) and eq. (36).

Eq. (35) suffers from the fact that the non-ideality, included through the parameter $\eta$, affects only the current flowing from the semiconductor to the metal but not that from the metal to the semiconductor. For large forward bias, only the exponential is important and it contains $\eta$; in reverse bias the current is $I_0$ which does not contain any ideality factor. To overcome this problem, a bias dependent SBH is introduced (whose origin, as image force lowering and interface states, has been discussed in the previous section):

$$\Phi_B(V) = \Phi_{B0} + \gamma\, e(V - R_S I), \qquad (37)$$

where $\gamma$ is a positive constant meaning that the barrier increases with increasing forward bias. By defining

$$\frac{1}{\eta} = 1 - \gamma = 1 - \frac{1}{e}\frac{\partial \Phi_B}{\partial V}, \qquad (38)$$

eq. (35) can easily be expressed as [48][54]

$$I = I_0 e^{\frac{e(V-R_S I)}{\eta kT}}\left(1 - e^{-\frac{e(V-R_S I)}{kT}}\right). \qquad (39)$$

We have not considered here the rapid increase of the reverse current (not shown in Fig. 3(c)) which happens when the reverse bias reach a critical value, $V_B$, called the breakdown voltage. Breakdown takes place when the high electric field in the depletion layer accelerate the charge carriers to a kinetic energy enough to cause ionizing collisions with the lattice atoms, *i.e.* generation of electron-hole pairs. The repetition of this phenomenon can result in an avalanche process which causes a sudden rise of the reverse current.

*(c) Measurements of ideality factor, Schottky barrier and series resistance.* When the current $I$ is plotted in semi-logarithm scale, the *I-V* characteristic of eq. (35) in forward bias (and for $V > kT/e$ and $V >> R_S I$) corresponds to the straight line,

$$\ln I = \ln I_0 + \frac{e}{\eta kT} V. \qquad (40)$$

For $V \to 0$ the linear behavior of the semi-log plot is lost due to the "-1" in eq. (35). However, according to eq. (39) a straight line all the way to $V = 0$ is obtained with a semi-log plot of $I/\left(1-\exp\left(-eV/kT\right)\right)$ *vs V*, provided that $V >> R_S I$ :

$$\ln \frac{I}{1-e^{-\frac{eV}{kT}}} = \ln I_0 + \frac{e}{\eta kT} V . \qquad (41)$$

In both cases the slope and the y-axis intercept of the straight line can be used to estimate $\eta$ and $I_0$, respectively. $I_0$, obtained extrapolating the forward *I-V* characteristic to $V = 0$, is called the zero bias (ZB) saturation current. Similarly, from eq. (33), a plot of $\ln\left(I_0/T^2\right)$ *vs* $1/T$ (or $1000/T$), known as Richardson plot, is a straight line whose slope and intercept allow the evaluation of $\Phi_B$ and $A^*$, respectively:

$$\ln\left(I_0/T^2\right) = \ln\left(AA^*\right) - \frac{\Phi_B}{k}\frac{1}{T} . \qquad (42)$$

According to eq. (33), the SBH can be obtained from the measurement $I_0$ at any reverse bias. Nevertheless, it is common practice to extracted the SBH from eq. (42) using the ZB saturation current and refer to it as the Schottky barrier height at zero bias. The SBH at ZB has the advantage of being independent of the product $AA^*$ (which is usually known with high incertitude) and avoids the problem of the choice of a reverse bias when dealing with devices having a bias depended reverse current (*i.e.* a bias dependent SBH) .

Finally, the series resistance $R_s$, which becomes important at high current, can be evaluated as follows (see Ref. [55]). In forward bias, and for $V$ high enough to neglect the "-1", taking the logarithm, eq. (35) can be written as $V = \frac{\eta kT}{e}\ln\frac{I}{I_0} + R_s I$ which, derived with respect to $I$ and using the equality $\frac{dI}{I} = d\left(\ln I\right)$, yields:

$$\frac{dV}{d\left(\ln I\right)} = \frac{\eta kT}{e} + R_s I , \qquad (43)$$

which allows the evaluation of $\eta$ and $R_s$ from the intercept and the slope of the straight line fitting $dV/d\left(\ln I\right)$ *vs* $I$ plot.

A voltage bias across the MSJ is able to modulate the depletion region width $w$ (eq. (21)) and to change the positive and negative charge located at the junction, making it behave as a parallel plate capacitor. The charge density per area of the depletion layer in the semiconductor can easily be evaluated using eq. (21):

$$Q_d = eNw = \sqrt{2e\varepsilon_s N\left(\phi_i - V - kT/e\right)} . \qquad (44)$$

According to eq. (44), when a small AC voltage $\delta V$ (so called small-signal) is added to the reverse DC bias ($V < 0$), the junction shows a capacitive behavior with capacitance per unit area expressed as

$$C = \left|\frac{\delta Q_d}{\delta V}\right| = \sqrt{\frac{e\varepsilon_s N}{2\left(\phi_i - V - kT/e\right)}} = \frac{\varepsilon_s}{w} . \qquad (45)$$

Frequencies in the range of the *kHz* are typically used for the small-signal. Eq. (45) suggests that a plot of the square of the reciprocal small-signal capacitance *versus* the reverse bias V is a straight line, with *x*-intercept equal to the built-in potential $\phi_i - kT/e$ and slope $2/(e\varepsilon_s N)$:

$$\frac{1}{C^2} = \frac{2(\phi_i - kT/e)}{e\varepsilon_s N} - \frac{2}{e\varepsilon_s N}V . \qquad (46)$$

Referring to Fig. 1(d), the measurement of $\phi_i$ can be used to estimate the Schottky barrier height as

$$\Phi_B = e\phi_i + E_c - E_F = e\phi_i + \frac{kT}{e}\ln\left(\frac{N_c}{N}\right), \qquad (47)$$

where, to express $E_c - E_F$, we used eq. (4) assuming that the charge density is equal to the doping density, $n \approx N$ (with *N* obtained for example from the slope of $1/C^2$ *vs* $V$ plot). Eq. (47) can be made more accurate by subtracting the image force lowering $\Delta\Phi_B$ of eq. (17) [56]. *C-V* measurements are often used as complementary technique to validate the $\Phi_B$ values obtained from *I-V* characteristics. Furthermore, the linearity of $1/C^2$ *vs* $V$ plot is an indication of a good-quality M/S interface, with no inadvertent interface layer and with low density of surface states [57].

3. **Applications of Schottky diodes**

Here we quickly go through some important uses of the Schottky diodes, relevant for the applications of the graphene/semiconductor junction proposed so far.

Compared to typical *p-n* junctions, Schottky diodes have higher current drive capability and, depending on the SBH, several order of magnitude larger $I_0$. For this reason Schottky diodes are the preferred rectifying devices in circuit applications requiring high current and low voltage. A great advantage of Schottky diodes is that they do not involve minority carriers and do not have the limitations related to minority carrier recombination time proper of a *p-n* junction. This makes Schottky diodes significantly faster and suitable in digital logic circuits as fast switches. High-speed Schottky photodiodes are found in optical communications or optical measurements. Schottky junctions are also used as low-cost photovoltaic cells.

*(a) Photodetectors.*

Photodetectors absorb light and convert it in an electric signal. Examples of Schottky photodetectors, fabricated with a metal layer deposited on top of a semiconductor wafer, are shown in Fig. 4. To increase the sensitivity of the device, it is a common practice to use either (Fig. 4(a)) a semitransparent thin film (10 *nm* or less and with an antireflection coating to reduce absorption and reflection light loss in the metal) or (Fig. 4(b)) a grating type (metal grid) structure. Transparent electrodes, as graphene, provided that can form a rectifying junction, are highly desirable in this context.

Often a pair of Schottky photodiodes in a so-called interdigital electrode structure is fabricated on the surface of the same semiconductor (Fig. 4(c)). Biases of opposite signs are applied to the two electrodes. In Fig. 4(c), where a n-type substrate is considered, the negative electrode forms with the semiconductor a reverse biased junction, while a forward biased junction is established by the other electrode (A and B junctions of Fig. 4(d)). The presence of a blocking junction suppresses the leakage current. With a sufficient large voltage, the depletion layer of the reverse junction can reach the other junction (reach-through configuration) and all the area between the electrodes contributes to detection (Fig. 4(d)). In such configuration, the electron-hole pairs photogenerated in the semiconductor region between the electrodes, drifts toward the electrodes and contribute to a photocurrent.

SB photodetectors are operated at reverse bias (usually much lower than the breakdown voltage, $V_B$), where the diode dark current is low. Depending on the energy $h\nu$ of the radiation, the photocurrent is generated by two different main mechanisms, which are shown in Fig. 5(a) and (b):

(1) when $\Phi_B < h\nu < E_g$, electrons are excited from the metal and injected into the semiconductor;

(2) when $h\nu \geq E_g$ , electron-hole pairs are generated into the depletion layer of the semiconductor.

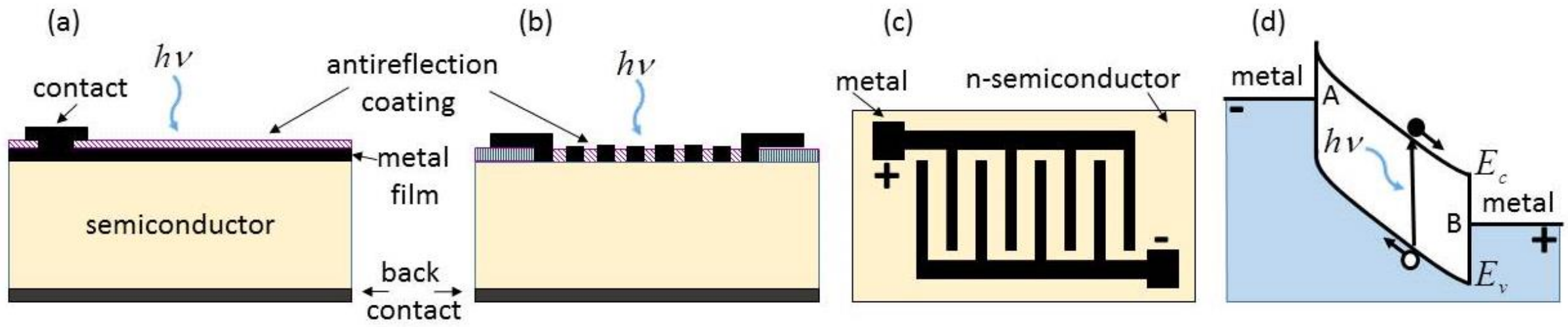

Fig. 4 - Schematic of Schottky barrier photodiodes. (a) Side view of a Schottky diode with semitransparent thin metal film and antireflection coating. (b) Side view of a grating type Schottky diode made with a metal grid. (c) Top view of an interdigital electrode structure corresponding to a pair of Schottky photodiodes. The positive (negative) electrode forms a forward (reverse) biased diode with the n-type semiconductor. (d) Band diagram of device (c) under high bias: the depletion layer of the reverse biased junction A extends and meets that of B (forward biased junction) so that the whole semiconductor between the electrodes is depleted.

With a SBH small enough, mechanism (1) can be used to detect long infrared radiation (IR). In this case the large reverse leakage current, caused by the low barrier, can be suppressed, by operating the device at cryogenic temperatures (T<77 *K*). Photodevices with low SBH are often used as test structure for the measurement of barrier height, bulk defects, interface states, etc.

Devices working with mechanism (2) are operated as a high-speed photodetector (100 *GHz* or more for Schottky barriers on GaAs are easily reached) since the response speed of the Schottky diode is limited only by the RC time constant and by the carrier transit time across the depletion region.

Fig. 5(c) shows a device operated in avalanche mode, with $|V| \approx |V_B|$ : in this case an internal gain (as high as $10^{+6}$) can be achieved and the Schottky photodetector can provides both high-speed and sensitivity to low intensity light.

The basic metric of a photodetector is the external quantum efficiency *EQE* (sometimes referred to as incident photon conversion efficiency *IPCE*), defined as the number of carriers produced per photon,

$$EQE = \frac{I_{ph}/e}{\Phi_{in}} = \frac{I_{ph}}{e}\frac{h\nu}{P_{in}}. \qquad (48)$$

In eq. (48) $I_{ph}$ is the photocurrent, $\Phi_{in} = P_{in}/h\nu$ is the incoming photon flux and $P_{in}$ is the incident optical power. The ideal quantum efficiency is unity. The reduction is due to current loss by recombination, incomplete absorption, reflection, etc. The internal quantum efficiency *IQE* is calculated in a similar way except that it considers the absorbed photon flux ($P_{in}\Phi_{abs} = Q_{in}A_{abs}$ , with $A_{abs}$ the absorbed fraction):

$$IQE = \frac{I_{ph}/q}{\Phi_{abs}}. \qquad (49)$$

Another similar metric is the responsivity, $R_I$ , which is the photocurrent divided by the incident optical power:

$$R_I = \frac{I_{ph}}{P_{in}} = \frac{EQE \cdot e}{h\nu} \qquad (A/W). \qquad (50)$$

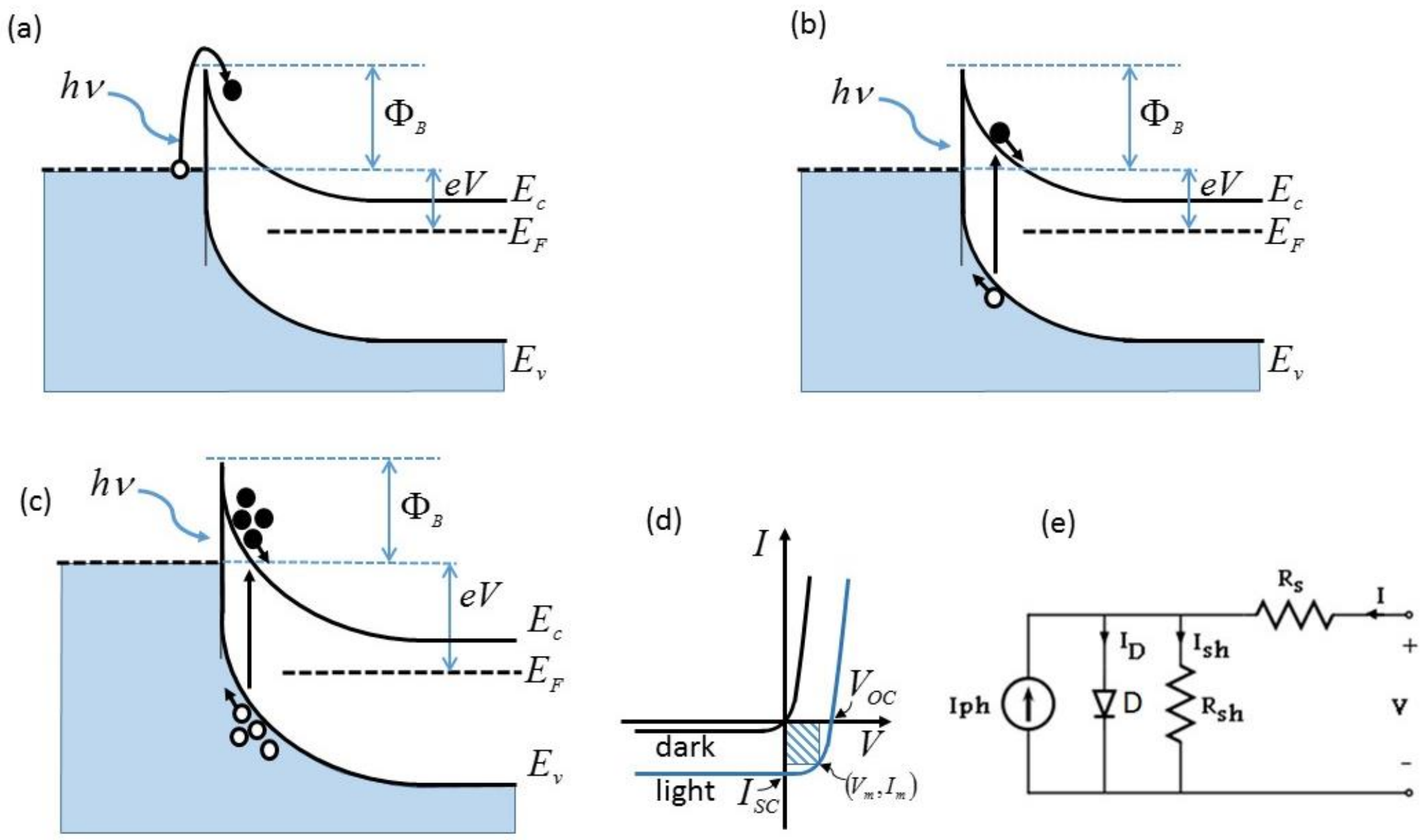

Fig. 5 - Detection modes of a Schottky barrier photodiode (metal on n-Si). (a) When $\Phi_B \leq h\nu < E_g$ and $|V| << |V_B|$, electrons are excited over the SB. (b) When $h\nu > E_g$ and $|V| << |V_B|$, *e-h* pairs are generated in the depletion layer of the semiconductor. (c) At $h\nu > E_g$ and $|V| \approx |V_B|$ the photogenerated *e-h* pairs gain enough energy to generate an avalanche multiplication process. (d) *I-V* characteristic of a Schottky diode in dark and under illumination. (e) Equivalent circuit of a solar cell Schottky diode.

In many applications, a photovoltage rather than a photocurrent is measured as sensitive method for photodetection. Measuring a voltage can have the advantage of eliminating the power consumption associated to Joule-heating. In such case, the responsivity (photovoltage responsivity) is defined as photovoltage $V_{ph}$ divided by the incident optical power:

$$R_V = \frac{V_{ph}}{P_{in}} \qquad (V/W). \qquad (51)$$

An internal gain mechanism can be exploited to increase the signal. Unfortunately, high gain also leads to higher noise. The noise, which should be kept as low as possible, determines the minimum detectable signal. Usually there are many factors which contribute to the total noise (dark current, thermal noise, shot noise, flicker noise, etc.). The noise-equivalent power (*NEP*) is the relevant figure of merit in this context. It corresponds to the incident RMS optical power required to produce a signal-to-noise ratio of 1 in a 1-Hz bandwidth. The *NEP* is expressed in units of $W\,Hz^{-0.5}$ and represents the minimum detectable power. If

$$S_I = \sqrt{\bar{I}^2_{noise}/1Hz} \qquad (52)$$

is the RMS dark noise density ($\bar{I}^2_{noise}$ is the average of the square of the current noise $I_{noise}$ measured with $0.5s$ integration time, which corresponds to a bandwidth of 1 Hz (Nyquist criterion)) then

$$NEP = \frac{S_I}{R_I}, \qquad (53)$$

with $R_I$ is the photocurrent responsivity of eq. (50) [58]. A similar definition of *NEP* can be given in the photovoltage mode:

$$NEP = \frac{S_V}{R_V}, \qquad (54)$$

with $S_V = \sqrt{\overline{V}^2_{noise}/1Hz}$ the analogous of $S_I$ for the voltage noise $V_{noise}$.

Finally the detectivity $D^*$ (measured in Jones – named after R.C. Jones) is defined as

$$D^* = \frac{\sqrt{AB}}{NEP} \qquad cmHz^{0.5}/W \quad (\equiv Jones), \qquad (55)$$

where *A* is the area of the photosensitive region and *B* is the frequency bandwidth of the detector. The detectivity takes into account simultaneously detector sensitivity, spectral response and noise.

*(b) Solar cells*. The built-in electrical field in the depletion region of a Schottky junction enables it as photovoltaic device. Similarly to the photodetector, the Schottky photovoltaic cell responds to light with $h\nu > \Phi_B$ and is sensitive to a large range of wavelengths. The Schottky barrier solar cell is easy to fabricate and has the potential for use as a low-cost photovoltaic power conversion device for large scale terrestrial power generation.

Electron-hole pairs generated in the depletion layer of the semiconductor are separated by the built-in electric field. In a metal/n-type Schottky junction, electrons are drifted toward the semiconductor making it more negative, while holes are moved toward the metal making it more positive. Because of this charge separation, when the device is illuminated and under open circuit condition a voltage $V_{oc}$ appears across the device with the metal end positive and the semiconductor end negative. Under short circuit condition a current $I_{sc}$ flows. The *I-V* characteristic of the Schottky junction under illumination is shown in Fig. 5(d).

A limitation of Schottky solar cells is the small $V_{oc}$. The open-circuit voltage is related to the SBH, and is often a few tenths of $eV$, smaller than that achieved with *p-n* junctions. The layout of a Schottky solar cell is the same as that of the photodetectors pictured in Fig. 4(a) and 4(b).

The photocurrent density $J_{ph}$ of a solar cell can be expressed as:

$$J_{ph} = \int_{\lambda_1}^{\lambda_2} eT(\lambda)\Phi_{in}(\lambda)\left(1-e^{-\alpha(\lambda)w}\right)d\lambda, \qquad (56)$$

where $\lambda$ is the radiation wavelength, $\lambda_1$ and $\lambda_2$ are the UV and the semiconductor cutoffs, $\alpha(\lambda)$ is the optical absorption coefficient of the semiconductor, $\Phi_{in}(\lambda)$ is the incident photon flux at wavelength $\lambda$, $T(\lambda)$ the transmission coefficient of the metal film and $w$ the width of the depletion layer [46].

The other main contribution to the photocurrent, which adds to that of eq. (56), comes from the minority carriers photogenerated in the substrate quasi-neutral region, able to diffuse to the depletion layer without recombining.

The generalized equivalent circuit of a Schottky solar cell is shown in Fig. 5(e). It includes the following parameters: 1) the shunt resistance $R_{sh}$ which takes into account all parallel resistive losses across the photovoltaic device including leakage current; 2) the series resistance $R_S$ (that corresponds to the bulk and contact resistances, and is the same that appears in eq. (35); 3) the forward current $I_D$ trough the M/S junction D; and 4) the photogenerated current $I_{ph}$ obtained integrating eq. (56) over the area of the device. From this model, it is straightforward to write the *I-V* equation of the photovoltaic cell:

$$I = I_0\left(e^{\frac{eV-R_S I}{\eta kT}} - 1\right) + \frac{V - R_S I}{R_{SH}} - I_{ph} . \qquad (57)$$

$I_{sc}$ and $V_{oc}$ can be obtained from eq. (57) for $V = 0$ and $I = 0$, respectively. When $R_S \to 0$ and $R_{SH} \to \infty$, simple expressions are obtained

$$I_{sc} \approx -I_{ph} \qquad (58)$$

and

$$V_{OC} \approx \frac{\eta kT}{e}\ln\left(\frac{I_{ph}}{I_0} + 1\right) \approx \frac{\eta kT}{e}\ln\left(\frac{I_{ph}}{I_0}\right) . \qquad (59)$$

Finally, replacing $I_0$ with eq. (33), a relation showing that $V_{oc}$ increases proportionally to $\Phi_B$ is obtained

$$V_{OC} \approx \frac{\eta kT}{e}\ln\left(\frac{I_{ph}}{AA^*T^2}\right) + \frac{\eta}{e}\Phi_B . \qquad (60)$$

The maximum power delivered by the a solar cell $P = I_m V_m$ corresponds to the maximum area of the rectangle formed by the *I-V* curve with *I* and *V* axes in the fourth quadrant, which is shaded in Fig. 5(d). A typical figure of merit of a solar cell is the so called fill factor

$$FF = \frac{I_m V_m}{I_{sc} V_{oc}} , \qquad (61)$$

which represents how close the *I-V* curve is to the ideal rectangular shape (*FF*=100%). Typical *FF* values are in the range 70-85% depending on the device structure. The external quantum efficiency *EQE* or incident photon conversion efficiency (*IPCE*) is defined as for photodetectors. The power conversion efficiency (*PCE*), the key metric of a solar cell, is defined as the ratio of the maximum output power to the solar energy input, under well-defined conditions:

$$PCE = \frac{I_m V_m}{P_{in}} = \frac{I_{sc} V_{oc} FF}{P_{in}} . \qquad (62)$$

The common standard condition is the so called air mass 1.5 solar radiation, *AM1.5*, which represents the sunlight at the Earth surface (integrated over all wavelengths) when the Sun is at an angle of ~48° from the vertical. At this angle, the incident power is about 963 $W/m^2$. The air mass zero condition, *AM0*, represents the sunlight outside the Earth atmosphere. The standard testing temperature is 25 °C. Compared to *p-n* cells with efficiencies ranging from 12-15% for polycrystalline Si to 15-20 % for monocrystalline Si (pushed up to 25% in laboratory cells), Schottky cells have lower conversion efficiency due to the limited $V_{oc}$, with typical values around 12-16% [59]-[62].

From the above discussion it is clear that an improved Schottky photodetector or solar cell can be achieved by replacing the metal electrode with graphene. Graphene, for his high transparency (high $T(\lambda)$), low reflection, high current transport capability (*i.e.* low $R_S$) and bias tunable SBH can serve simultaneously as transparent electrode and anti-reflecting coating as well as active layer for electron-hole separation and transport. Metallic electrodes such as Cu, Ag, Pd, and Au thin films are the common top electrodes in M/S solar cells. However, compared to graphene, they have inferior optical and mechanical properties, higher sheet resistance and require more complicated high-vacuum deposition process.
Cheap screen-printed metal contacts can be used with silicon *p-n* solar cells and this is one of the main reasons why *p-n* cells remain dominant over their Schottky counterparts, which require an evaporated metal grid at high temperature. Another drawback of the Schottky solar cells is the time drift due to M/S contact instabilities. However, replacing metals with CVD graphene, which can be transferred on the semiconductor cheaply and at room temperature, can give a boost to Schottky solar cells.

**4. Graphene**
Graphene is a one-atom thick, planar layer of carbon atoms arranged in a 2D hexagonal honeycomb lattice where each atom is covalently bond to its three nearest neighbors at a distance $a = 1.42$ Å. The remaining $p_z$-electron per atom, not involved in the covalent bonding, is delocalized over the whole graphene lattice, and is responsible for the electric conductivity.
Isolated graphene was discovered in 2004 by Geim and Novoselov [1]-[2], who made graphene accessible with a technique as simple as the mechanical exfoliation. Owing to the strong covalent bonds, graphene has an extraordinary mechanical strength [6][63].

*(a) Electronic and transport properties.* The hexagonal lattice (formally a triangular lattice with two atoms per unit cell), which can be considered as made of two interpenetrating triangular sub-lattices A and B (Fig. 6(a)), provides graphene with an unique electron band structure. The conduction and valence bands touch each other at six points in the reciprocal space, of which the two $\vec{K}$ and $\vec{K}'$ corners of the hexagonal Brillouin zone are inequivalent (Fig. 6(b)). This makes graphene a zero bandgap semiconductor (semimetal).
The dispersion relation $E(\vec{k})$ is obtained from the tight-binding Hamiltonian

$$H_{\vec{k}} = \begin{pmatrix} 0 & -t\left(1+e^{i\vec{k}\cdot\vec{a}_1}+e^{i\vec{k}\cdot\vec{a}_2}\right) \\ c.c. & 0 \end{pmatrix}, \qquad (63)$$

and corresponds to two eigenvalues, one positive and one negative, for each value of $\vec{k}$, resulting in the two branches illustrated in Fig. 6(b) [3][9][11][64]-[65]:

$$E_{\pm}(\vec{k}) = \pm t\sqrt{1+4\cos\frac{\sqrt{3}ak_x}{2}\cos\frac{ak_y}{2}+4\cos^2\frac{ak_y}{2}}\ . \qquad (64)$$

In eq. (63) and (64), $\vec{k}$ is the crystal momentum, $\vec{a}_1$ and $\vec{a}_2$ are the primitive lattice vectors, $t \approx 2.7\, eV$ is the nearest-neighbor hopping interaction between A and B sites, $a$ is the bond length and c.c. indicate the complex conjugate of the off-diagonal matrix element.
Close to the 6 corners of the Brillouin zone, within ~1 $eV$, the bands have the shape of intersecting vertical cones with common apices, corresponding to the Dirac energy $E_D$ (which is usually assumed as the reference energy); the crystal momenta $\vec{K}$ and $\vec{K}'$ are known as the Dirac points. Close to the Dirac points the dispersion relation is linear; around $\vec{K}$ for example it can be written as

$$E = \hbar v_F \left|\vec{k}-\vec{K}\right|, \qquad (65)$$

where $v_F \approx 10^{+6}\, m/s$ is the Fermi velocity. This energy corresponds to the spectrum of a 2D Dirac-like Hamiltonian for massless fermions:

$$H_K = \hbar v_F \begin{pmatrix} 0 & k_x - ik_y \\ k_x + ik_y & 0 \end{pmatrix} = \hbar v_F \vec{\sigma} \cdot \vec{k} , \qquad (66)$$

with $\vec{\sigma} = (\sigma_x, \sigma_y)$ the 2D vector of the Pauli matrices and $\hbar v_F = 3ta/2$. Eq. (66) is formally obtained by expanding the tight-binding Hamiltonian close to $\vec{K}$ with $\hbar v_F = 3ta/2$ (close to $\vec{K}'$ the Hamiltonian is the same as eq. (66) with $\vec{\sigma}$ replaced by the complex conjugated $\vec{\sigma}^* = (\sigma_x^*, \sigma_y^*)$). At low energy, carriers in graphene are described by the equation

$$-i\hbar v_F \vec{\sigma} \cdot \vec{\nabla} \psi(\vec{r}) = E\psi(\vec{r}) , \qquad (67)$$

and behave as relativistic particles traveling at constant speed

$$v = \frac{1}{\hbar} \frac{\partial E}{\partial k} = v_F , \qquad (68)$$

which is $\sim 1/300$ of the speed of light in vacuum and is independent of $\vec{k}$. This behavior is responsible for much of the research attention that graphene has been receiving as platform for investigating the properties of the Dirac fermions and for perspective high-speed electronic applications.

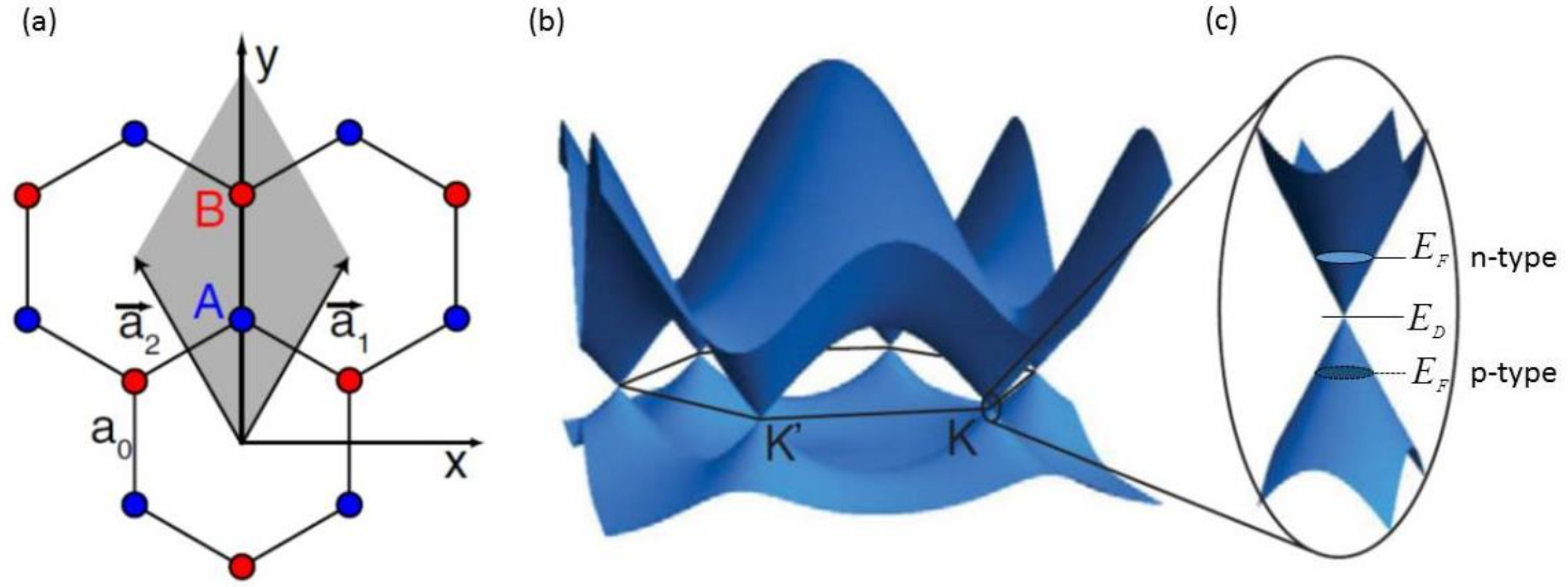

Fig. 6 - (a) Hexagonal honeycomb lattice of graphene. The blue and red circles indicate atoms belonging to sub-lattices A and B (see text); the lines between the circles indicate the chemical bonds. $a_0$ is the nearest-neighbor distance. $\vec{a}_1$ and $\vec{a}_2$ are primitive lattice vectors and the gray cell is the unit cell. (b) Band structure of graphene showing the zero bandgap inequivalent Dirac points $\vec{K}$ and $\vec{K}'$. (c) Zoom of the dispersion relation close to a *K*-point for small energies showing an intersecting double cone. A Fermi level above (below) $E_D$ corresponds to n (p)-type graphene. Figure adapted from Ref. [11].

Another important property of graphene is a special feature of the carrier wavefunction which leads to other unusual properties. Due to the two interpenetrating sub-lattices, A and B, carriers near the Dirac point $\vec{K}$, can be described by a two component wavefunction [3]:

$$\psi_{\pm,\vec{K}}(\vec{k}) = \frac{1}{\sqrt{2}} \begin{pmatrix} e^{-i\theta_k/2} \\ \pm e^{i\theta_k/2} \end{pmatrix} , \qquad (69)$$

where $\theta_k = \arctan\left(\frac{q_x}{q_y}\right)$ and $\vec{q} = \vec{k} - \vec{K}$ is the momentum measured from the Dirac point $\vec{K}$. A similar expression with opposite ± choice is obtained at $\vec{K}'$. This wavefunction has some interesting implications. For example, electrons along $+k_x$ and $-k_x$ have orthogonal wavefunctions, so there is no probability of backscattering at 180 degrees which favors mobility and enables near ballistic transport at room temperature. The mobility $\mu$, in $Vcm^{-2}s^{-1}$, is defined both in a diffusive or ballistic regime as the ratio between the conductivity $\sigma$ and the carrier charge density and is commonly used to characterize the graphene structural quality:

$$\mu = \frac{\sigma}{en}. \qquad (70)$$

The conductivity (which for a two-dimensional system coincides with the so called sheet conductance) and the mobility of graphene depend on the microscopic scattering processes that occur in graphene at a given temperature [10][66]-[67]. Scattering in graphene is still an active field of research, and we will only mention some of the main features in this section.
The carrier density $n$ sets the position of the Fermi level, $E_F$, with respect to the Dirac point. For ideal neutral graphene without free carriers, $E_F$ is located at the Dirac point, where $n \approx 0$ ($n = 0$ at $T = 0K$). Graphene becomes an n- or p-type conductor when the Fermi level shifts above or below the Dirac point and $n$ corresponds to an excess of electrons or holes, respectively (Fig. 6(c)).
The relation between $n$ and $E_F$ can be easily derived considering that the density of states in graphene depends linearly on the energy $E$. In fact (see for example Ref. [66]), considering $\vec{q} = \vec{k} - \vec{K}$, for a 2D system, the number of states between $q$ and $q + dq$ is

$$N(q)dq = \frac{2\pi q dq}{(2\pi/L)(2\pi/W)} \times 2 \times g, \qquad (71)$$

where $LW = A$ is the area of the sample on which the number of states is calculated, the denominator is the *q*-space area occupied by a 2D *q*-state, the factor 2 is the spin degeneracy and g (=2 in graphene) is the valley degeneracy. Using the dispersion relation, eq. (65), results in

$$N(q)dq = N(E)dE = Ag\frac{EdE}{\pi(\hbar v_F)^2}. \qquad (72)$$

By defining $D(E) = N(E)/A$ as the density of states per energy and area, and recognizing that the energy in graphene with respect to the Dirac point can be greater or less than zero, we obtain

$$D(E) = 2|E|/\pi\hbar^2 v_F^2 = D_0|E| \qquad (73)$$

and

$$n = \int_0^{+\infty} D(E) f(E) dE = \frac{2}{\pi}\left(\frac{kT}{\hbar v_F}\right) F_1\left(\frac{E_F}{kT}\right) \approx \frac{1}{\pi}\left(\frac{E_F}{\hbar v_F}\right)^2. \qquad (74)$$

In eq. (74), $f(E)$ is the Fermi function of eq. (3), $F_1(E_1/kT)$ is the so-called Fermi-Dirac Integral and the last expression rigorously holds at $T = 0K$. Eq. (74) shows that the Fermi level changes as the square root of the carrier density:

$$E_F = \mp \frac{h}{2\sqrt{\pi}} v_F \sqrt{n} \qquad (75)$$

(the - and + sign correspond to p and n-type graphene, respectively).
The position of $E_F$ can be varied experimentally either by chemically doping the graphene (by K, Ca, NH3,$O_2$, Organic molecules, etc. [68] or by inducing an excess of carriers by means of an electric field generated by an applied bias or a gate [69]. The possibility of controlling the position of the Fermi energy, the sign of the excess carries and hence the conductivity by doping or by a bias/gate is a remarkable feature of graphene and opens the possibility of a new class of electronic devices. This feature plays an important role in the G/S Schottky junction, where the modulation of $E_F$, differently from the M/S case, can be used to tune the Schottky barrier height $\Phi_B$ and the rectification properties of the junction.
At the Dirac point, even at cryogenic temperatures when $n$ and $\sigma$ should tend to zero, the conductivity of graphene remains finite. This is a consequence of the intrinsic properties of the 2D Dirac fermions, which set a limit on the minimum attainable conductivity. In short and wide strips (width to length ratio $W/L >> 1)$) of ideal graphene, with no impurities or defects and for $T \to 0K$, transport at the Dirac point is explained as propagation of charge carriers via evanescent waves (tunneling between the leads) [70]. Under these conditions, the minimum of conductivity can reach the universal minimum value

$$\sigma_{Dirac} = 4e^2/(\pi h), \qquad (76)$$

regardless of the edges of the graphene strip [70]. Experimental confirmation of the ballistic transport and the universal minimum conductivity in graphene was provided by low-temperature transport spectroscopy on single-layers and bilayers [71] and through measurements of shot noise at low frequency in field effect devices ($200\ nm$ long and with $W/L = 24$) at temperatures around $4.2K$ [72]. When the effect of graphene edges cannot be neglected ($W/L < 3$) or at the presence of disorder which locally affects the density of carriers, the evanescent states are accompanied by propagating states and the minimum conductivity rapidly increases [73]. Due to the fabrication process, graphene usually contains various sources of disorder, as defects, impurities, strong interaction with charges in surrounding dielectrics, phonons, etc. This disorder causes spatial inhomogeneities in the carrier density. Local accumulations of charge carriers, so called electron-hole puddles [74], produce percolation paths for carrier transport and prevent the transition to the ideal minimum conductivity state at the Dirac point. Hence, for real graphene the measured conductivity at cryogenic temperatures is much higher than the universal minimum value, changes from sample to sample and is typically in the range $2 \div 5e^2/h$ on good quality samples [75]-[77].
At higher carrier density, *i.e.* away from the Dirac point, the mentioned disorder sources, acting as scattering centers, reduce the electron mean free path. Two transport regimes are often considered depending on the mean free path length $l$ with respect to the graphene length $L$. When $l > L$, transport is ballistic since carriers can travel through graphene at Fermi velocity $v_F$ without scattering. On the other hand, when $l < L$, transport is diffusive since carriers undergo elastic and inelastic collisions. In both cases, transport can be described by the Landauer formalism [78] and the conductivity can be expressed as

$$\sigma = \frac{L}{W}\frac{2e^2}{h}\int_0^\infty T(E)M(E)\left(-\frac{\partial f}{\partial E}\right)dE, \qquad (77)$$

with $T(E)$ the transmission function and $M(E)$ the number of conducting channels.
For ballistic transport

$$T(E) = 1, \qquad (78)$$

while for diffusive transport

$$T(E) = \frac{\lambda(E)}{\lambda(E) + L}, \qquad (79)$$

where L is the length of the sample and $\lambda(E)$ is the energy-dependent scattering mean free path. $M(E)$ can be calculated [66] from the dispersion relation eq. (65), and, similarly to the density of states, has a linear dependence on the energy *E*:

$$M(E) = W\frac{2|E|}{\pi\hbar v_F}. \qquad (80)$$

From eq. (77)-(80), under the approximation that $-\frac{\partial f}{\partial E} \approx \delta(E - E_F)$ valid for $T \to 0K$, a simple expression of graphene conductivity can be easily obtained:

$$\sigma = \frac{2e^2}{h}\left(\frac{2E_F}{\pi\hbar v_F}\right)\lambda(E_F). \qquad (81)$$

In eq. (81), $\lambda = L$, independent of the energy *E* in the ballistic regime, and

$$\lambda(E) = \frac{\pi}{2} v_F \tau(E) \qquad (82)$$

in the diffusive regime, where $\tau(E)$ is the momentum relaxation time, *i.e.* the average time between scattering events.
Recalling eq. (70) and the $n$ *vs* $E_F$ relation of eq. (75), eq. (82) implies that $\sigma \propto \sqrt{n}$ and $\mu \propto 1/\sqrt{n}$ in the ballistic regime. This dependence, which is sketched in Fig. 7(a), has been experimentally observed on clean graphene [79]. Experimental values of $\mu$ as high as $2\times10^{+5}\ cm^2V^{-1}s^{-1}$ for $n < 5\times10^{+9}\ cm^{-2}$ have been measured on suspended graphene at liquid helium temperature [80]-[85]. In contrast, on $SiO_2$ where the ballistic regime is more difficult to observe because of several scattering mechanisms, a value of $\mu \approx 10 \div 15\times10^{+3}\ cm^2V^{-1}s^{-1}$ is typically measured [82].
In the diffusive regime, $\lambda(E)$ and $\tau(E)$ depend on the scattering mechanism. In graphene, three main scattering mechanisms are considered: Coulomb scattering by charged impurities (long range scattering), short-range scattering (defects, adsorbates), and electron-phonon scattering.
Charged impurity scattering is a very important scattering mechanism [79]. It is caused by the presence of charged impurities close to the graphene sheet. These impurities can be trapped ions in the top or bottom dielectric or ions on the graphene surface. Coulomb scattering is more relevant at low energies and the relaxation time corresponding to it varies linearly with energy, $\tau(E) \propto E$ [83]-[84]. According to eq. (70), (75), (81) and (82), $\sigma \propto n$ and the mobility is independent of $n$. The observation of a linear $\sigma\ vs\ n$ plot (Fig. 7(a)) is frequently taken as evidence for the presence of charged impurity scattering. Coulomb scattering can be reduced either by removing the substrate or by using a substrates (as boron nitride BN) which are less prone to charge trapping [85]. Also helpful is a current-induced cleaning and a high-temperature annealing which remove ionized atoms from the graphene. As a remark, we notice that experimental studies on intentionally damaged graphene have shown that even uncharged defects, if able to produce strong scattering, may result in a linear $\sigma\ vs\ n$ behavior [86].
Short range scattering potential due to localized defects as vacancies and cracks [83][87] is usually approximated by a delta function. The resulting scattering rate is proportional to the final density of states, so $1/\tau(E)\propto E$, and is independent of temperature. Hence, for this scattering mechanism, the conductivity does not depend on $n$ and $\mu \propto 1/n$.

Deformation potential scattering by acoustic phonons [86]-[88]) is another important scattering mechanism. Phonons can be considered an intrinsic scattering source since they limit the mobility at finite temperature even when there are no defects. Longitudinal acoustic (LA) phonons are known to have a higher electron-phonon scattering cross-section. The scattering of electrons by LA phonons can be considered quasi-elastic since the phonon energies are negligible in comparison with the Fermi energy of electrons. Optical phonons in the graphene can also scatter carriers – especially at temperatures above 300K and are believed to be responsible for the decrease in conductivity at high temperatures [89]. The decrease of conductivity at high temperature is also ascribed to polar optical phonon in the underlying $SiO_2$ [90]. Phonon scattering is usually invoked to explain the temperature dependence of $\sigma$ but it does not introduce any dependence on $n$ (Fig. 7(a)) .

In suspended graphene, in addition to in-plane phonons, carbon atoms can oscillate in the out-of-plane direction leading to a new class of low energy phonons, known as flexural branch [91], which can constitute the main limitation to electron mobility. Differently from in-plane phonons which have a linear dispersion relation, in absence of strain, the rotational symmetry makes flexural phonons to obey a quadratic dispersion relation. This implies that there is a high number of these low energy phonons and that the graphene sheet can be easily deformed in the out-of-plane direction. The dispersion relation becomes linear at long wavelengths when the sample is under tension due to the rotational symmetry braking [92]. Compared to in-plane phonons, flexural phonons have a different coupling to charge carriers. Out of plane displacements can enter only quadratically into the Dirac Hamiltonian and consequently charge carriers can excite flexural phonons only in pairs. The contribution of flexural phonons to the resistivity of suspended graphene scales with temperature as $T^{5/2}\ln T$, while the in-plane phonons contribution is proportional to $T^4$. The unusual $T^{5/2}$ scaling implies that scattering from flexural modes dominate the phonon contribution to the resistivity below a crossover temperature which for graphene is ~130 $K$ [91]. Strains with not too large values, as those induced by a back gate, can suppress significantly the flexural phonon scattering, a property which opens the door to the possibility of locally modifying the resistivity of suspended graphene by strain modulation.

Other scattering mechanisms can affect the conductivity. For graphene on $SiO_2$ we have already mentioned the surface polar phonons (SPP) of the $SiO_2$ substrate which produce an electrical field that strongly couples to electrons in graphene.

Different scattering mechanisms add up to produce a total conductivity $\sigma_{Tot}$ given by

$$\frac{1}{\sigma_{Tot}}=\frac{1}{\sigma_1}+\frac{1}{\sigma_2}+\ldots, \qquad (83)$$

where $\sigma_i$ is the conductivity corresponding to a given scattering mechanism. According to eq. (83), the smaller $\sigma_i$ limits the total $\sigma_{Tot}$. An example is given in Fig. 7(a) where acoustic phonons and charged impurities are considered.

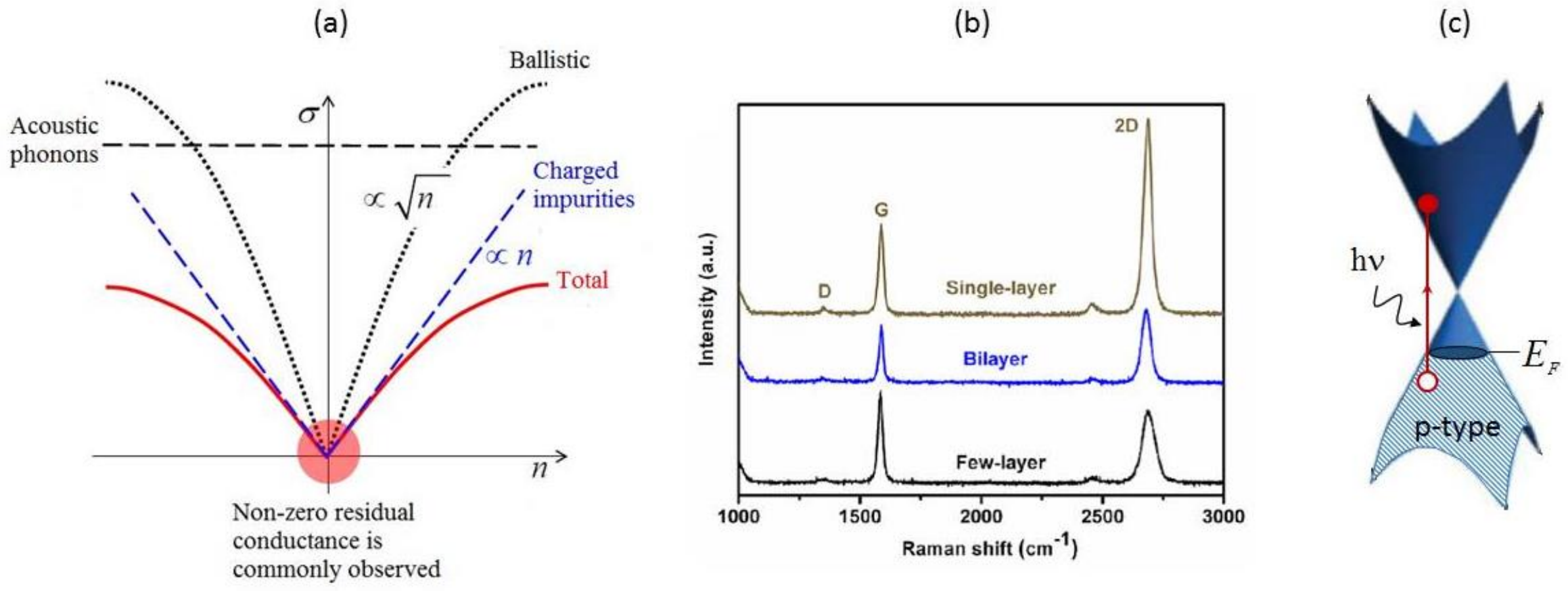

Fig. 7 - (a) Conductivity *vs* carrier density ($\sigma$ *vs* $n$) for graphene. Acoustic phonons (short range) and ionized impurity (long range) scattering are considered. Adapted from Ref. [66] (b) Raman spectra of single layer, bilayer and few layer graphene. (c) Optical generation of an *e-h* pair in p-type graphene corresponding to a momentum-conserving process with energy gain $h\nu$.

*(b) Optical properties*. Graphene has also remarkable optical properties [8][93]. The gapless energy band enables charge carrier generation by light absorption over a very wide energy spectrum, unmatched by any other material (Fig. 7(c)). This includes the ultraviolet, visible, infrared (*IR*) and terahertz (*THz*) spectral regimes. In the near-*IR* and visible, light transmittance $T$ of graphene does not depend on frequency, being controlled by the fine structure constant $\alpha = e^2/(4\pi\varepsilon_0\hbar^2 c)$ [94] . At normal incidence, the transmittance can expressed as [95]:

$$T = (1-0.5\pi\alpha)^2 \approx 1-\pi\alpha \approx 0.977. \qquad (84)$$

Considering the thickness of 0.334 nm, a single-layer of suspended graphene has an unusually high absorption of $A = 1-T \approx 2.3\%$, corresponding to an absorption coefficient of about $7\times10^{+5}cm^{-1}$. This is about 50 times higher than for example the absorption of GaAs at $\lambda = 1.55\mu m$ and demonstrates the strong coupling of light and graphene, which can be exploited for conversion of photons into electrical current or voltage [96]-[97]. Because graphene sheets behave as a 2-D electron gas, they are optically almost noninteracting in superposition, and the absorbance of few-layer graphene sheets is roughly proportional to the number of layers. The proportionality is gradually lost and the transparency remains quite high while adding further layers: graphene layers corresponding to a thickness of $1\mu m$ still have a transparency of approximately 70% [98].
In addition, the reflectivity of graphene, $R$, is very low: $R = 0.25\pi^2\alpha^2(1-A) = 1.3\times10^{-4}$, though it increases to 2% for 10 layers [99].

*(c) Synthesis*. Graphene can be produced by four main processes: micro-mechanical exfoliation of highly ordered pyrolytic graphite, chemical vapor deposition (CVD), epitaxial growth on silicon carbide (SiC), and reduction of graphene oxide (GO).
Mechanical exfoliation of highly ordered pyrolytic graphite (HOPG) is the method originally developed by Geim and Novoselov in 2004 [1] and consists in the use of scotch tape to repeatedly peel a thin flake of HOPG till fading layers of graphite are left on the tape. The tape is then pressed onto a substrate, typically a $SiO_2$/Si where single and few layer graphene is transferred. After some cleaning, graphene can be identified using the contrast difference in an optical microscope [100] or through Raman spectroscopy [101]. Graphene and graphitic materials present three main peaks in the Raman spectrum: the band D at ~1350 $cm^{-1}$, the G band at ~1580 $cm^{-1}$ and the 2D band at ~2680 $cm^{-1}$ (Fig. 7(b)). The D band is induced by defects in the graphene lattice and is not usually seen in highly ordered graphene layers. The G band results from in-plane vibration of sp$^2$ carbon atoms and is peculiar of most graphitic materials. The 2D band is due to a two phonon resonance process, involving phonons near the $\vec{K}$ point, and is dominant in graphene as compared to bulk graphite [102]. The intensity ratio of the G and D band can be used to characterize the number of defects in graphene sample. The shape of the 2D peak and its intensity relative to the G peak is a standard way to characterize the number of layers of graphene. A sharp and symmetric Lorentzian 2D peak with an intensity more than twice the G peak is typical of monolayer graphene. The 2D peak becomes broader and less symmetric and decreases in intensity when the number of layers increases [103]-[104].

Catalytic and epitaxial growth of graphene on supporting surfaces are able to produce a wafer scale graphene layer and are important for large scale manufacturing and integration of graphene into existing device fabrication processes [105].

In CVD carbon is supplied in gas form and a metal is used as catalyst and substrate to grow graphene [106]. A carbon containing gas, as $CH_4$ or $C_2H_2$, is transported in a quartz chamber using an inert gas carrier as Ar (some other gases, as $H_2$, are added for specific purposes). The growth of graphene takes place at temperature of ~1000 °*C*. The decomposition of the carbon containing gas on the metal surface creates a concentration gradient between the surface and the bulk, causing carbon atoms to diffuse into the metal and form a solid solution. Upon cooling, C atoms dissolved in the metal at high temperature precipitate out and segregate at the metal surface, forming one or more layers of graphene. Ni, Cu, Ru, Ir are the most commonly used substrates. After the formation of graphene, a thin layer of poly(methyl methacrylate) (PMMA) is spin-coated on it. The metal below the graphene is then chemically etched (ammonium persulfate is typically used for Cu) and the graphene remains supported on the PMMA membrane and can be deposited on a substrate. The PMMA is finally dissolved with acetone.

Graphene layers can be epitaxially formed by heating silicon carbide (SiC) in ultra-high vacuum to temperature between 1000 °*C* and 1500 °*C* [107][108]. Hexagonal 4H-SiC ( $E_g = 3.26\,eV$ ) and 6H-SiC ( $E_g = 3.03\,eV$ ) polytypes are commonly employed. Either of the two polar faces of SiC, the $(0001)$ silicon-terminated face (the Si-face) or the $(000\overline{1})$ carbon-terminated face (the C face) are suitable to grow graphene. The high temperature causes Si to sublimate and leaves behind a carbon rich surface which is graphitic in nature and can be used to form graphene. SiC is semi-insulating, so that devices can be constructed by patterning graphene formed on SiC, without concern for electrical current flowing through the substrate. Generally it is found that the electrical properties of graphene grown on SiC are somewhat inferior to those of exfoliated or CVD graphene and several issues still remain, as controlling the number of layers produced, repeatability of large area growth, and interface effects with the SiC substrate.

While the first three methods can produce graphene with a relatively perfect structure and excellent properties, GO has the important advantage of being produced using inexpensive graphite as raw material by cost-effective chemical methods with a high yield [109]-[110]. In addition GO has the characteristic of being highly hydrophilic and so able to form stable aqueous colloids to facilitate the assembly of macroscopic structures by simple and cheap solution processes. To partially restore the structure and properties of graphene, GO has to be reduced. A review of the reduction techniques is given in [111]. Different reduction processes result in different properties of reduced GO (rGO), which in turn affect the final performance of materials or devices composed of rGO. Hydrazine is the commonly used reducing agent.

## 5. Experimental aspects of the graphene/semiconductor junction

In this section we review some significant experimental investigations on the G/S Schottky junction properties, by trying to keep their chronological order of publication. The special case of graphene with two-dimensional layered semiconductors will be treated in section 8.

Two simplified layouts of a GSJ are shown in Fig. 8 (other setups will be presented while reviewing specific applications). The top side of the semiconductor substrate is covered by an insulating layer (typically $SiO_2$) which is patterned and etched to create trenches and expose the semiconductor surface. Graphene is deposited along the edge of the insulator in a way to cover the semiconductor and part of the insulator. The part of graphene on the semiconductor forms the GSJ, while the part on the oxide is contacted by a metal. In Fig. 8(a) metal is deposited after graphene, while in Fig. 8(b) metal is pre-deposited. Particular care has to be taken to clean the semiconductor surface to achieve a good G/S interface. Si substrate are usually treated with a buffered oxide etch solution (as $NH_4F/HF$) just before graphene deposition to prevent formation of native oxide. Most of the device reported used graphene synthesized by CVD on a metal substrate (Ni, Cu, etc.). Post-growth, a thin layer of PMMA is spin-coated on top of graphene. After that, the metal is chemically etched leaving graphene supported on the PMMA membrane. Graphene is then cleaned by rinsing in baths of deionized (DI) water and transferred onto the substrate, usually without any extra-drying. The use of this wet transfer process typically results in graphene films with high concentrations of p-type dopants ($4 \cdot 10^{+12} cm^{-2}$) and unpredictable electronic mobility ($500 \div 10^{+4}\, cm^2 V^{-1} s^{-1}$) [82][112]-[115]. Alternatively, to minimize chemical contamination, a dry transfer procedure reported by Petrone et al. [116] can be used. In this procedure, which minimizes the introduction of impurities present in the DI water at the G/S interface, the graphene/PMMA sample is rinsed in isopropanol which is then blown dry with nitrogen so graphene is transferred under dry conditions onto the target substrate.
To fabricate G/S devices of given area, a litho-patterning can be done after graphene deposition and transfer PMMA removal. Finally the part of graphene on the insulator and the top surface of the substrate are contacted by a metal. In some applications, metal is deposited and patterned on the substrate prior to graphene transfer. In such case, graphene adheres on metal as well as on the semiconductor substrate by van-der-Waals forces. In good devices, a low resistance ohmic contact is formed with metal. Electrical measurements are performed by applying a bias (positive for n-type semiconductor or negative for p-type semiconductor) to the graphene contact with respect to the substrate.

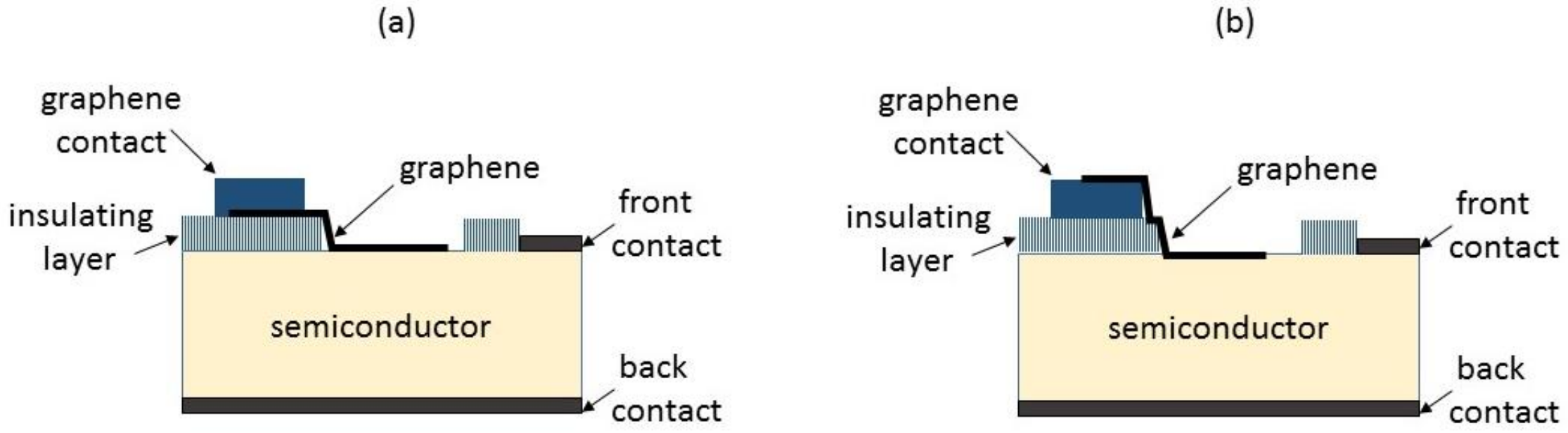

Fig. 8 - Schematic diagram of a G/S Schottky diode. The insulating layer on top of the substrate is etched to form a window and allow graphene to contact the semiconductor. Adhesion of graphene is by van-der-Waals forces. Ohmic contacts are formed on the back side or on the front side of the semiconductor. (a) Metal contact deposited on graphene. (b) Graphene deposited on metal contact.

*(a) Experimental results* – The formation of graphite/semiconductor Schottky barriers was first reported in 2009 by S. Tongay et al. [117]. HOPG was employed over Si, GaAs and 4H-SiC. The substrates were carefully cleaned to remove any native oxide and/or contaminants and ohmic contacts were formed on the substrate using well established contact recipes [118]-[119]. Three different techniques were used to make HOPG/semiconductor junctions: (1) spring loaded bulk HOPG, (2) van der Waals adherence of cleaved HOPG flakes, and (3) HOPG "paint". In the first technique, relatively large (~ $1mm^2$) HOPG pieces were gently pressed onto the substrate; in the second technique, mechanically exfoliated ($\sim 0.5 mm^2$) HOPG sheets were landed on the semiconducting substrates and in some cases observed to flatten out with strong adherence to the substrate for effect of the van der Waals attraction; in the paint technique, graphite flakes were sonicated in 2-butoxyethyl and octyl acetate, "painted" on the substrate and dried in air. These three techniques allowed the fabrication of devices with similar current density-voltage characteristics. The experiment demonstrated that the graphite/semiconductor contact has a rectifying behavior, near room temperature, (Fig. 9(a)-(c)) with the forward-bias characteristic well described by thermionic emission and fitted by eq. (35), over 2 to 6 current decades. The series resistance is responsible for the deviations from linearity at higher forward voltages. Remarkably, rectification was preserved at temperatures as low as $20\,K$. The zero-bias SBHs were extracted from eq. (42) with measurements in the temperature range $250 \div 300\,K$ (Fig. 9(d)): 0.40 $eV$ for Si, 0.60 *eV* for GaAs and 1.15 for 4H-SiC. Slightly higher values were estimated by small-signal capacitance *vs* voltage (*C-V*) measurements at 1 kHz using eqs. (46) and (47). The term $kT/e$ was neglected and no other correction effects, as the image force lowering, which would have probably reduced the discrepancy, were used. Tongay et al. [117] attributed the higher values to the possible presence of a very thin oxide layer a the G/S interface. The linearity of $1/C^2$ *vs* $V$ (Fig. 9(e)) was taken as an indication of a small surface density of states, which allowed the use the Schottky-Mott relation of eq. (8) to estimate a workfunction in the range of 4.4-4.8 *eV* for the HOPG, in agreement with values reported in literature [120]-[122]. The ideality factors of the devices fabricated with the cleaved flakes ($1.12 \le \eta \le 1.50$) were found to be typically smaller than those of the samples prepared by the paint and pressure-contact methods which had $\eta \le 2.0$.

Since the outermost layer of the HOPG electrode is a single graphene sheet, S. Tongay et al. [117] correctly predicted that graphene/semiconductor (G/S) junction would form a Schottky barriers and would manifest a similar rectifying behavior.

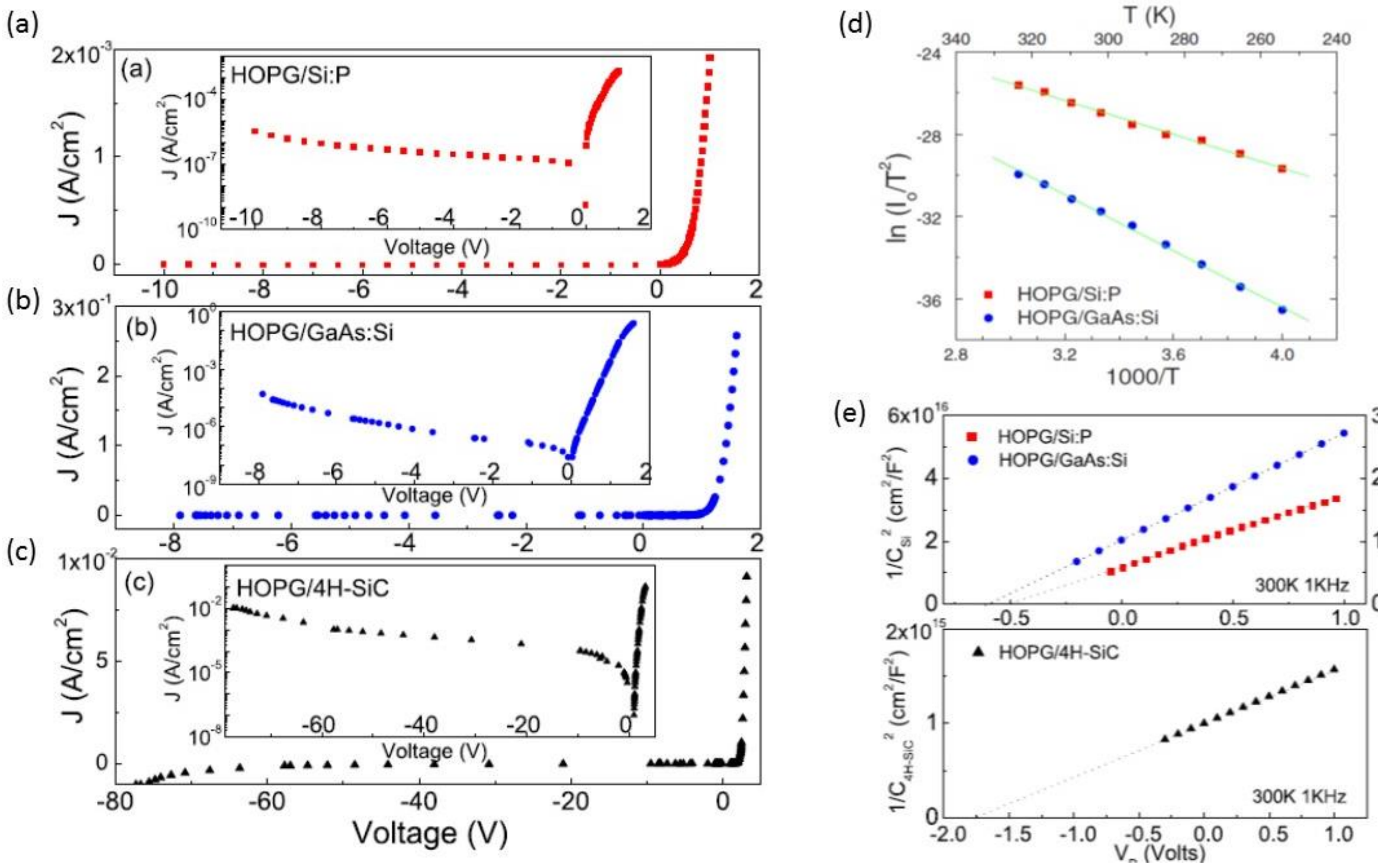

Fig. 9 - Plots of the current density *J* *vs* applied bias *V* at room-temperature for (a) graphite/n-Si (red squares), (b) graphite/n-GaAs (blue circles), and (c) graphite/n-4H-SiC junctions (black triangles) obtained with the "paint" method. Insets: *J-V* plots on semi-logarithmic axes.(d) Plot of $\ln\left(I_0/T^2\right)$ *vs* $1/T$ (scaled by a factor of 1000) from 250 *K* up to 330 *K* for diodes in (a) and (b). (d) Inverse square of the capacitance per unit area measured at 1kHz as a function of reverse bias at room temperature, showing a linear behavior. Figure adapted from Ref. [117]

A Schottky barrier between graphene and SiC was first observed by S. Sonde et al. [123] who measured it in a scanning probe microscopy study where they compared graphene epitaxially grown on 4H-SiC(0001) with single layer graphene deposited after mechanical exfoliation. The adherence of graphene, in both cases is similar, dominated by van der Waals attraction. The local *I-V* characteristics obtained by the probe were strongly rectifying. Using a method developed in Ref. [124] and based eqs. (32) and (33), the forward bias exponential characteristics were used to estimate the SBH for Pt on pure SiC and for the systems Pt/exfoliated-graphene/SiC and Pt/epitaxial-graphene/SiC (Fig. 10(a) and (b)). Assuming a workfunction of 4.5 *eV* for graphene and considering the electron affinity of 3.7 *eV* for 4H-SiC [125]), eq. (8) would imply a $\Phi_B = 0.8 eV$. Sonde et al. [123] measured a SBH of 0.85 *eV* for deposited graphene and a substantially reduced value of 0.36 $eV$ for epitaxial graphene (Fig. 10). Epitaxial graphene synthesis on SiC occurs through a series of complex surface reconstructions [126]-[127] and includes a C-rich buffer, with large percentage of $sp^2$ hybridization, before the real graphene [128]-[130]. This buffer layer may be more or less defective with more or less Si dangling bonds at the interface and has a strong influence on the electronic properties of epitaxial graphene on SiC(001). As shown in Fig. 10(d), the Fermi level is supposed to be pinned (and raised) by donor centers in this buffer layer. The pinning of the Fermi level results in n-doping of graphene and in a reduction of the SBH. On the contrary, no doping occurs with deposited graphene (Fig. 10(c)). The charge transfer from SiC to graphene was studied in Ref. [131] where the saturation density of n-type doping of single layer graphene was estimated around $10^{+13}\,cm^{-2}$ (corresponding to $E_F \sim 0.4eV$) for a $0.3nm$ SiC surface-to-graphene spacing and a bulk donor density $10^{+19}\,cm^{-3}$ in the SiC buffer layer or a surface state density of $1\times10^{+13}\,cm^{-3}eV^{-1}$.

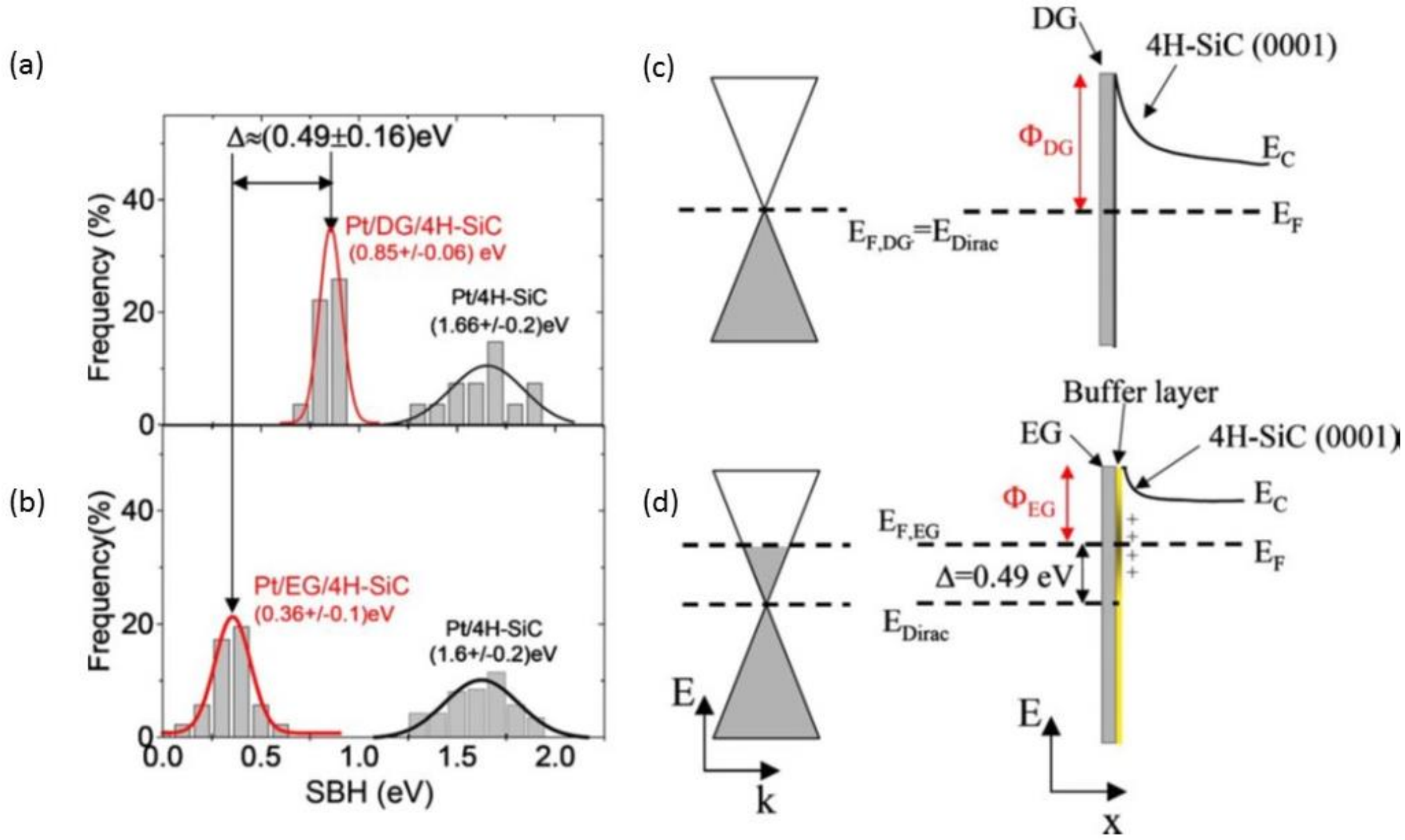

Fig. 10 - Histograms of SBHs evaluated for Pt/graphene/4H-SiC and Pt/4H-SiC in the cases of (a) exfoliated and deposited graphene (DG) and (b) epitaxially grown graphene (EG). Corresponding band diagrams for DG and EG Schottky junction on SiC are shown in (c) and (d), respectively. EG has lower SBH due to Fermi level pinning to donor states in the carbon-rich buffer layer on the SiC surface. $E_{F,DG}$ and $E_{F,EG}$ are the graphene Fermi-level with respect to the Dirac point. Figure reproduced from Ref. [123].

C.-C. Chen. et al. [132] studied *I-V* characteristics of exfoliated graphene flakes, consisting mainly of bilayer and few layer graphene, on both n-Si and p-Si. They included temperature dependence and the local effect of light absorption. Their devices were fabricated with graphene flakes deposited in part on top of a $Si/SiO_2/Si_3N_4/Cr/Au$ stack and in part on bare Si, as shown in Fig.11. Gold formed an ohmic contact with graphene (the contact resistance measured with a four-probe method was $212\cdot10^{-6}\Omega\cdot cm^2$), while the graphene-silicon (G/Si) contact included a Schottky barrier and exhibited a rectifying behavior. Interestingly, the contact resistance between silicon and graphene was estimated to be $74\cdot10^{-6}\Omega\cdot cm^2$. The lower contact resistance indicates the bonding energy of graphene/silicon interface ($\sim151\pm28\,mJ/m^2$) is slightly higher than that of graphene/Au interface [133]-[134].

Particular care was taken to clean and passivate the Si surface before graphene deposition. $SiO_2$ was first wet etched, then DI rinsed and dried on hot plate at 120°C. A final dry etching using $CF_4$ RIE was proved crucial to achieve devices with finite series resistance and rectifying *I-V* characteristics. Since, graphene flakes are gas impermeable [135] no further oxidation of the Si surface is expected to occur at the G/Si interface after the deposition of graphene. This was confirmed by the time stability of the devices whose *I-V* characteristics showed no change or degradation when measurements performed soon after the fabrication were compared to measurements taken weeks later.

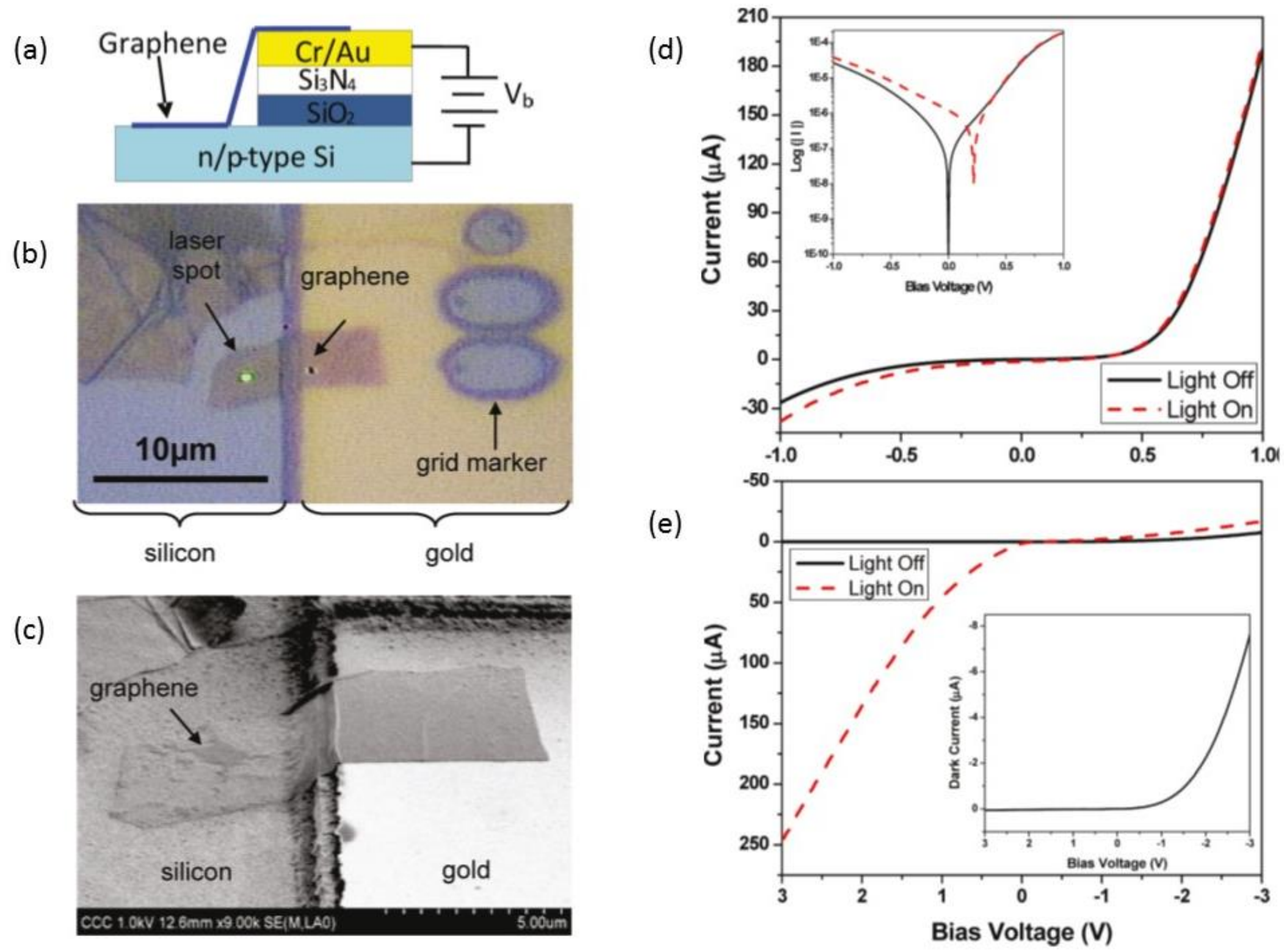

Fig. 11 - Graphene/Si Schottky diode of Ref. [132]. (a) Schematic diagram. (b) Optical image of bilayer graphene on Si and Cr/Au contact. The approximately 0.5 μm diameter 532 *nm* wavelength laser spot was used for spatially resolved photocurrent measurements. (c) SEM image of bilayer graphene on Si and Cr/Au contact. (d) Room temperature *I-V* characteristic of bilayer graphene/n-Si Schottky diode with and without uniform illumination of the graphene flake by a 30 *mW* laser (the inset shows the current on log scale). (e) Room temperature *I-V* characteristic of bilayer graphene/p-Si Schottky diode with and without uniform illumination of the graphene flake with a 30 *mW* laser (the inset shows the enlarged dark current-voltage characteristic). Figure adapted from Ref. [132].

Fig. 11(d) and (e) show the measured *I-V* characteristics at room temperature, with "on" state at positive bias for G/n-Si and at negative bias for G/p-Si. In both devices, graphene was a bilayer. Similar behavior was obtained at 100 and 400 *K* but with reduced or increased current, mainly due to the temperature dependent conductance of the silicon substrate. Also shown in Fig. 11 is the modification of the *I-V* curve when a 30 *mW*, 532 *nm* laser was uniformly shining on graphene. The increase of the current and the appearance of an open circuit voltage (more clearly shown in the inset of Fig. 11(d)) demonstrate photoexcitation of charge carriers and photocurrent generation. Thanks to the optical transparency of the thin graphene layer, incident laser light is expected to be absorbed in the Si substrate, where it generates the photocurrent observed under reverse conditions. Furthermore, using a 10 *mW* laser focused to a 0.5 $\mu m$ spot size (as shown in Fig. 11(b)), C.-C. Chen et al. [132] were able to measure 2D maps of photocurrent, by registering an *I-V* curve for each position of the spot onto and around the G/Si contact surface. The maps show very weak photocurrent generation in regions without graphene (where photogenerated *e-h* pairs mainly recombine without contributing to current) and higher photocurrent in the zone with graphene closer to the Au electrode. The photocurrent tend to linearly increase toward the Au electrode, a result that was interpreted in terms of series resistance: the in-plane resistance creates an effective load resistance in the circuit, which causes the measured photocurrent to be reduced when illuminating away from the gold electrode under the same laser intensity.
The dark *I-V* characteristics were interpreted in the framework of the Schottky theory eq. (33) and (35). From a fit of the forward characteristics, the ideality factors of several devices were found in the range 4.89-7.69 for n-Si and 29.67-33.50 for p-Si devices at room temperature (these high values suggest a "dirty" interface) . No obvious dependence of the ideality factor on the number of graphene layers was found. Using eq. (33), with $I_0$ at ZB, the SBH was estimated to be 0.41 *eV* on average for the bilayer graphene n-Si devices and 0.44 *eV* for the bilayer graphene p-Si devices at 300 K, which is consistent with the higher current density (at least one order of magnitude) measured for n-Si devices. The barrier height on n-Si device is consistent with the value measured in reference [117] for graphite/n-Si devices. A comparison of Fig. 11(d) and (e) shows that, unlike the dark current, the photocurrent at the same reverse bias is more pronounced on p-silicon devices. This is easily explained considering that the photocurrent increases with the width $w$ of the depletion layer at the G/Si interface (the larger *w*, the more is the light absorbed). According to eq. (21), *w* was larger in the p-Si since it had a lower doping level.
Remarkably, a decrease in the photocurrent was observed after an anneal in vacuum at 200 °C for 20 hours. The anneal is proven to transform the graphene, which was originally p-type because of exposure to air, into n-type [136]. This n-type doping was confirmed by a change in the Raman spectrum (upshift and line width narrowing of the G-band), and caused an increase of the SBH of 0.036 eV. This result implies that the G/Si SBH can be modified by intentional doping of graphene, an important feature that can be exploited for sensor applications.
Ref. [137] reported similar rectifying behavior for few-layer graphene/Si junctions both on n-Si and p-Si (Fig. 12). However, in this case, graphene flakes (with 4 layers on average) were synthesized by CVD of acetylene on a Mg-supported Fe-Co catalyst, purified with hydrochloric acid, washed and dispersed in a solution of N-methyl-2-pyrrolidone. The solution was then sprayed on Si substrates by airbrushing.

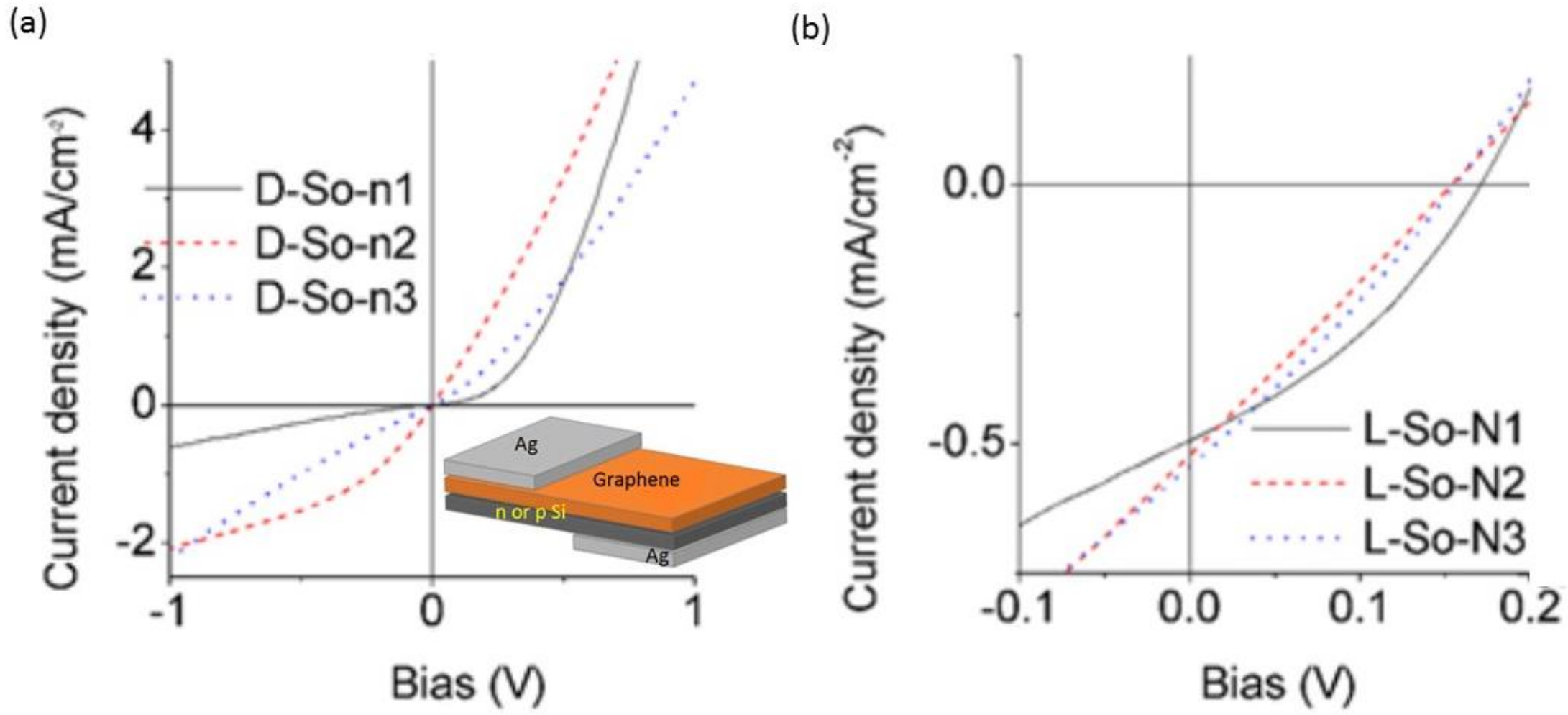

Fig. 12 - J-V characteristic for 3 few-layer-graphene/n-Si devices in dark (a) and under AM1.5 illumination (b). A schematic layout of the devices is shown in the insert of (a). Figure adapted from Ref. [137].

Rectification with an "on" to "off" current ratio of 10 to $10^{+3}$ at ±1V were achieved. Higher current densities were obtained with n-Si devices which had more doped substrate. The ZB SBHs, as extracted using eq. (33) and (35), were 0.52 to 0.67 *eV* for the G/n-Si and 0.61 to 0.73 *eV* for the G/p-Si devices, confirming the tendency to higher values of the SBH on p-Si reported by Ref. [132]. Moreover, upon white light illumination, the *I-V* curve shifted in the fourth quadrant (Fig. 12(b)), thus exhibiting a photovoltaic effects. At *AM1.5* a power conversion efficiency (PCE) of about 0.02% was observed on n-Si and less than 0.005% on p-Si. To improve this relatively low efficiency, Ref. [137] suggested increasing the conductivity of graphene by chemical doping.
The photovoltaic effect in the G/n-Si junction had been previously reported and studied in Ref. [138], with films of graphene sheets, synthesized by CVD on Ni, coated on substrates with pre-deposited Au contacts (see Fig. (21); we will review these devices later in the section on GSJ solar cells). The graphene films consisted of multiple layers of overlapped and interconnected graphene and formed on n-Si a diodes with time-stable *I-V* characteristics well described by the thermionic theory and with SBH ~0.78 *eV*, ideality factors ~1.5 and rectification ratio of $10^{+4} - 10^{+6}$. The average conversion efficiency under standard condition was ~1.5% with possible improvement through a better balancing of the conductivity and transparency of graphene films.

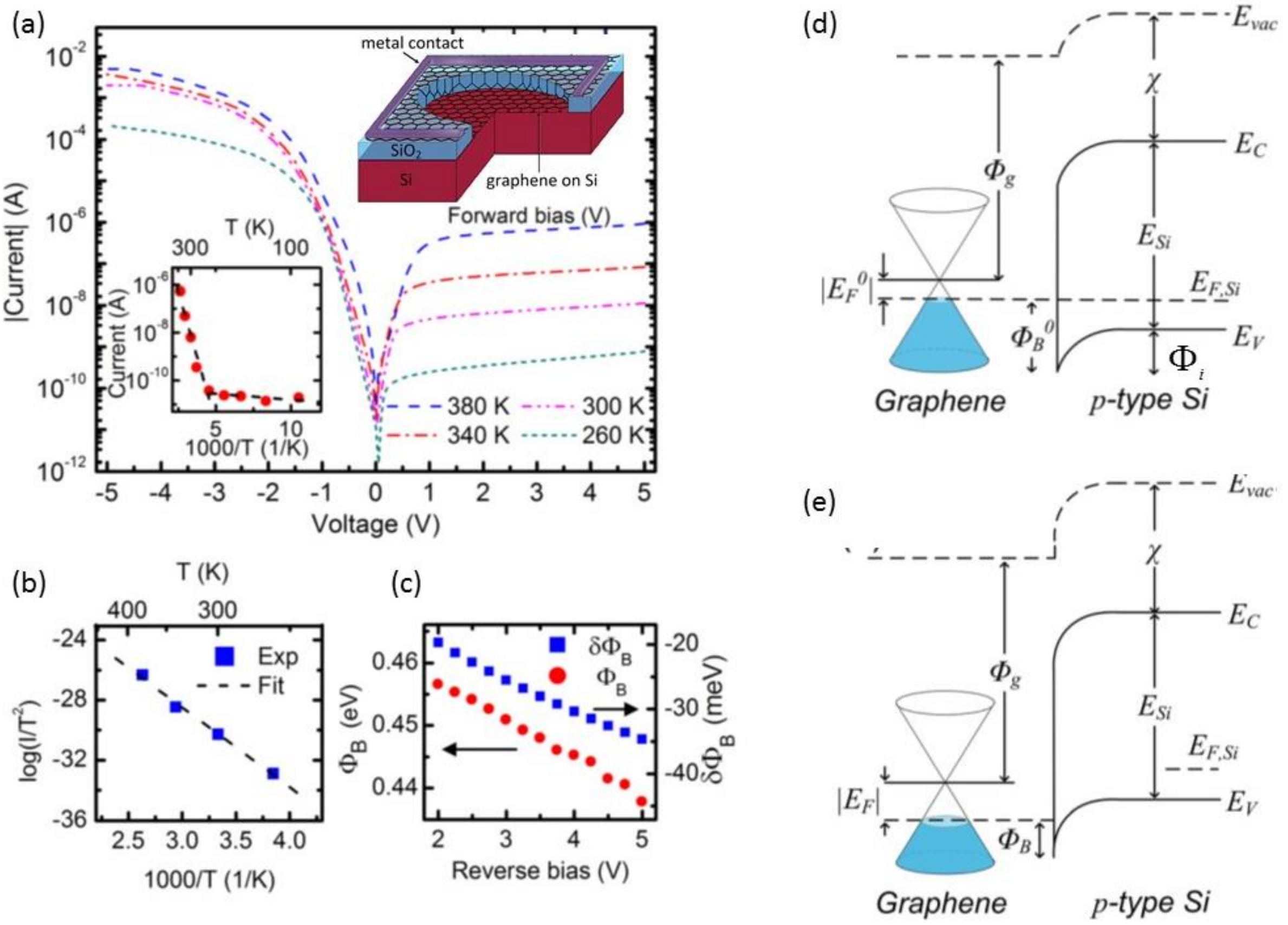

Fig. 13 - (a) Current-voltage characteristics of a G/p-Si Schottky junction ($2.5\times10^5\,\mu m^2$ area) at various temperatures ranging from 260 *K* to 380 *K*. The upper inset shows a schematic of the device. The lower inset is the Arrhenious plot (eq. 42) of the reverse saturation current at *2V* bias in the temperature range 95 *K* to 380 *K*. The change of slope shows that thermionic emission is replaced by tunneling at *T*<260 *K*. (b) Richardson plot for the device in part (a) with *I* at a reverse bias of 2V, in the thermionic emission regime. (c) Calculated SBH $\Phi_B$ (left y-axis) and the calculated change $\delta\Phi_B$ (right y-axis) in the SBH due to the Fermi level shift in graphene as a function of reverse bias V (see text). (d) Energy band diagram of the G/p-Si Schottky junction at zero bias and (e) under reverse bias *V*. $E_F$ is the graphene Fermi-level with respect to the Dirac point. Figure adapted from Ref. [139].

The properties of the G/p-Si junction (as well as their photoresponse, see next section) were studied in Ref. [139] where CVD graphene grown on Cu was used to form G/Si devices on moderately doped Si ($3\times10^{+16}\,cm^{-3}$). Graphene was transferred onto the Si/$SiO_2$ substrate and patterned by plasma etching into G/p-Si/G structures or G/p-Si junctions. Graphene lying on $SiO_2$ was contacted with Ti/Au (5

nm/50 nm). A schematic of the G/p-Si device is shown in the inset of Fig. 13(a). Fig. 13(a) shows the temperature-dependent *I-V* behavior of this device, studied in the range of 95 to 380 K.

The device has rectifying *I-V* characteristics, confirming the Schottky nature of the G/p-Si junction. For $T > 260$ K, the low forward-bias current and the reverse saturation current show the temperature dependence predicted by thermionic emission. However, below 260 K, as seen in the lower inset of Fig. 13(a), the temperature dependence becomes very weak. This is interpreted by Y. An et al. [139] as a change of transport regime, from thermionic to tunneling through the barrier. The SBH as extracted from the slope of the Richardson plot (eq. 42) of the saturation current at a reverse bias of *2V* and for T > 260 *K* (Fig. 13(b)), results to be $\Phi_B \approx 0.46$ eV. Considering the different bias, this value is in agreement with the one measured by [132] on low doped p-Si ($1.25\times10^{+14}$ $cm^{-3}$), indicating that a possible dependence of the SBH from the substrate doping is very weak. Furthermore, Y. An et al. [139] found that the SBH is dependent on the reverse voltage with an almost linear decrease with increasing reverse bias, as shown in Fig. 13(c). The dependence of the SBH on the reverse bias is a peculiarity of the G/S Schottky junction and will be reconsidered in the next section. Here, we anticipate that the variation of the SBH, $\delta\Phi_B(V)$, is due to the usual image-force barrier lowering observed in any M/S junction as well as to the dominant contribution stemming from the a-typical band shape of graphene. As shown in Fig. 13(d), representing the G/p-Si band diagram at zero bias, the graphene and p-Si Fermi levels aligned trough injection of holes from the semiconductor to graphene (graphene has a workfunction $\Phi_g$ lower than p-Si, $\Phi_g < \Phi_{p-Si}$, and the more energetic holes at the p-Si interface are transferred to graphene to establish thermal equilibrium). Due to the low density of states (see eq. (72) - (75)), this charge transfer causes a down-shift of the Fermi level of graphene, $E_F$. Such down shift is increased upon application of a reverse bias, and, as shown in Fig. 13(e), results in a bias dependent lowering of the SBH.

After the pioneering study on the graphite/semiconductor interface, S. Tongay et al. [140] too reported on the rectifying properties of the G/S junction. They used graphene synthesized on Cu foils via a multistep, low-vacuum process, transferred by PMMA over different n-type substrates: n-Si, n-GaAs, n-4H-SiC, and n-GaN (Gallium Nitride) with pre-deposited Cr/Au contact on the insulating layer (Fig. 14(a)). Fig. 14(b) shows the G and 2D peaks of Raman spectrum for graphene over GaN at different biases. A shift of the G band to higher and of the 2D band to lower wave numbers as well as a 2D/G peak ratio decrease is evident when a high reverse bias (*V=-10V*) is applied. The G and 2D peaks of graphene Raman spectrum are sensitive to the carrier density, and hence to the Fermi level [141]. In reverse bias, electrons are injected into graphene, which becomes electron-doped and its $E_F$ shifts up. The observed Raman shifts confirm the bias-induced changes of the graphene Fermi level ( $\Delta E_F \approx 0.2 - 0.5$ eV in this study). As previously noted, the control of the graphene Fermi level by the bias is an important feature of the GSJ.

*I-V* and *C-V* measurements were taken separately and confirmed the formation of junctions with strong rectification on all substrates. For all devices, both *I-V* and *C-V* characteristics followed the thermionic transport model. An example is given in Fig. 14(c)-(e) for a G/GaAs device. In forward bias, *I-V* follows eq. (39) over several decades (Fig. 14(d)), with deviations from linearity at higher bias due to the series resistance. Fig. 14(d) shows that the current increases with the temperature as expected since at higher temperature the probability of charge carriers to overcome the barrier increases. The zero bias saturation current $I_0$, extracted from the forward bias *I-V* curves, follows eq. (42) (Fig. 14(e)). Also $(1/C)^2$ *vs V* (not shown here) exhibits the linear dependence predicted by eq. (46). From the temperature-dependent measurements the following Schottky barrier heights were extracted at zero bias: 0.86 *eV*, 0.79 *eV*, 0.91 *eV*, and 0.73 *eV* for Si, GaAs, SiC, and GaN, respectively. When these values are used in Schottky-Mott eq. (8) to estimate the graphene workfunction $\Phi_g$ at zero bias, $\Phi_g \approx 4.8 - 4.9$ *eV* is obtained. These values are slightly higher that the commonly accepted value of 4.5-4.6 *eV* [142]-[145]. As explanation of the discrepancy, Ref. [140] invokes possibly lowering of $E_F$ caused by hole doping of graphene during the etching-transfer process or due to contact with the gold electrodes or the likely Fermi level pinning due to a high density of surface states.

Ideality factors in the range of 1.2–5.0 were reported with no obvious correlation to the type of substrate. (According to Ref. [140], ideality values exceeding unity might be associated with enhanced image-force lowering across the G/S interface.)

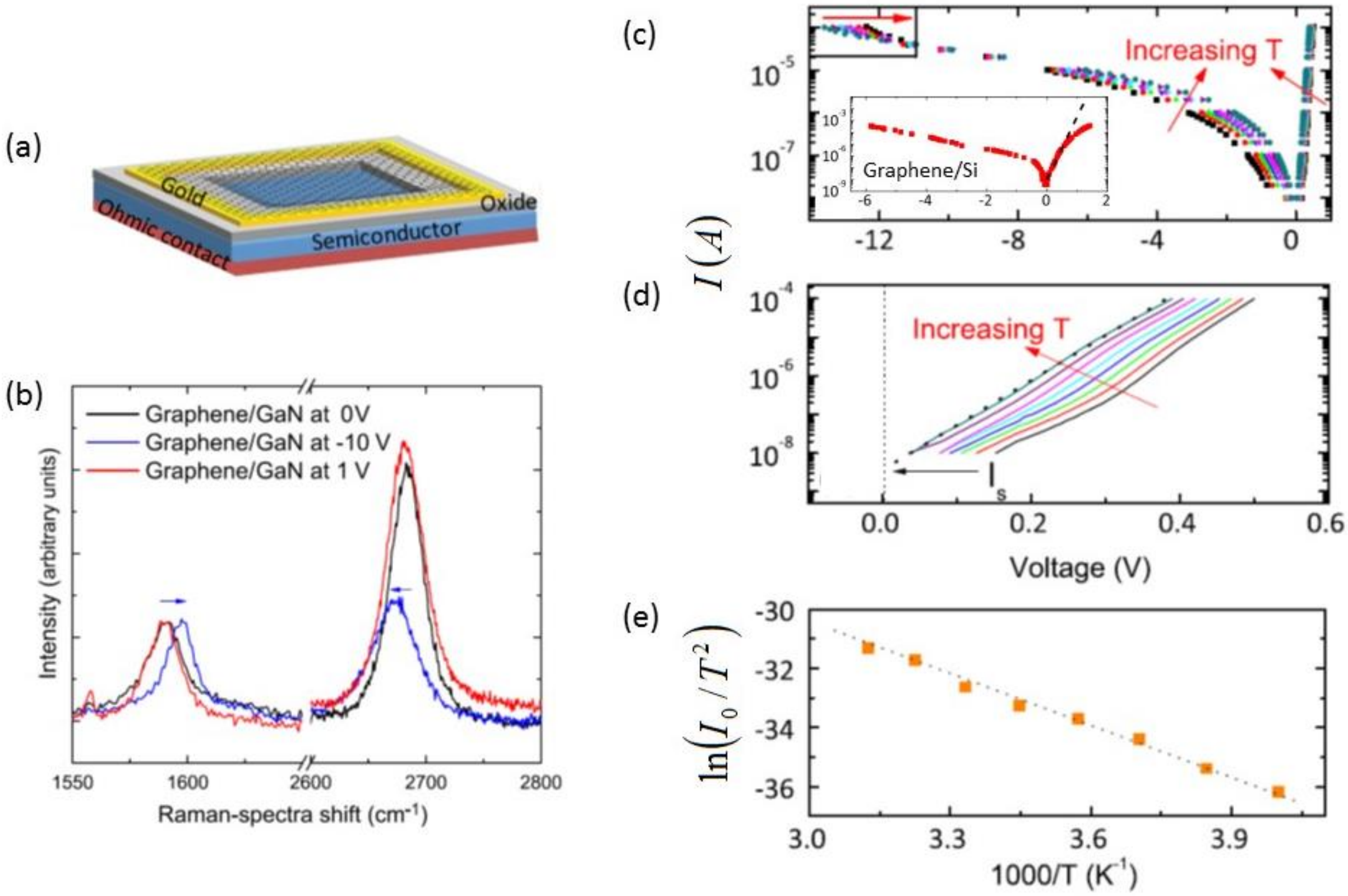

Fig. 14 - (a) Layout of G/S devices with predeposited Au contacts (b) *In-situ* Raman spectra taken on G/GaN device as a function of the applied bias. When the diode is reverse biased, the G and 2D peak positions shift relative to zero bias position (the direction of the shift is indicated by the arrow) (c) G/GaAs diode *I-V* characteristics for temperatures from 250 *K* up to 320 *K* (the direction of increasing temperature is indicated by the arrow). The inset shows the typical deviation from the linearity at high forward bias observed in all diodes. (d) Zoom in of the forward bias *I-V* curves. (c) Arrhenius plot with the extracted saturation current at zero bias plotted according to eq. (42). Figure adapted from Ref. [140].

Fig. 14(c) confirms that G/S diodes exhibit a stronger dependence of the reverse current on the reverse bias when compared with *p-n* or M/S Schottky diodes. As already underlined, and reported also in other experimental studies [139][146], the pronounced bias dependence of the reverse current is a peculiarity of G/S junction. Finally, Ref. [140] pointed out that a G/S Schottky junction can be used as chemical sensor since a small change in the SBH due to adsorbates on graphene can have a dramatic impact on the exponential forward current.

By focusing on the quality of the interface, D. Sinha and J. U. Lee [146] realized G/nSi Schottky diodes with $\eta \approx 1$. They fabricated G/S devices on n-Si substrates by transferring graphene grown on Cu by low-pressure CVD (Fig. 15(a)). To achieve ideal Schottky diodes, they reduced the metallic impurities brought by graphene at the G/Si interface. Impurities such as Ca, Si, Ru, Pt and Ce are present on the surface of the Cu film used in the CVD technique and can remain on graphene after the Cu is etched post growth. To minimize such metal contaminants D. Sinha and J. U. Lee etched the topmost ~700 *nm* of Cu using an ammonium persulfate solution for 90s. Fig. 15 shows the *J-V* characteristics of 3 types of Schottky diodes that were fabricated: the first (SB01) is the device with initial etching of Cu, and the second (SB02) is a device fabricated exactly as the previous one, but with no etching of Cu; a Cr/n-Si Schottky diode was included as reference.

Compared to the other devices, SB01 had an almost ideal behavior with the lowest leakage current at ZB; however, differently from the other devices, which have almost constant reverse saturation current, SB01 showed a clear increase of the reverse saturation current with the reverse bias.

The fit of eq. (35) in the forward bias range where the series resistance $R_S$ is negligible gives a $\eta = 1.08$ for SB01, indicative of a defect-free interface. The SBH and the Richardson constant $A^*$ were extracted from eq. (33) by varying the temperature in the range 300-380 $K$ with $J_0$ measured at different reverse voltages. $A^*$ was found significantly lower than the theoretical value of $112\,Acm^{-2}K^{-2}$ and dependent on the device type (Fig. 15(b)), while the results for the SBH are shown in Fig. 15(c). For Cr/n-Si and SB02 devices, $\Phi_B$ is ~0.37 $eV$ and ~0.57 $eV$ and remains almost constant with the reverse bias, indicating that the Fermi level in the semiconductor is pinned due to impurities at the interface. The small decrease with increasing reverse bias can be attributed to the image charge barrier lowering, as given by eq. (17). SB01 device instead shows a noticeable lowering of the Schottky barrier with reverse bias, suggesting that the cleaner interface does not pin the Fermi level and that some other mechanism takes place there. As seen in Fig. 15(b) the Schottky model clearly fails in accounting for the leakage increase in reverse bias of graphene devices, and in particular of SB01. To explain the unusually low value of $A^*$ in graphene devices and the bias dependence of the reverse current, Sinha and J. U. Lee suggested to combine the finite density of states (DOS) of graphene with a new transport model based on Landauer theory as we will see it in next section.

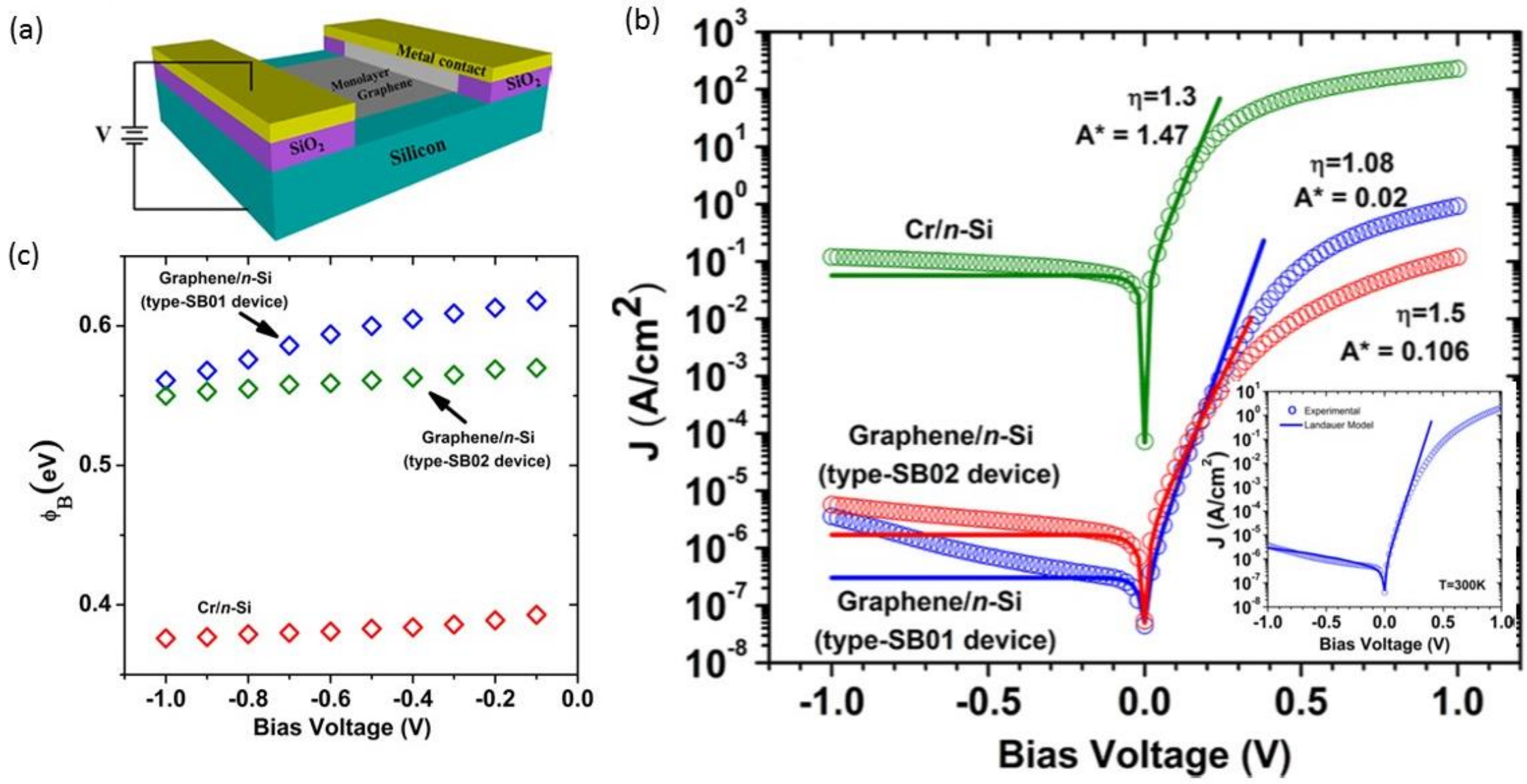

Fig. 15 - (a) Layout of the device of Ref. [146] (b) *J-V* characteristics of 3 types of Schottky diodes. SB01 was fabricated with initial etching of Cu to reduce remnants of metal contaminants. Such etching was not applied for SB02, which, apart that, was fabricated exactly as SB01. A reference Cr/n-Si diode was fabricated as well. The fits correspond to the Schottky model and shows poor agreement to the experimental data in reverse bias. The inset shows a better fit to the *I-V* characteristic of SB01 with a model based on Landauer equation, which will be discussed in the next section. (c) Extracted Schottky barrier height $\Phi_B$ as a function of the reverse bias. Figure adapted from Ref. [146].

The combination graphene-GaN has attracted considerable attention. GaN is a binary *III-V* direct bandgap semiconductor, with a bandgap of 3.4 *eV*, largely used since the 1990s for bright light-emitting diodes, high-power and high-frequency devices [147]-[149]. It has low sensitivity to ionizing radiation which makes it suitable for solar cells as well as for military and space applications where radiation tolerant devices are required. A number of studies have focused on graphene used as transparent conducting electrode on GaN and we will review some of these studies in the G/S application section. Despite the higher mobility of n-GaN (< 2000 $cm^2V^{-1}s^{-1}$) with respect to p-GaN (<150 $cm^2V^{-1}s^{-1}$) [150] which obviously make it best suited for device applications, the combination

G/n-GaN was considered less interesting since it was expected to form a Schottky barrier as low as ~0.4 *eV* and have a poor rectifying behavior because of the small graphene/GaN workfunction difference. Ref. [151] reported few layer graphene/n-GaN devices with excellent temperature stability. Their devices showed a stable rectifying behavior up to temperatures of 550 *K* with SBH~0.74 *eV* and ideality factor ~2.9. At higher temperatures, the reverse leakage current was observed to gradually increase until rectification was lost a $T > 650\,K$, as expected, since at high enough temperatures, electrons acquire enough energy to overcome the barrier in both directions. Furthermore, the same devices survived an extended (2 days) anneal at 900 *K* showing full recovery and improved rectification after anneal (SBH~0.70 *eV* and ideality factor ~2.7). The improvement was attributed to the anneal-induced removal of process-related residues from the graphene surface. Also, the reduction of non-thermionic emission current components, as evidenced by the lower ideality factor, was an indication of the removal of impurities at the G/n-GaN interface. The reduced SBH was the consequence of a change in the interface band due to dedoping which caused a shifts in graphene Fermi level as confirmed by Raman measurements. This experiment demonstrated the already mentioned thermal stability of graphene and its ability to be integrated in high performance analog devices operating at elevated temperatures. The findings in Ref. [140] and [151] ignited further interest in the G/n-GaN contact.

The electrical characteristics and the carrier transport mechanisms of the graphene on undoped GaN junction have been recently investigated by S. Kim et al. [152]. They fabricated Schottky diodes by transferring Cu synthesized graphene onto a GaN wafer, followed by oxygen plasma etching to accurately define the active junction region (Fig. 16(a)). A clearly defined GSJ area can help with the extraction of more accurate Schottky parameters. Fig. 16(b) shows the semi-logarithmic *I–V* characteristics of the device at room temperature and under dark conditions. Excellent rectifying behavior with a rectification ratio as high as $\sim 10^{+7}$ at ±2.0 *V* and low reverse leakage current ($1.1\times10^{-8}$ $A/cm^2$ at -5*V)*, lower than the previously reported values [140] and [151], were achieved, confirming the good fabrication process. Over a voltage range of 0.1–0.5 V, the theoretical fits of the forward *I–V* curves by eq. (33) and (35), with $R_S = 0$ and $A^* = 26.4A/(cm^2K)$, yielded $I_0 = 9.8\times10^{-14}A$, $\eta = 1.32$, and $\Phi_B = 0.90\ eV$. Although $\eta$ has the best value quoted so far, it was extensively studied and its dependence on voltage, as obtained from eq. (39) and (41), was used to extract the density of surface states originating from native crystal defects, interfacial layer, residual chemical contaminants, etc. The density of surface states was estimated in the range of ~ $10^{13}$ $states/(cm^2 eV)$, and with a peak at $E_c - 0.8\,eV$ (inset of Fig. 16(a)), which is the location where the Fermi level is expected to be pinned and is at the origin of the observed high Schottky barrier.

The SBH was estimated from the fit to the *I-V* curves at different temperatures, in the range 300-400 K, and was found to monotonically grow with temperature (table in Fig. 16(c)), while the same measurements used in a Richardson plot (Fig. 16(c)) gave a SBH barrier height $\Phi_B = 0.72eV$. To explain the temperature dependence of $\Phi_B$, S. Kim et al. [152] suggested the barrier inhomogeneity model [47][54][153]-[154], which considers a Gaussian distribution of the barrier heights around a mean value $\phi_{BM}$ and with standard deviation $\sigma$:

$$\phi_B = \phi_{BM} - \frac{q\sigma^2}{2kT}. \qquad (85)$$

Since transport across a G/S junction is a temperature activated process, for an inhomogeneous interface, at low temperature, transport is dominated by the low Schottky barrier patches. With increasing temperature more electrons have sufficient energy to surmount the higher barriers, hence the effective barrier increases with *T*.

Fitting eq. (85) to the measurements of $\phi_B$ at different temperatures, $\phi_{BM} = 1.24\,eV$ and $\sigma = 0.13eV$ were estimated (the $\sigma$ value is in agreement with the value obtained for a conventional metal contact on GaN).

Finally, S. Kim et al. [152] performed measurements under low frequency, weak illumination and observed a photocurrent (at reverse bias voltage) and an open circuit voltage, indicating photoexcitation of carriers and generation of photocurrent. They used light with energy $h\nu$ much lower than $E_g = 3.4eV$ of GaN; therefore the photocurrent that they measured was caused by electrons excited from the graphene over the barrier into the GaN.

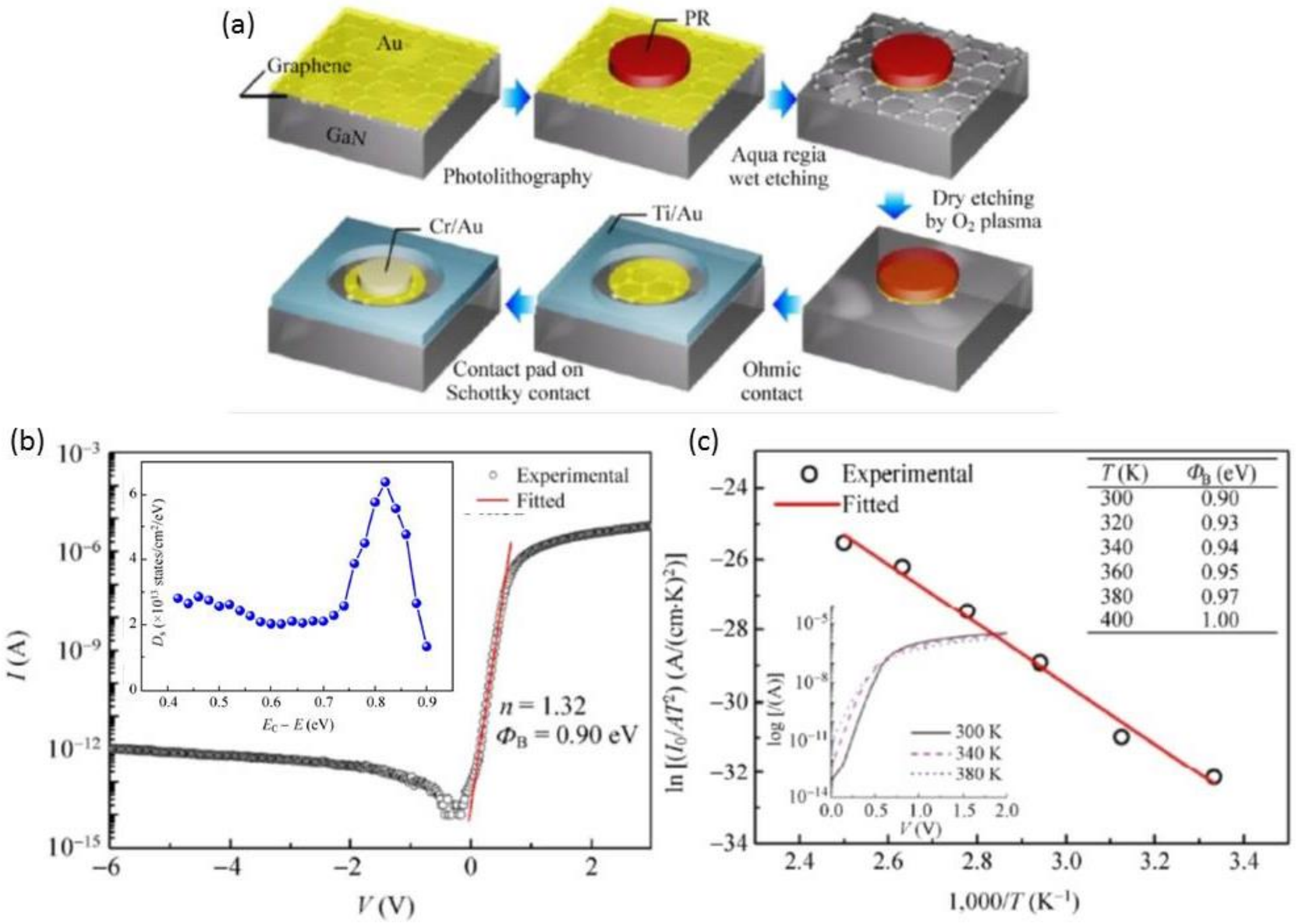

Fig. 16 - (a) Fabrication process of graphene-GaN Schottky diodes, showing that a photoresist (PR) is used for the patterning of graphene by dry etching for devices of small, definite area. Ti/Au forms ohmic contact on bare GaN while graphene, which is covered by Au, is accessed through a Cr/Au pad. (b) The *I–V* characteristics of a G/nGaN Schottky diode. The fit line refers to the thermionic model. The serious resistance (here $R_S \approx 7.6M\Omega$) is responsible for the saturation behavior at V > 0.6 *V*. The inset shows the interface state density *vs* energy, with a peak at $E_c - 0.8eV$. (c) Richardson plot of the Schottky diode over a temperature range of 300-400 *K*. The upper inset shows a table of $\Phi_B$ *vs T* as obtained by fitting the forward *I–V* curves using Eqs. (33) and (35). Figure adapted from reference [152].

The barrier inhomogeneity model of eq. (85) was used also by S. Parui et al. [154] to account for the temperature dependence of the SBH of the highly rectifying (more than $10^{+6}$) G/n-Si diodes that they fabricated both with exfoliated and large area CVD graphene. They measured *I-V* characteristics in the temperature range 80-300 *K* and extracted zero bias SBHs, which showed a monotonic increase from 0.34 to 0.69 $eV$ with increasing temperature, well fitted by eq. (85). Furthermore they reported a gradual shift of the threshold voltage (*i.e.* the forward voltage at which the current starts to increase sharply) towards lower biases when increasing the temperature, consistent with the increase of thermal energy of the electrons. S. Parui et al. [154] suggested that the SBH inohomogeneity could arise from local modification of the graphene workfunction by non-uniform interface charge distribution caused by potential fluctuations due to ripples, relative high conductance at the edges and graphene grain boundaries.

| Substrate | Device type | SBH (eV) | Ideality factor | Substrate doping ($cm^{-3}$) or resistivity (Ω·cm). Notes | Reference |
|---|---|---|---|---|---|
| n-Si | Graphite/n-Si | 0.40 | 1.25-2.0 | $1\times10^{15}$ $cm^{-3}$ | [117] |
| | Bilayer graphene/n-Si | 0.41 | 4.89-7.69 | $2.4\times10^{15}$ $cm^{-3}$ | [132] |
| | 4-layer graphene/n-Si | 0.52-0.67 | 1-2 | 0.295 Ω×cm | [137] |
| | Few-layer graphene/n-Si | 0.78 | 1.57 | $1.5\text{-}3\times10^{15}$ $cm^{-3}$ | [138] |
| | Few-layer graphene/n-Si | 0.6 | 1.94 | 4-6 Ω×cm. Si nanowires. | [209][183] |
| | Graphene/n-Si | 0.86 | 1.2-5 | $2\text{-}6\times10^{15}$ $cm^{-3}$ | [140] |
| | Graphene/n-Si | 0.62 | 1.08 | $1\times10^{16}$ $cm^{-3}$<br>CVD graphene on etched Cu | [146] |
| | Graphene/n-Si | 0.57 | 1.5 | $1\times10^{16}$ $cm^{-3}$ | [146] |
| | Graphene/n-Si | 0.69<br>0.83 | 1.46<br>2.53 | $1\times10^{15}$ $cm^{-3}$ , exfoliated graphene<br>$1\times10^{15}$ $cm^{-3}$ , CVD graphene | [154] |
| | Graphene/n-Si | 0.79 | na | 1-10 Ω×cm. $E_F$ pinned | [169] |
| | Graphene/n-Si | 0.32 | na | $0.1\text{-}1\times10^{16}$ $cm^{-3}$ | [174] |
| | Graphene/n-Si | 0.79 | 1.6-2 | $0.8\text{-}1\times10^{15}$ $cm^{-3}$ | [204] |
| | Graphene/n-Si | 0.89 | 1.3-1.5 | $0.8\text{-}1\times10^{15}$ $cm^{-3}$<br>Graphene doped with TFSA | [204] |
| | Graphene/n-Si | 0.833-0.858 | 1.56-1.58 | 0.05-0.2 Ω×cm | [216] |
| | Graphene/n-Si | 0.79 | 1.41 | $5\times10^{14}$ $cm^{-3}$ | [234] |
| | Graphene/n-Si | 0.71 | 3.7 | 1-10 Ω×cm | [236] |
| | Graphene/n-Si | 0.407 | 1.1 | $1\times10^{16}$ $cm^{-3}$ | [238] |
| | | | | | |
| p-Si | Bilayer graphene/p-Si | 0.44 | 29.67-33.50 | $1.25\times10^{14}$ $cm^{-3}$ | [132] |
| | 4-layer graphene/n-Si | 0.61-0.73 | 1-2 | 21 Ω·cm | [137] |
| | rGO/p-Si | 0.47 | na | 10 nm thick rGO flakes on p-Si | [221] |
| | Graphene/p-Si | 0.48 | na | $3\times10^{16}$ $cm^{-3}$ | [139] |
| | Graphene/p-Si | 0.51 | na | $0.1\text{-}1\times10^{16}$ $cm^{-3}$ | [174] |
| | Graphene/p-Si | 0.44-0.47 | 1.3-2.1 | 1-10 Ω×cm | [176] |
| | Graphene/p-Si | 0.74 | 1.31 | $1.5\times10^{15}$ $cm^{-3}$ | [234] |
| | Graphene/p-Si | 0.652 | 4.88 | 1-10 Ω×cm | [236] |
| | | | | | |
| GaAs | Graphite/n-GaAs | 0.60 | 1.25-2.0 | $3\times10^{16}$ $cm^{-3}$ | [117] |
| | Graphene/n-GaAs | 0.79 | 1.2-5 | $3\text{-}6\times10^{16}$ $cm^{-3}$ | [140] |
| | | | | | |
| GaN | Few-layer graphene/n-GaN | 0.74 | 2.9 | $1\times10^{16}$ $cm^{-3}$ | [151] |
| | Few-layer graphene/n-GaN | 0.70 | 2.4 | $1\times10^{16}$ $cm^{-3}$ .Two days 900K anneal | [151] |
| | Graphene/n-GaN | 0.73 | 1.2-5 | $1\text{-}3\times10^{16}$ $cm^{-3}$ | [140] |
| | Graphene/GaN | 0.72<br>0.90* | 1.32 | Undoped substrate.<br>*Eq (33)-(35) fit in forward bias | [152] |
| | Graphene/n-GaN | 0.33 | na | $6\times10^{18}$ $cm^{-3}$<br>barrier analysis with self-developed model with interface states | [155] |
| | Graphene/p-GaN | 0.40 | na | $4.5\times10^{17}$ $cm^{-3}$<br>barrier analysis with self-developed model with interface states | [155] |
| | | | | | |
| CdS | Graphene/n-CdS | 0.4* | 1.37 | n-CdS nanowires. * turn-on voltage | [202] |
| | | | | | |
| SiC | Graphite/n-4H-SiC | 1.15 | na | $1\times10^{16}$ $cm^{-3}$ | [117] |
| | Graphene/4H-SiC(0001) | 0.85<br>0.36 | na | Exfoliated graphene on SiC<br>Epitaxial graphene on SiC | [123] |
| | Graphene/n-4H-SiC | 0.91 | 1.2-5 | $1\text{-}3\times10^{17}$ $cm^{-3}$ | [140] |

Table 1: Schottky barrier height and ideality factor of graphene/semiconductor Schottky diodes

A summary of the SBH and the ideality factor of Schottky diodes made with graphene on various substrates is reported in Table I. Typically, the substrates used are lightly doped, with resistivity in the range $<10\,\Omega cm$. For a given semiconductor, the table shows substantial variability, indicating that the properties of the G/S junction are still strongly dependent on the fabrication technique and vary from sample to sample. Furthermore, different measurement techniques and sometimes different models are used to extract the SBH, which make the direct comparison of published results often meaningless. A manufacturing and measurement standardization is still far from achieving.

## 6. Modeling the graphene/semiconductor junction

An exhaustive theory of the G/S Schottky junction does not exist to-date, although few phenomenological models, explaining particular experimental features, have been proposed.
A first attempt to model the G/S junction was done by S. Tongay et al. in Ref. [140] who proposed a simple modification of the thermionic emission theory to include the mentioned dependence of the graphene Fermi level $E_F$ on the voltage bias and explain the bias-driven increase of the reverse saturation current. In metal/semiconductor diodes, the Fermi level of the metal stays constant upon application of a bias due to the high density of states of the metal. When the metal is replaced by graphene, a charge exchange between the 2D graphene, with limited density of states, and the 3D semiconductor induces a shift of the graphene Fermi level (a transfer of 0.01 electrons per unit cell shifts the Fermi level by 0.47 $eV$ [143]).
To show how this works, let us consider graphene on a n-type semiconductor and let us assume that at zero bias the Fermi level of graphene is at the Dirac point (Fig. 17 (a)). In forward bias, $E_F$ shifts down since the negative charge in graphene required to mirror the positive charge of the depletion layer of the semiconductor is reduced (Fig. 17(b)). This effect is small since a low forward bias is usually applied. In reverse bias, the depletion layer of the semiconductor increases substantially and so does the negative charge in graphene. This shifts the Fermi level upwards (Fig. 17(c)). The variation of the graphene Fermi level (and hence of the workfunction $\Phi_g$) modifies the SBH. As shown in Fig. 17, $\Phi_B$ is slightly increased in forward bias and decreased in reverse bias. The decrease of $\Phi_B$ in reverse bias causes the dependence on $V$ of the reverse current.

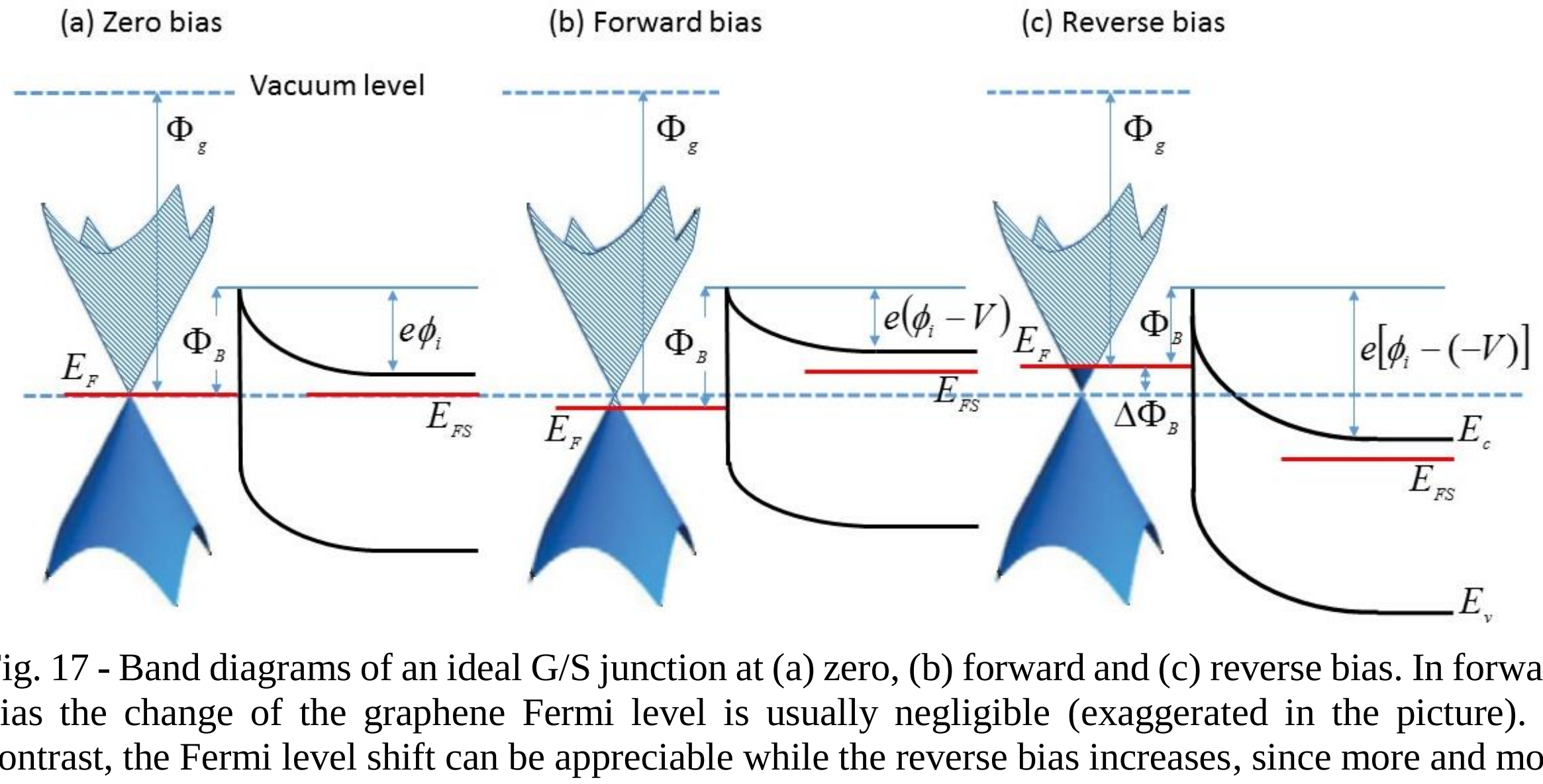

Fig. 17 - Band diagrams of an ideal G/S junction at (a) zero, (b) forward and (c) reverse bias. In forward bias the change of the graphene Fermi level is usually negligible (exaggerated in the picture). In contrast, the Fermi level shift can be appreciable while the reverse bias increases, since more and more charge is induced in graphene to mirror the immobile charge of the semiconductor depletion layer.

The bias induced variation of the graphene Fermi level, $\Delta E_F$, can be easily included in the thermionic model of the Schottky diode, by replacing the constant SBH in eq. (33) and (35) with a voltage dependent $\Phi_B(V)$. In such way, the original functional form of the diode equation is preserved and is generalized to allow the estimation of the SBH at any given voltage.

If the band alignment of Fig. 17(a) holds at zero bias, the density of carriers per unit area induced in graphene by the application of a voltage *V*, $n_{in}$, can be expressed considering that this density is opposite to the variation of the density of the positive donor ions per unit area in the depletion layer of the n-type semiconductor. Therefore, according to eq. (44), $n_{in}$ is given by

$$n_{in}(V) = -\Delta n_{depl.} = -\left(\sqrt{2\varepsilon_s N(\phi_i - V - kT/e)/e} - \sqrt{2\varepsilon_s N(\phi_i - kT/e)/e}\right). \qquad (86)$$

When $V > 0$, $n_{in}$ is positive, and graphene is hole doped (Fig. 17(b)); in reverse bias, $V < 0$ and $n_{in} < 0$, and the doping is n-type (Fig. 17(c)). From eq. (75), $n_{in}$ corresponds to the Fermi level variation, $\Delta E_F(V)$, given by

$$\Delta E_F(V) = \mp \frac{h}{2\sqrt{\pi}} v_F \sqrt{|n_{in}(V)|}, \qquad (87)$$

where the -/+ signs correspond to the down/up shift of $E_F$ in forward or reverse bias, respectively.

If graphene has a doping with density $n_0$ at zero bias ($n_0$ is negative for electrons and positive for holes), then the zero bias Fermi level is initially shifted with respect to the Dirac point by an amount given by eq. (87) with $n_{in}$ replaced by $n_0$. This doping can be due to the synthesis/transfer process, to interaction with the substrate, to exposure to air, etc. In this case, referring to the situation of Fig. 17 with positive spatial charge region for the semiconductor, the total carrier density per area in graphene caused by the application of a bias is

$$n_g(V) = n_0 + n_{in}(V) = -\sqrt{2\varepsilon_s N(\phi_i - V - kT/e)/e}, \qquad (88)$$

where $n_{in}(V)$ is the bias induced contribution to carriers. For simplicity, in eq. (88) and following we consider an ideal junction and neglect possible interface states (the inclusion of interface states would add an extra term proportional to $D_{it}(E_F - \Phi_0)$ to the right-hand side of eq. (88), with $D_{it}$ and $\Phi_0$ the density of interface states and the neutral level, as discussed by H. Zhong et al. [155]).

$n_{in}$ results in a variation of the Fermi level, which according to eq. (75), can be expressed as

$$\Delta E_F(V) = \frac{h}{2\sqrt{\pi}} v_F \left(\sqrt{|n_g(V)|} - \sqrt{|n_0|}\right) = \frac{h}{2\sqrt{\pi}} v_F \left(\sqrt{|n_0 + n_{in}(V)|} - \sqrt{|n_0|}\right). \qquad (89)$$

Obviously, $\Delta E_F = 0$ at zero bias, when $n_g = n_0 = -\sqrt{2\varepsilon_s N(\phi_i - kT/e)/e}$. Under the assumption that $n_{in} << n_0$, $\Delta E_F$ can be expressed as

$$\Delta E_F(V) = \frac{h}{4\sqrt{\pi}} v_F \frac{n_{in}}{\sqrt{|n_0|}}, \qquad (90)$$

which, using eq. (88), becomes

$$\Delta E_F(V) = \frac{h}{4\sqrt{\pi}} v_F \frac{1}{\sqrt{n_0}}(n_g - n_0) = a\left(\sqrt{(\phi_i - V - kT/e)} - \sqrt{\phi_i - kT/e}\right) \qquad (91)$$

with $a \equiv \frac{h}{4\sqrt{\pi}} v_F \sqrt{\frac{\varepsilon_s N}{2en_0}}$ .

The variation of the graphene Fermi level corresponds to an opposite variation of the SBH, $\Delta\Phi_B(V) = -\Delta E_F(V)$, as can easily be seen from Fig. 17. If $\Phi_B^0$ and $\Delta\Phi_B$ are the zero bias SBH and the correction to the ZB SBH due to the applied voltage $V$ (whether forward or reverse), then the voltage dependent Schottky barrier height can be written as:

$$\Phi_B(V) = \Phi_B^0 + \Delta\Phi_B(V) = \Phi_B^0 - \Delta E_F(V) = \Phi_B^0 - a\left(\sqrt{(\phi_i - V - kT/e)} - \sqrt{\phi_i - kT/e}\right). \qquad (92)$$

In the Schottky model, the constant $\Phi_B^0$ appears in the expression of the reverse saturation current. By replacing there $\Phi_B^0$ with $\Phi_B(V)$, the Schottky model equations (33) and (35) can be finally rewritten for the GSJ as

$$I_0 = AA^*T^2 e^{-\frac{1}{kT}\left(\Phi_B^0 + \Delta\Phi_B(V)\right)} = AA^*T^2 e^{-\frac{1}{kT}\left[\Phi_B^0 - a\left(\sqrt{\phi_i - V - kT/e} - \sqrt{\phi_i - kT/e}\right)\right]} \qquad (93)$$

and

$$I_0 = I_0\left(e^{\frac{e(V - R_S I)}{\eta kT}} - 1\right) = AA^*T^2 e^{-\frac{1}{kT}\left[\Phi_B^0 - a\left(\sqrt{\phi_i - V - kT/e} - \sqrt{\phi_i - kT/e}\right)\right]}\left(e^{\frac{e(V - R_S I)}{\eta kT}} - 1\right). \qquad (94)$$

According to eq. (93), since $\Delta\Phi_B(V = 0) = 0$, $\Phi_B^0$ can still be evaluated from the reverse saturation current $I_0$ at zero bias (extrapolation to *V=0* of forward bias current); however, once $\Phi_B^0$ is known, measurements of the reverse current at given *V* can be used to estimate the correction $\Delta\Phi_B(V)$.

Eq. (94) maintains the general form of the diode equation and includes the bias dependence of the reverse saturation current resulting from the Schottky barrier variation caused by the limited number of states of graphene around the Dirac point.

To analyze the *I-V* characteristics of their G/n-Si diodes (Fig. 15), trying in particular to explain the anomalously low $A^*$ and the strongly bias dependent reverse current, D. Sinha and J. U. Lee [146] realized that the details of the whole diode characteristics were best explained using the Landauer formalism and suggested that the injection rate from graphene ultimately determines the transport properties.

They wrote the Landauer transport equation (77) for the density of current $J$ [156] as:

$$J = -\frac{e}{\tau}\int_{-\infty}^{+\infty} T(E)D(E)\left(f_g - f_S\right)dE, \qquad (95)$$

where $\tau$ is the time scale for carrier injection from the contact, *T(E)* is the transmission probability over the zero bias barrier $\Phi_B^0$, $D(E) = D_0|E|$ is the graphene density of states of eq. (73), $f_g$ and $f_s$ are the Fermi functions of graphene and the semiconductor, respectively. Eq. (95) can be quite easily evaluated under the simplifying assumptions that $\Phi_B^0 >> kT$ (the condition needed for rectification) and that carriers with energy greater than the Schottky barrier are transmitted, while those with energy lower than $\Phi_B^0$ are reflected, *i.e.*

$$T(E) = \begin{cases} 1 \text{ for } E \geq \Phi_B^0 \\ 0 \text{ for } E < \Phi_B^0 \end{cases}. \qquad (96)$$

With these assumptions eq. (95) can be written as

$$J = -\frac{e}{\tau} D_0 \int_{\Phi_B^0}^{+\infty} E \left( \frac{1}{e^{(E-E_F)/kT}+1} - \frac{1}{e^{(E-E_{FS})/kT}+1} \right) dE . \qquad (97)$$

As can be seen in Fig. 17, for transmitted electrons $E - E_F \approx E - E_{FS} \geq \Phi_B^0 >> kT$. Then, the Fermi distributions can be approximated by a Boltzmann distribution, $\left(e^{(E-E_F)/kT}+1\right)^{-1} \approx e^{-(E-E_F)/kT}$, and

$$J = -\frac{e}{\tau} D_0 (kT)^2 \left( \frac{\Phi_B^0}{kT} + 1 \right) e^{-\Phi_B^0/kT} \left( e^{E_F/kT} - e^{E_{FS}/kT} \right). \qquad (98)$$

When a bias $V$ is applied to graphene with respect to semiconductor, then $E_F - E_{FS} = -eV$ and eq. (98) ultimately yields an equation formally identical to the ideal diode one:

$$J = J_0 \left( e^{eV/kT} - 1 \right), \qquad (99)$$

where the reverse current $J_0$ depends on $T$, $\tau$ and $V$:

$$J = \frac{e}{\tau} D_0 (kT)^2 \left( \frac{\Phi_B^0}{kT} + 1 \right) e^{-\Phi_B^0/kT} e^{E_F/kT} = \frac{e}{\tau} D_0 (kT)^2 \left( \frac{\Phi_B^0}{kT} + 1 \right) e^{-\Phi_B/kT} . \qquad (100)$$

Here, as in eq. (92), $\Phi_B = \Phi_B^0 - E_F(V)$ is the bias dependent SBH. According to eq. (100), the Richardson constant is now

$$A^* = \frac{e}{\tau} D_0 k^2 \left( \frac{\Phi_B^0}{kT} + 1 \right), \qquad (101)$$

and depends on $\tau$, on the temperature as well as on the ZB SBH. $\tau^{-1}$ is the injection rate of carriers from the contact to graphene (which is also the injection rate from graphene to the semiconductor) and is related to the coupling energy which controls the contact resistance (larger coupling energy corresponds to lower contact resistance). According to this model the injection rate from contact to graphene plays a key role in the transport properties of a G/S junction.

## 7. Applications of graphene/semiconductor Schottky diodes

*(a) Photodetectors.*

A general overview of state-of-the-art photodetectors based on graphene (and other two-dimensional materials) was recently published by Koppens et al. [157]. Here we solely focus on photodetectors based on the graphene/semiconductor junction.

G/S Schottky diodes, when operated under reverse bias, can be used as photodetectors. In these devices, optical absorption takes place mainly in the semiconductor, and graphene (which is the exposed side of the junction) acts as optically transparent and anti-reflecting carrier collector. Absorption in graphene may become important at lower energies, *e.g.* for $IR$ radiation.

Devices with graphene as optical absorber, as metal-graphene-metal (MGM) or FET structures, although highly appealing for ultrafast applications, suffer the limitations of the weak absorption ( $A \approx 2.3\%$ ) and the short lifetime of the order of picoseconds of photogenerated carriers in graphene [158]-[159] and the small effective photodetection area. These limitations result in low quantum efficiency and photocurrent responsivity, which in absence of a gain mechanism, remains limited to few tens $mA/W$ ($10-20\,mA/W$ in the wavelength range $400\ nm \le \lambda \le 1550\ nm$) [160]-[168]. In MGM devices, photo-generated carriers are captured by the electric field of the graphene-metal contacts, so only the small fraction of carriers generated very close to the contacts can be collected by the external circuit, while the rest of the carriers generated in graphene quickly recombines without any contribution to the external photo current. As a result, the effective photo detection area in MGM detectors is restricted to narrow regions adjacent to the graphene-metal interface. G/S Schottky diodes can effectively address these issues and produce a higher responsivity with the further benefit of a semiconductor compatible technology. In GSJs, light is absorbed in the thicker depletion layer of the semiconductor and the effective photo detection area is only restricted by the G/S contact area. The separation and transport of photo-generated carriers happen in the depletion layer of the semiconductor; if the photogenerated carriers are able to reach the graphene, they will contribute to the photocurrent, independently of their excitation location along the sensible area and without the issue of the fast of *e-h* pair recombination in graphene. The important point here is that once photogenerated carriers are injected into graphene by the built-in electric field, these carriers can survive much longer than intrinsically photoexcited pairs in graphene. Their lifetime is related to the probability of being back injected into the semiconductor. Furthermore, if the semiconductor has a slow intrinsic recombination time, this can further reduce the recombination rate. As results the lifetime of carriers photogenerated at the G/S junction can be several order of magnitude higher (up to milliseconds) than the lifetime in graphene. Consequently, G/S Schottky junctions can have a higher quantum efficiency that MGM devices; the external quantum efficiency is in the range of 50% and 65% depending on the wavelength. Actually, as we will discuss in this section, GSJ can even have an intrinsic gain mechanism resulting in giant responsivity. Another important and unique feature of the GSJ is that responsivity can be tuned by the applied reverse voltage bias which makes these devices an ideal platform for fast and sensitive photodetection at variable brightness.

An extensive study on the photodetection properties of the G/n-Si Schottky junction was performed by X. An and coworkers [169] who investigated devices with CVD-grown monolayer and few-layer graphene on lightly doped n-Si (Fig. 18(a)), built with a scalable and CMOS-compatible fabrication process.

In the dark or under low power illumination (less than few $\mu W$), the *I-V* characteristics of the junction follow the conventional photodiode behavior, as shown in Fig. 18(b), while at higher power some anomalies are observed (Fig. 18(c)). From detailed measurements of the Schottky barrier heights made using graphene, doped graphene and Ti/Au on Si, the Fermi level of the substrate was found pinned to the charge-neutrality level by its own surface states, with a Schottky barrier height $\Phi_B \approx 0.8\,eV$. Fig 18(d) shows the energy band diagram at thermal equilibrium and in dark condition. Here graphene is assumed neutral and the Fermi levels of graphene and n-Si are denoted as $E_F$ and $E_{FS}$, respectively. Incident photons generate *e-h* pairs in Si. In steady state, these excess carriers can be accounted for by introducing quasi-Fermi levels, separately for holes and electrons, near the valence and conduction band edges. The quasi-Fermi levels in Si, indicated as $E^q_{FS,n}$ and $E^q_{FS,p}$, are defined such that, in the steady state, the concentrations of electrons *n* and holes *p* (including the photogenerated carriers) are:

$$n = N_c e^{-\frac{E_C - E^q_{FS,n}}{kT}}, \qquad (102)$$

$$p = N_v e^{-\frac{E^q_{FS,p} - E_V}{kT}}. \qquad (103)$$

At low power, the electron quasi-Fermi level of n-Si is slightly above the dark Fermi level $E_{FS}$ and below $E_c$, $E_{FS} \le E^q_{FS,n} < E_C$, while the *quasi-Fermi* level for holes is below $E_{FS}$ and above $E_V$: $E_{FS} > E^q_{FS,p} > E_V$. The separation of $E^q_{FS,n}$ and $E^q_{FS,p}$ increases with power, since the photogenerated *e-h* pairs (excess carriers) move $E^q_{FS,n}$ and $E^q_{FS,p}$ towards $E_C$ and $E_V$, respectively.

Excess (photogenerated) holes from Si are injected into graphene and, similarly to Si, can be described by a quasi-Fermi level $E_{F,p}^{q} < E_F$ ($E_F$ is the dark Fermi level). $E_F$ is lowered with respect to the bulk Si bands if a forward bias is applied to graphene, and so is $E_{F,p}^{q}$ (Fig. 18(e)). However $E_{F,p}^{q}$ cannot go lower than the quasi-Fermi level of silicon $E_{FS,p}^{q}$. This limits the number of accessible states for the injection of photogenerated holes from Si to graphene. At low incident powers, all the photogenerated holes can find accessible states in graphene to inject into, resulting in the conventional photodiode-like response. $E_{F,p}^{q}$ lies between $E_F$ and $E_{FS,p}^{q}$, $E_{FS,p}^{q} \le E_{F,p}^{q} \le E_F$ and no limiting condition is reached. Something different happens at high incident power, since there are not enough states in graphene for the photogenerated hole injection and the *I-V* behavior deviates from that of a conventional photodiode. Fig. 18(c) shows that for high light powers (up to 6.5 $mW$) there is a strong suppression of photocurrents close to $V = 0$, and a sharp rise and rapid saturation of photocurrents at low reverse biases. At forward bias, the lowering of the graphene Fermi level reduces the range for $E_{F,p}^{q}$ which is easily brought to the limit $E_{FS,p}^{q}$. This greatly diminishes the number of accessible states for the photoexcited holes to inject into graphene from Si (the available states are those for which $E_{F,p}^{q} \ge E_{FS,p}^{q}$, corresponding to the small red area of Fig. 18(d)). Hence, under a forward bias, an increasing incident power quickly aligns $E_{F,p}^{q}$ to the quasi-Fermi level of holes in Si, $E_{F,p}^{q} = E_{FS,p}^{q}$, and this condition stops further injection. Increasing the incident light power beyond this point disallows any further photogenerated holes to inject into graphene, and no change in the forward *I-V* characteristic is observed.

Vice-versa, a reverse bias on graphene lifts $E_F$ with respect to the bulk silicon bands, opens up a larger number of accessible states for the holes to inject into (corresponding to the wider red part of the cone in Fig. 18(e)) and allows the collection of all the injected holes. As a result, the photocurrent, can completely recover under small reverse biases. At high reverse bias (with enough density of states to accommodate all photoexcited holes), the photocurrent is limited by the photogeneration rate which results in a saturated current, as shown in Fig. 18(c)).

At high power, in the region of low reverse bias, the maximum allowed photocurrent (related to the condition $E_{F,p}^{q} \ge E_{FS,p}^{q}$) can be controlled by the reverse bias itself. Fig. 18(c) shows that in this region the current is proportional to the reverse bias. This bias tunable photocurrent responsivity is an attractive feature of a GSJ photodetector which can be adapted for variable-brightness imaging.

An alternative explanation of the current suppression shown in Fig. 18(c) at low forward biases was proposed by Y. Song et al. [170], who considered the interfacial native oxide as a key knob to control the junction current, following a previous work on CNT/Si devices [171]. They suggested a model where the native oxide and the relatively low work function of graphene cause significant carrier recombination that suppresses the photocurrent and results in a "s-shaped kink" at low bias as that observed in Fig. 18(c). We will review this model in the section on solar cells.

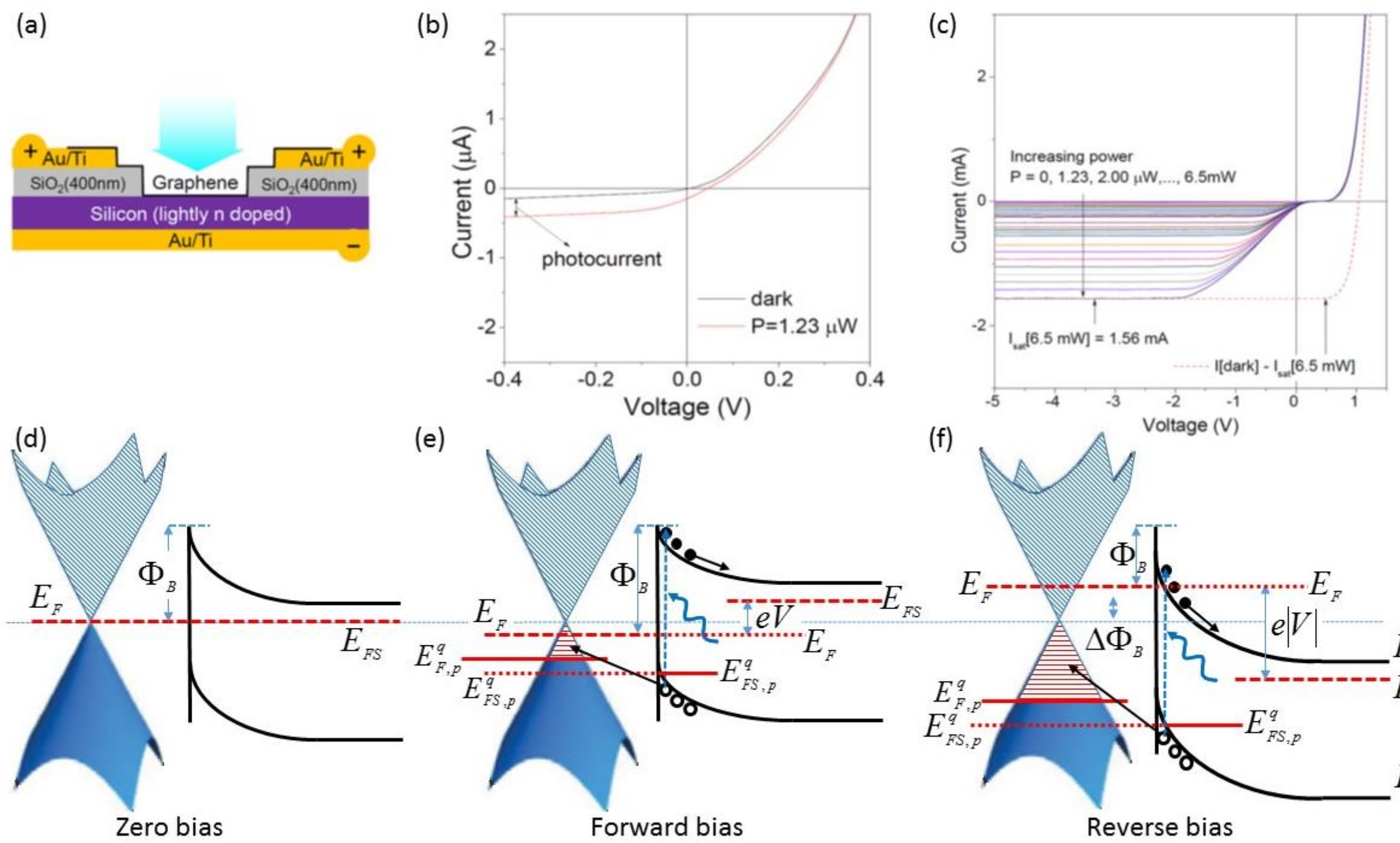

Fig. 18 - (a) Schematic of the monolayer G/Si junction device of Ref. [169]. (b) *I-V* curves of the device under darkness and weak illumination (P = 1.23 μW, λ = 488 *nm*) showing a conventional photodiode-like behavior. (c) Deviation of the *I-V* curves from a conventional photodiode response as the incident light power is increased up to *P = 6.5 mW* (the red dashed line corresponds to the expected behavior of a conventional M/S diode). Figure adapted from reference [169]. (d) Thermal equilibrium energy band diagram in darkness (the Fermi level of n-Si is pinned to the charge neutrality level of its own surface states and $\Phi_B \approx 0.8eV$ ). Band diagrams and Fermi level $E_F$ in darkness (dashed line), and quasi-Fermi level at high irradiation power ( $E_F^q$ , continuous line) under (e) forward and (f) reverse bias (the subscript S is used for silicon).

X. An et al. [169] demonstrated devices with high photovoltage responsivity $R_V$ (see eq. (51)) exceeding $10^{+7}\,V/W$ at low power (~10 *nW*). This very high responsivity renders G/Si Schottky diodes competitive devices for weak signal detection. At the lowest power of 10 *nW*, a $NEP \sim 1pW/Hz^{0.5}$ was measured, confirming the low detection limit of these devices (a power as low as $1pW$ can be detected above the noise level, when integrated over 0.5*s*). The corresponding detectivity (eq. 55) was $D^* = 7.69 \cdot 10^{+9}\,Jones$. To further characterize the sensitivity of the device to small changes of incident power, the photovoltage responsivity (or contrast sensitivity defined as $dV_{ph}/dP_{in}$ ) was used, largely resulting independent of the device area and with the remarkable value of $10^{+6}\,V/W$ at low light intensities. In the photocurrent mode, at a given negative bias (-2V), the response was found to remain linear over at least six decades of incident power, with a photocurrent responsivity (see eq. 50) up to 225 *mA/W*, which is at least 1-2 order of magnitude higher than that reported for Ge or Si photodetectors [172]-[173].

The time required to switch, when the incident light is turned on or off, was a few milliseconds, which is quite appealing for applications such as high speed photography, videography, etc.

The measured maximum quantum efficiency was *EQE* ~ 57% over at least 4 order of magnitude of the incident power ($10^{-3}$ to 1 *mW*).

According to the proposed model, the device performances can be further improved by controlling the number of graphene layers and the graphene doping. Layer-thickening provides more states for injection of holes while doping the graphene can increase the sheet conductance thus improving the photocurrent responsivity.

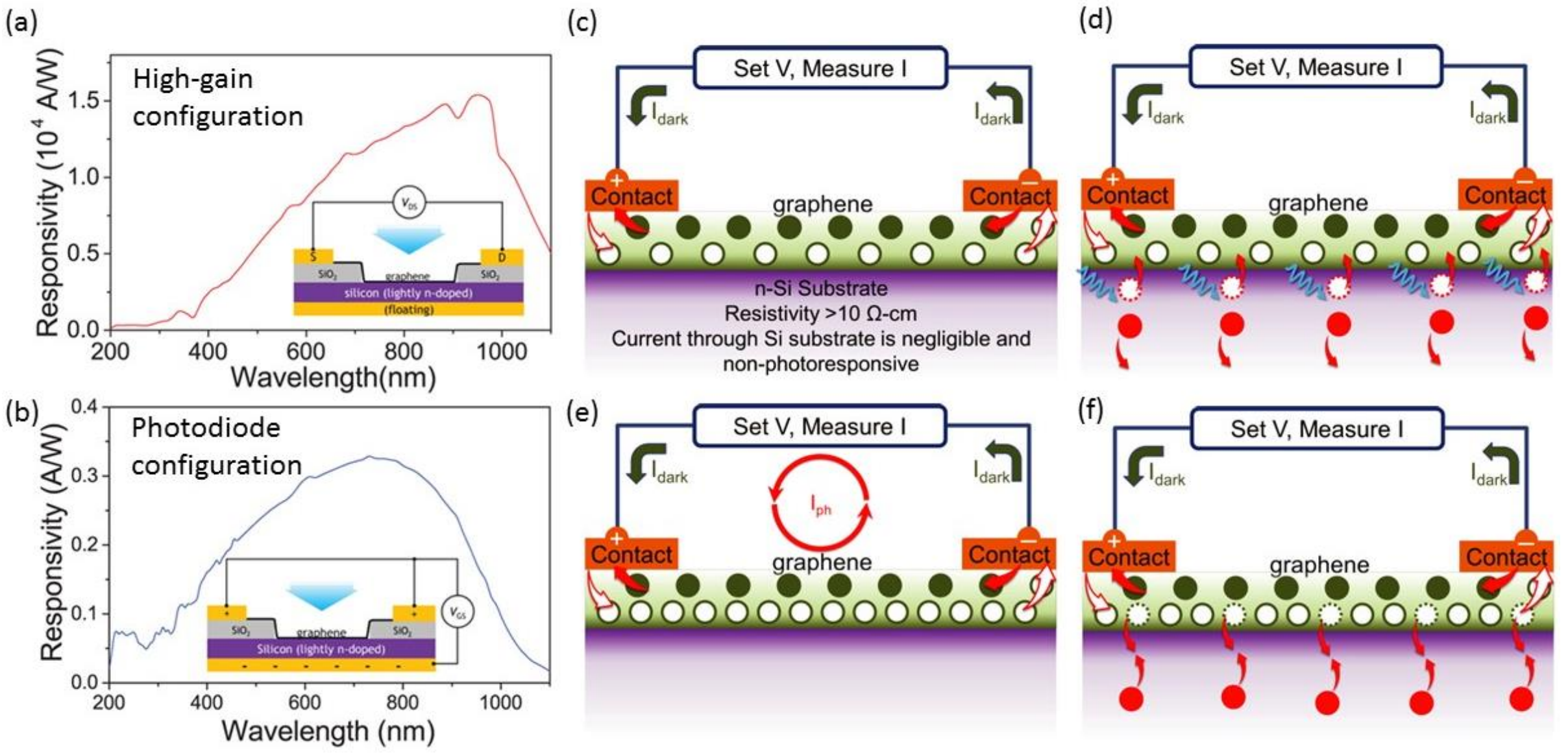

Fig. 19 - Spectral responsivity of the device (shown in the inset) of Ref. [169] in (a) high-gain mode and (b) photodiode mode. A dramatic difference in their magnitudes is observed, even though the spectral shapes are similar. (c)-(e) Schematics outlining the gain mechanism at the G/Si junction (see text for explanation). Electrons and holes are denoted by dark and light circles, respectively. In (c) the dark current $I_{dark}$ is due to intrinsic carriers in graphene. (d) Incident photons generate *e*-*h* pairs in the (lightly n-doped) silicon. Holes are swept into graphene by the built-in electric field of the junction and contribute to the current in the external circuit. Due to the fast transit time of graphene and the low probability of back injection in silicon, a single injected photocarrier can circulate several times (e) and substantially contribute to the current before recombine (f). This mechanism generate an internal quantum gain. Figure adapted from [174]

F. Liu and S. Kar [174] recently demonstrated that devices similar to those of Ref. [169] can operate with ultra-high responsivity, up to $10^{+7}\ A/W$, also in current mode. The devices, when operated in a "horizontal" (or high gain) configuration (see inset of Fig. 19(a)) rather than in vertical photodiode mode (see inset of Fig. 19(b)), are capable of an internal gain mechanism which they call Quantum Carrier Reinvestment (QCR). The QCR exploits the ultrafast transition of photogenerated carrier within graphene and the relatively large recombination time scale in the G/Si system, which can exceed a millisecond, to achieve ultrahigh quantum gain values. F. Liu and S. Kar [174] achieved quantum gains greater than $10^{+6}$ electrons per incident photon and responsivities approaching $\sim 10^{+7}\ A/W$.

Fig. 19(a) and (b) compare the responsivity of the same device when operated in high gain mode (a) and in photodiode mode (b). As we have already pointed out, the intrinsic photocurrent in graphene is limited to a few $mA/W$, while the photocurrent responsivity seen in Fig. 19(a) is many order of magnitude higher. This implies that the photocurrent of the device under study (no matter the configuration) has insignificant contributions from photocarriers generated in graphene. At the same time, the spectral shape of the responsivity curve in Fig. 19(a) is similar to that of Fig. 19(b), which is obtained in photodiode mode, clearly indicating that most of the photoinduced carriers originate in silicon and then get injected into graphene. The model developed to account for the far higher efficient photon-to-charge conversion in the gain-mode configuration is sketched in Fig. 19(c)-(f). This qualitative model is based on the photoinduced injection of carriers from Si into graphene. Compared to graphene, the lightly n-doped Si is highly nonconductive and under an applied external bias $V_{DS}$ (as in Fig. 19(a)), and in darkness, a dark current $I_{dark}$ flows through the external circuit due to the intrinsic carriers in graphene, as shown in Fig. 19(c). When light is shined on the device, photogenerated *e*-*h* pairs are separated by the built-in electric field: electrons move away into the body of silicon while holes get injected across the junction into graphene, as shown in Fig. 19(d). These additional holes, injected into graphene, add a current in graphene and in the external circuit, denoted as $I_{ph}$ in Fig. 19(e). Due to the extremely rapid transit of carriers within graphene, a single injected hole can "circulate" (in the sense that a hole can be removed by an electrode and replaced by a hole injected by the other electrode) many times before another "equivalent" hole reverse-injects across the junction into silicon (Fig. 19(f)). During the lifetime of an injected hole carrier, $\tau_r$, a time scale determined by the quantum-mechanical probability of recombination, the hole can be "reinvested" several times into the external circuit, adding to the net photocurrent, and leading to a quantum gain (QCR mechanism).

The quantum gain is easily estimated by introducing the source-to-drain transit time $\tau_t$. The ratio of the recombination time $\tau_r$ and $\tau_t$ is the number of times that each injected carrier is "circulated" or "reinvested". Reminding that the external quantum efficiency $EQE$ (see eq. (48)) is the number of carriers generated in Si and injected in graphene per incident photon, the quantum gain, $QG$, can be expressed as the $EQE$ times the number of "reinvestments" of each hole:

$$QG = EQE \times (t_r / t_t). \qquad (104)$$

On the other hand, in the simple diffusion transport model, the transit time can be related to mobility of graphene and to the external bias as

$$t_\tau = \frac{L}{v_d} = \frac{L}{\mu E} = \frac{L^2}{\mu V}, \qquad (105)$$

where L is the inter-electrode distance. The photocurrent responsivity $R_I$ (eq. 50) can be obtained as

$$R_I = EQE \frac{e}{h\nu} \tau_r \left( \frac{\mu V}{L^2} \right). \qquad (106)$$

According to eq. (106) the photocurrent response is linearly dependent on the applied drain-source bias, inversely proportional to the square of the device size, and directly proportional to the recombination time scale of the system. Each of these dependences were experimentally tested and verified in Ref. [174].

The photodetecting properties of Cu-CVD graphene on p-type Si (heavily doped $3 \times 10^{+16} cm^{-3}$) were reported in Ref. [139]. Photocurrent measurements were performed on interdigitated structures graphene/p-Si/graphene (Fig. 20).

The dark *I-V* characteristics at room temperature for the graphene/p-Si/graphene structure in the bias range from -3V to 3V, reported in the inset of Fig. 20(d), shows the typical behavior expected for two back-to-back Schottky diodes. The symmetry of the *I-V* indicates also that the Schottky junctions formed by opposite fingers are uniform.

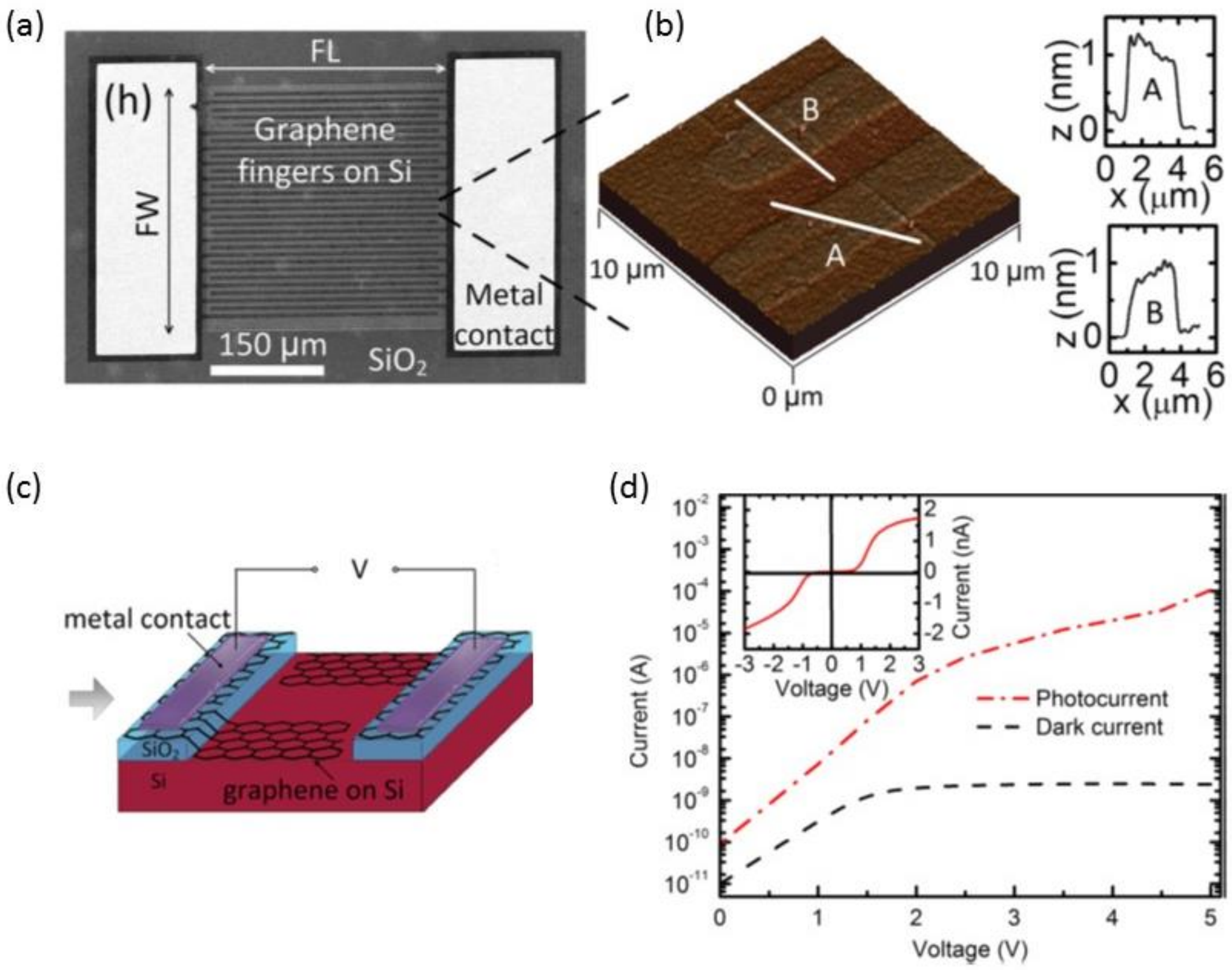

Fig. 20 - (a) SEM image of a fabricated graphene/p-Si/graphene interdigitated device (finger width and spacing is $5\mu m$, active area is $300\times300\,\mu m^2$). (b) AFM image of patterned graphene fingers on Si. The height profiles along "A" and "B" lines are shown on the right (*x* is the distance along the lines and *z* is the height) (c) schematic of the interdigitated device. (d) Dark current and photocurrent as a function of bias voltage measured on an interdigitated photodetector with finger width and spacing of $10\mu m$ and active area $400\times400\,\mu m^2$. The photocurrent is measured under 633 *nm* He-Ne laser illumination with 5.1 *mW* power and $\sim830\,\mu m$ spot size. The inset shows the dark *I-V* characteristic at positive and negative bias. Adapted from Ref. [139]

To characterize the photoresponse of such photodetectors, the device was irradiated with a He-Ne laser (*633 nm* wavelength, *5.1 mW* power, and $\sim830\,\mu m$ spot size) at room temperature. Fig. 20(d) shows the dark and the photocurrent of the same device as a function of voltage bias. Under illumination the current increases by about five orders of magnitude at 5V bias, from which a current responsivity $R_I \approx 110\, mA/W$ was evaluated. This value is of the same order of magnitude of that obtained for graphene on n-Si, thus confirming the great potential of the interface G/Si for low power photodetection. An often reported metric for MSM (metal/semiconductor/metal) photodiodes is the normalized photocurrent-to-dark current ratio (*NPDR*) defined as

$$NPDR = \frac{I_{ph}}{I_{dark}\cdot P_{inc}} = \frac{R_I}{P_{inc}}, \qquad (107)$$

which resulted $4.55\times10^{+4}\, mW^{-1}$, a value higher than that reported for similar devices fabricated with carbon nanotubes or Ti/Au on p and n-Si [175], due to a lower dark current.

Graphene-Si Schottky junctions have been proposed and demonstrated also as sensitive infrared (*IR*) detectors [176]. Infrared radiation in the C band of 1528-1561 *nm* and the L band 1561-1620 *nm* is of great interest and importance for optical communications. Differently from the detection of visible light where graphene is used as a transparent electrode and photoconversion happens in the Si depletion layer, at *IR* wavelengths the photocurrent is generated in graphene.
The great advantage of the G/Si diode is that all the photo-generated carriers in graphene, independent of their excitation location, have a similar chance to be separated and transferred to Si. In fact, whenever the generated carriers pass through the thin layer of graphene, they are swept by the electric field of Si depletion layer and contribute to the photocurrent. Furthermore, the relatively short lifetime of the photo-generated carriers of graphene is not a limitation in this case because the thin graphene layer is shorter than the mean recombination length of the carriers and the photo-generated carriers in graphene in all parts of the junction have high probability to be separated at the interface with Si before recombining.
Ref. [176], in particular, studied the effect on *I-V* characteristics of an exfoliated G/p-Si Schottky junction under 1550 *nm* excitation laser. The reverse current of the junction (with a ZB barrier in the range 0.44-0.47 *eV*) increased under irradiation, corresponding to a photocurrent responsivity $R_I = I_{ph}/P_{inc} = 2.8-9.9 mA/W$ respectively at -5 *V* and -15 *V* reverse bias (the responsivity increases with reverse bias, but so does the level of noise). For comparison, all-Si detectors at the same wavelength have responsivity at least an order of magnitude lower [177].
Monolayer G/Si junctions were studied as near-infrared photodetectors also in Ref. [178], by fabricating a device able to operate at zero external voltage bias because of a strong photovoltaic behavior of the G/Si Schottky junction. A responsivity and detectivity $R_I = 29\, mA/W$ and $D = 3.9\times10^{11}\, cm\sqrt{Hz}/W = 29\, mA/W$ were measured at room temperature. The device showed great potential for low light detection with intensity down to $\sim 1\, nW\, cm^{-2}$ at 10 *K* and a fast time response with speed of $100\, \mu s$, which allowed the device following a fast varied light with frequency up to $2100\, Hz$.
The possibility of using G/Si junctions for near to mid-infrared detection was further investigated in Ref. [179], which demonstrated the use of in-plane absorption in a graphene-monolayer structure and the feasibility of exploiting indirect transitions in G/Si junction waveguides for mid-infrared detection. A graphene/silicon photodiode was formed by integrating graphene onto a silicon optical waveguide on a silicon-on-insulator (SOI). The waveguide enabled absorption of evanescent light that propagates parallel to the graphene sheet, and resulted in a responsivity as high as $R_I = 130\, mA/W$ at 1.5 *V* bias for 2750 *nm* light at room temperature. A photocurrent dependence on bias polarity was observed and attributed to two distinct mechanisms for optical absorption, that is, direct and indirect transitions in graphene at 1550 *nm* and 2750 *nm*, respectively.
The potential of G/Si junctions for detection of *THz* frequencies is at the moment under study.

*(b) Solar cells*
Solar cells have become very popular in recent decades as alternative energy source, due to enormous energy obtained by earth from the sun ($\sim 1.2\times10^{+17} W$). The current Si-based photovoltaic devices require high-quality raw materials and quite costly processing techniques to increase power conversion efficiency (*PCE*). Improvement in efficiency and cost reduction are relentlessly pursued to promote extensive solar cell applications. In this context, graphene is considered primarily as the conductive transparent electrode far long sought to replace the expensive ITO (indium tin oxide) (or FTO, fluorine tin oxide) electrode commonly used in solar cells for charge injection or collection. ITO suffers from many drawbacks, such as high cost, limited resource of indium, brittleness, etc., while graphene has the advantage of a better transparency to UV/blue light, a higher thermal conductivity as well as chemical stability, mechanical strength and flexibility. Furthermore, graphene has a low sheet resistance $R_{sh}$, which although depending on the production method, can be tuned by doping or by the number of layers. Thus, graphene enables easy tradeoff between high transparency and low $R_{sh}$ requested in solar cell [180]-[184]. Graphene films with $R_{sh}$ of few hundred ohm per square at about 80% optical transparency have been demonstrated on photovoltaic devices [185].
Other than as electrode, graphene microsheets have been dispersed into conjugated polymers to improve exciton dissociation and charge transport in organic solar cells [186]-[187]. Recent reviews on graphene applications in solar cells are found in Ref. [188]-[190].

In this section we focus on specific applications of the G/S junction as photovoltaic power generator. Photons absorbed in the semiconductor depletion layer generate *e-h* pairs that are separated by the built-in electric field and can be collected at the graphene and semiconductor contacts, thus generating power. In addition to the high transparency and to the controllable electrical conductivity, graphene allow the work function, and hence the characteristics of the GSJ, to be tuned as desired to optimize the solar cell *PCE*. G/Si solar cells are considered as among the lower-cost candidates in photovoltaics due to the simple fabrication process and to the already competitive conversion efficiency, with values around ~10% or ~15% with the addition of antireflective coatings. The techniques developed for G/Si can be applied to other semiconductors (as GaAs, CdSe, etc.) with enhanced optical properties.

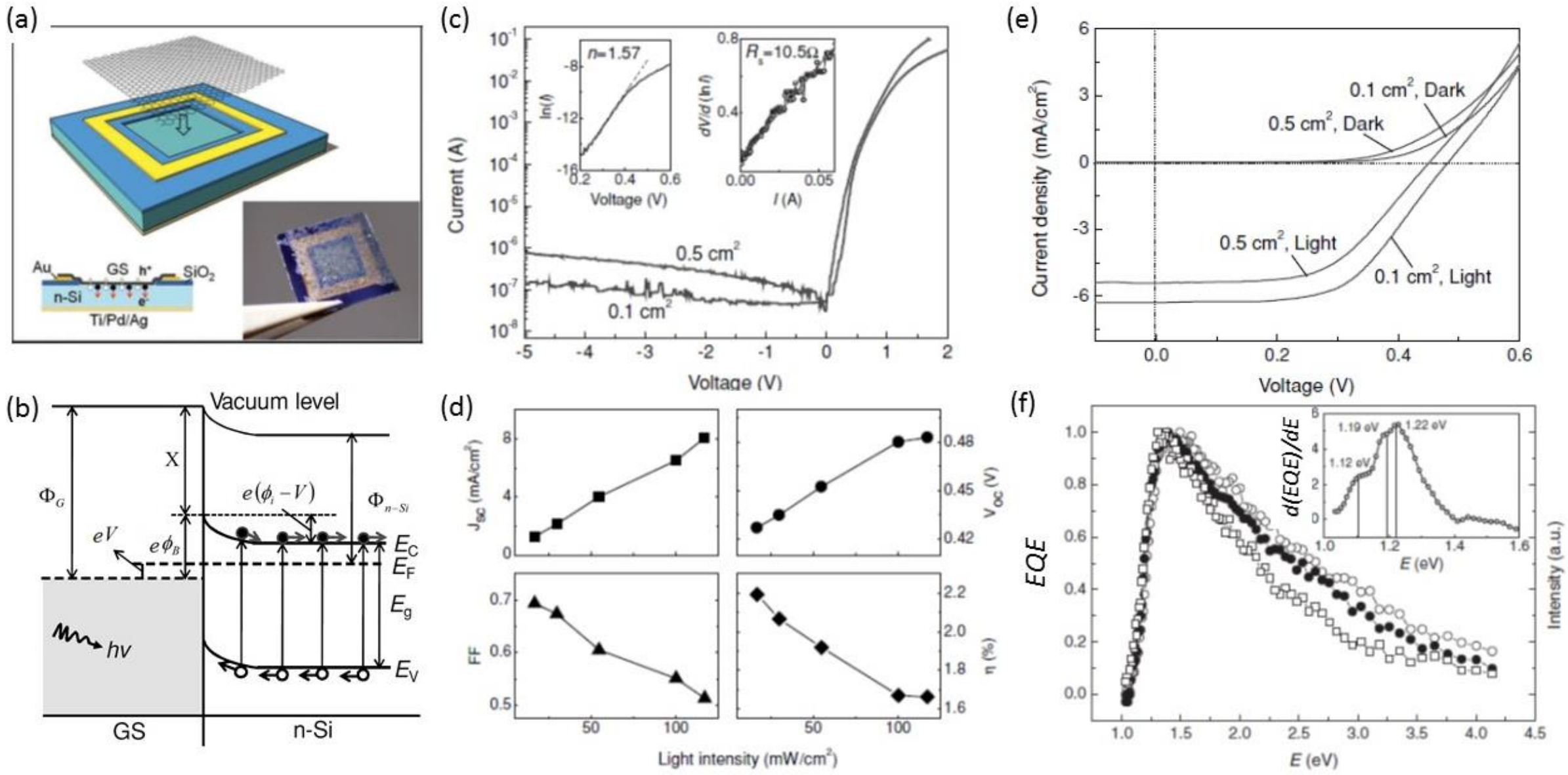

Fig. 21 - Solar cell with films of graphene on n-Si. (a) Layout and a photograph of the devices. (b) Energy diagram of forward-biased G/n-Si Schottky junction upon illumination (c) *I-V* characteristics of two devices (0.1 $cm^2$ and 0.5 $cm^2$) showing excellent rectification. The insets show the ideality factor and the series resistance of the 0.1 $cm^2$ cell extrapolated from the forward linear region d) Solar cell parameters ( $J_{SC}, V_{OC}$ , *FF*, and *PCE*) *vs* light intensity for the 0.1 $cm^2$ G/n-Si cell (e) *J–V* curves of cells illuminated at *AM1.5* equivalent light (f) External quantum efficiency (*EQE*) *vs* photon energy (the inset shows the differential *EQE* spectrum) for cells with different area. Figure adapted from Ref. [138].

The photovoltaic properties of the carbon/Si interface were initially studied with the successful formation of p-type a-C/n-Si heterojunctions (a-C denotes a diamond-like amorphous carbon film) which showed efficiencies less than 1% at *AM1.5* [192]-[193]. Further development involved carbon nanotubes (CNTs)/Si junctions. The photovoltaic properties of films of CNTs on Si have been extensively studied with efficiencies varying from few percents to 14% [194]-[202]. Typically films of CNTs combines two types of heterostructures (Schottky and p-n), since both metallic and semiconducting nanotubes coexist in as-grown materials. A drawback of films composed of CNT networks is the interspace between bundles which reduces the conductivity of the film and the effective absorbing area, even though this can be advantageous for transparency. In this regard, graphene films are advantageous since they can easily guarantee better electrical connection and full area coverage with superposing flakes.

One of the first studies on G/n-Si Schottky junction for solar cells was reported by X. Li et al. [138] with devices consisting of a conform and continuous film of graphene sheets coated onto a patterned n-Si/$SiO_2$ substrate with Au contacts (Fig. 21(a)). Individual sheets were formed mostly by mono-layer, bilayer and few layer graphene. The graphene film served as transparent and anti-reflecting electrode for light illumination (reflection was reduced by ~70% in the visible region and ~80% in the near-*IR*) as well as active layer for *e-h* separation and hole transport. The built in field responsible for the photogenerated charge separation was estimated correspond to a barrier $\Phi_i = 0.55 - 0.75\,eV$ . The corresponding band-structure is shown in Fig. 21(b). The *I-V* characteristics of the devices were highly rectifying (rectification ratio of $10^{+4} - 10^{+6}$) (Fig. 21(c)) and with $\ln I$ nearly linear in the range of 0.1-0.4 *V*, corresponding to a diode ideality factor $\eta = 1.57$ .

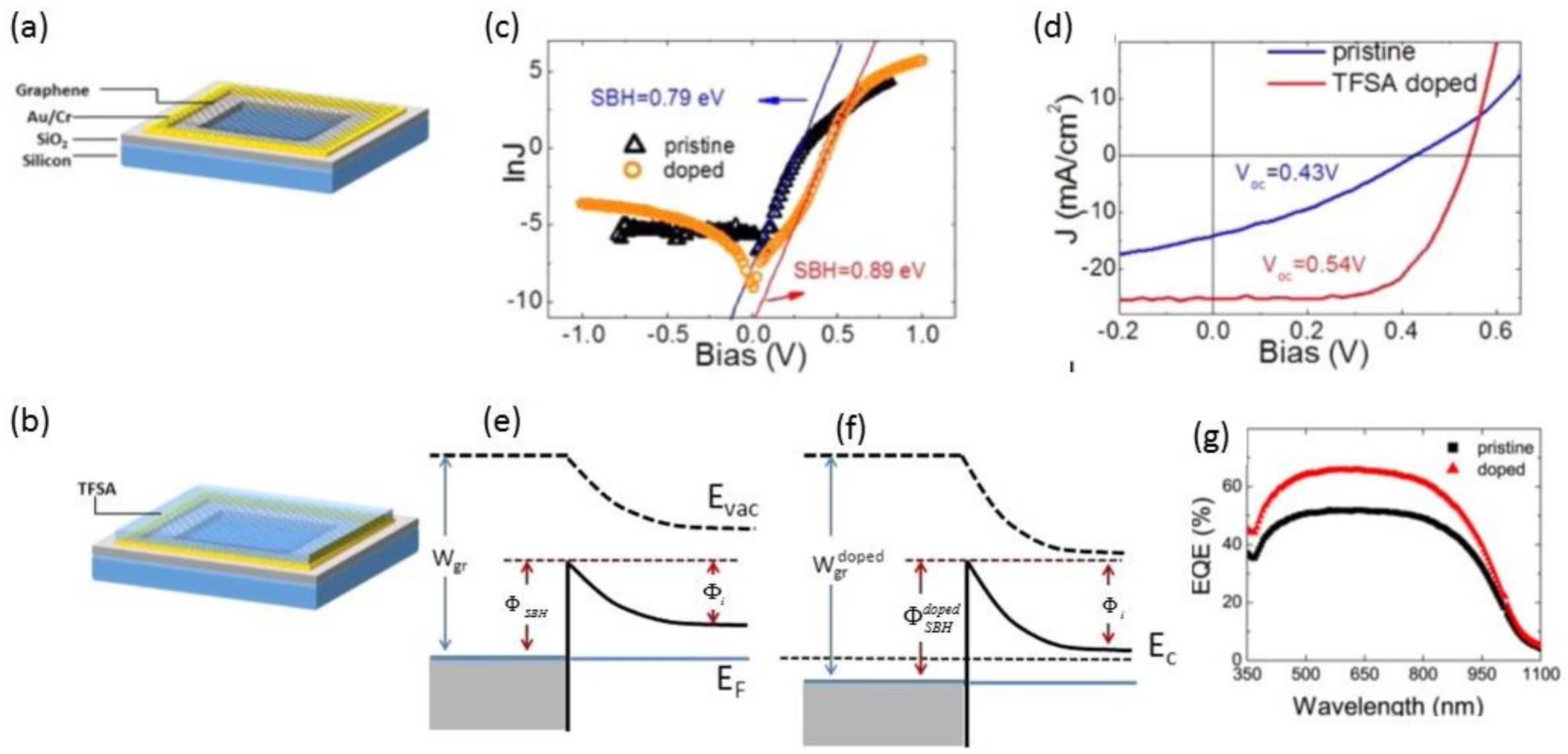

Fig. 22 - Graphene/n-Si solar cell (a) without and (b) with TFSA (c) *J-V* characteristics of pristine and doped solar cell. (d) Zoom in of *J-V* characteristics. (e) band diagram at the G/n-Si interface before (e) and after doping of graphene (f). External quantum efficiency *vs* wavelength (g). Fig. adapted from Ref. [205].

The reverse leakage current is nearly proportional to the area of the contact as the leakage current is restricted to the area of n-Si directly under the graphene sheet. The ZB Schottky barrier was estimated as 0.75-0.8 *eV*. The photovoltaic properties of the cell were characterized at *AM1.5* illumination (Fig. 21(e)) yielding an open-circuit voltage $V_{OC} = 0.42 - 0.48V$ , a short-circuit current density $J_{SC} = 4 - 6.5\,mAcm^{-2}$ and a *FF* of 45-56%, which corresponds to an overall *PCE* =1-1.7% (Fig. 21(d)). Both $V_{OC}$ and $J_{SC}$ depend linearly on the light-intensity incident on the cell, consistent with a systematic increase in photogenerated carriers. The *EQE vs* photon energy is shown in Fig. 21(f); its first derivative, displayed in the inset, presents a sharp peak around 1.2 eV, which represents the fastest photon-to-electron conversion and can be assigned to the bandgap of silicon.

The *PCE* of these devices was almost an order of magnitude lower than that for reported for CNTs on Si, but it was obtained without any balancing of the conductivity and the transparency of graphene sheets and no optimization of the G/Si interface. Ref. [138] also reported the successful series and parallel combination of the cells to multiply $V_{OC}$ and $J_{SC}$ respectively, thus checking the feasibility of arrays of solar panels.

The photovoltaic effect of a single graphene sheet on p-or n-Si was observed by C.C. Chen et al. in Ref. [132]. Solar cells were also demonstrated with graphene on CdS [203] and CdSe [204] nanowires or nanobelts, with *PCE* up to 1.65% at *AM1.5*.

Single layer G/n-Si solar cells with higher *PCE* were reported by X. Miao et al. [205]. Such cells exhibited a *PCE* of ~1.9% under *AM1.5* illumination, which was increased till 8.6% with chemical doping by bis(trifluoromethanesulfonyl)-amide (TFSA), $((CF_3SO_2)_2NH)$. TFSA, which has the advantage of environmental stability due to its hydrophobic nature, provides graphene with p-doping, thus reducing its sheet resistance and increasing its work function, without changing its optical properties [206]. The result is an increase of $V_{OC}$ , $J_{SC}$ , *FF* and *EQE* which benefit from increased hole SBH and built-in potential, and reduced series resistance $R_s$ (see the model of Fig. 5(e)).

The devices were fabricated by transferring CVD graphene grown on Cu foils on Si/$SiO_2$ substrates with pre-deposited Au/Cr contacts (Fig. 22(a) and (b)). Processing included exposure of bare Si to ambient air, for up to 2 h, to take advantage from oxygen passivation of dangling bonds that reduces surface states and improves the cell performance, as reported for conventional MIS cells [207]. The role of interfacial native oxide on Si is further discussed at the end of the section. Doping of graphene was accomplished by spin-casting TFSA.

Fig. 22(c) shows the *I-V* curves of pristine and TSFA-doped solar cells. The zero bias Schottky barrier, obtained from eq. (33) and eq. (35) in the forward region of negligible $R_s$, is increased of about 0.1 *eV* by TFSA doping (from 0.79 *eV* to 0.89 *eV*). The doping down-shits the graphene Fermi level and increases the SBH and the built-in barrier, $\Phi_i$ (Fig. 22(e) and (f)). Greater $\Phi_i$ and widened depletion layer allow a more efficient collection of photogenerated charge. Furthermore, according to eq. (60), the increase of the SBH results in a larger $V_{OC}$ which is beneficial for *EQE*. Fig. 22(d) also shows an increase of $I_{SC}$ for doped device and a higher *FF*, which can be partially attributed to the improvement in graphene electrical conductivity and the associated reduction of ohmic losses. The *PCE* is increased by doping from 1.9 to 8.6%. Doped devices also show a reduced ideality factor (from 1.6−2.0 to 1.3−1.5 range). Higher ideality factor is likely caused by charge puddles on the graphene that are unintentionally formed during the process and which give rise to associated Schottky barrier inhomogeneities. The controlled doping of graphene by TFSA can possibly yield more uniformly doped regions, thus reducing the Schottky barrier inhomogeneity and the ideality factor.

Finally Fig. 22(g) shows the external quantum efficiency before and after doping. The *EQE* of the pristine cell is similar to state-of-the-art Si solar cells as expected since only Si absorbs and converts photons in *e-h* pairs [208]. After TFSA-doping, the *EQE* is increased over 60%, representing a ∼30% enhancement compared to the pristine cell as result of the more efficient charge separation and charge collection and the reduced $R_s$, as we have pointed out.

The doping approach to enhance the *PCE* has been pursued in several other works. Some of them use nitric acid ($HNO_3$) for its p-type chemical doping effect on graphene. Ref. [209] reported devices consisting of a heterojunction of graphene-based woven fabric on n-Si, where the graphene-based woven fabric served as transparent conducting window electrode, hole collector and anti-reflection layer. It was prepared by interlacing two sets of graphene micron-ribbons with ribbons passing each other essentially at right angles. By using a woven copper mesh as template, the graphene-based woven fabric grown from chemical vapor deposition retains the network configuration of the copper mesh. Such devices showed a *PCE* of ~3% which after $HNO_3$ doping was pushed to 6.1% . $HNO_3$ is also used to dope graphene in Ref. [210] and Ref. [183] for devices made of graphene films on array of Si nanowires or nanoholes.

While doping is an attractive method to improve the cell efficiency, its impact is limited by the *PCE* of the starting junction. For this reason, it is highly desirable that the most ideal junction configuration is first established by optimizing key parameters as the optical transmittance and the sheet resistance of graphene sheets, before the doping.

A systematic effort in quantifying the impact of the sheet resistance and optical transmittance on the *PCE* of graphene-based solar cells was done by X. An et al. [211]. By varying the number of graphene layers, they studied the effects of optical transmittance and sheet resistance of graphene on $V_{OC}$, $J_{SC}$, *FF*, *EQE*, etc., and through optimization of the cell they achieved *PCE* >3% without any doping and with near-100% internal quantum efficiency. Layer thickening favors carrier collection till the point it prevents a large fraction of light from reaching the junction, resulting in a worsening of the cell performance.

Moreover, they p-doped graphene via solutions of 1-pyrenecarboxylic acid (PCA) dissolved in methanol and powder of $AuCl_3$ dissolved in Nitrometane to achieve a maximum *PCE* of ~7.5% and a short-circuit current density exceeding $J_{SC} = 424\, mAcm^{-2}$.

To enhance light absorption Ref. [210] and Ref. [183] proposed devices made of graphene films on nanostructured Si prepared as Si nanowire (SiNW) and Si nanohole (SiNH) arrays on planar Si substrates (Fig. 23(a) and (b)). In comparison with planar Si, nanostructured silicon provides increased light absorption due to multiple reflections and more efficient charge separation/transport due to the larger surface area. Furthermore they tried to overcome some inherent limitations of Schottky type solar cells, as the large leakage current resulting from the low barrier height, by the addition of a thin organic layer to form graphene/organic/inorganic hybrid heterojunctions. Few layers CVD graphene films, doped with $HNO_3$, were deposited on passivated n-Si, with surface consisting of methyl-group terminated Si (CH3-Si). Methyl groups have been demonstrated to be highly effective in reducing carrier recombination velocity [212]-[213], with the benefit of higher photocurrent.

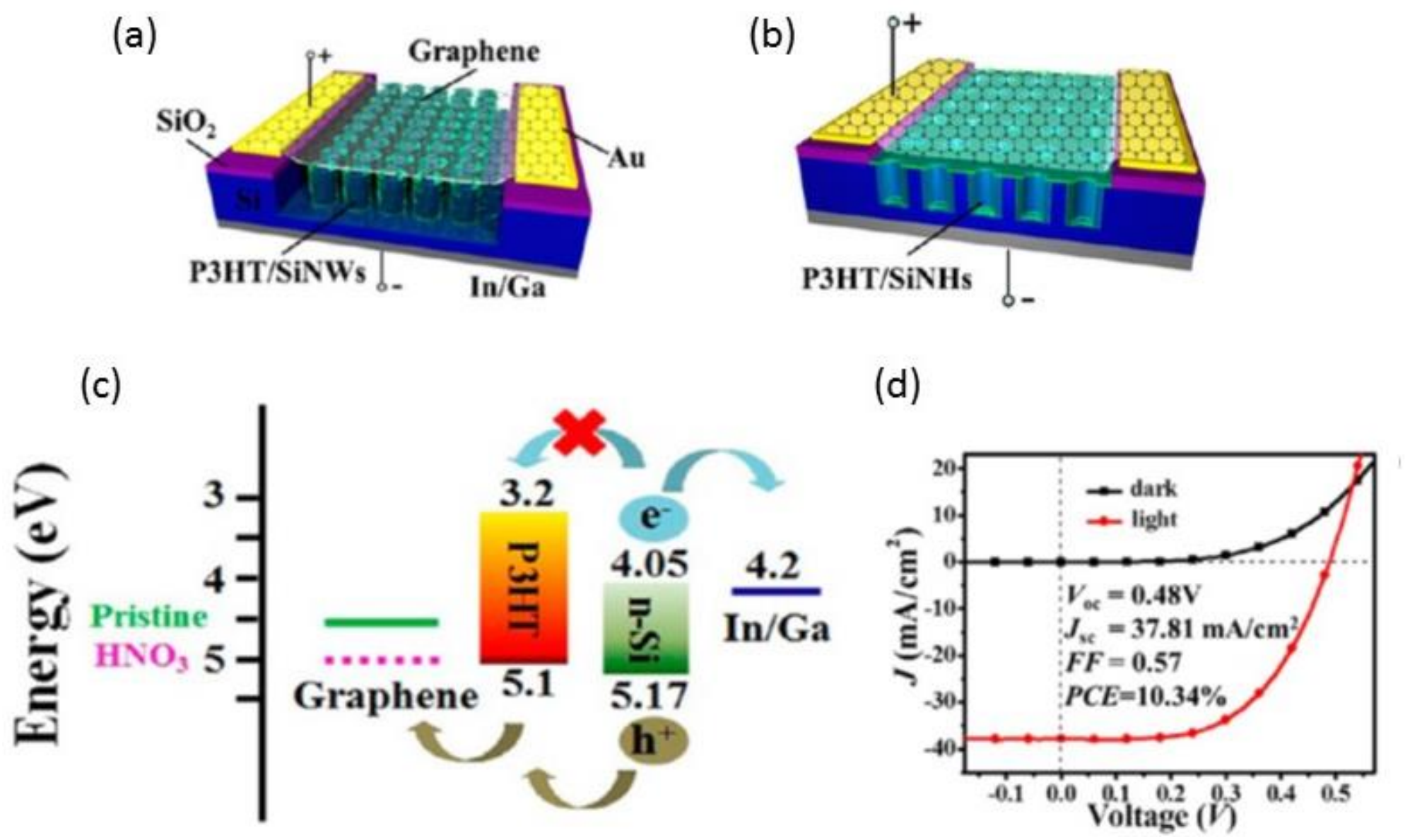

Fig. 23 - Schematic illustrations of (a) graphene/P3HT/Si nanowire array and (b) graphene/P3HT/Si nanohole array hybrid devices. (c) Band diagram of the hybrid solar cell. (d) *J−V* curve of a 5-layer Gr/P3HT (10 *nm*)/CH3−Si nanohole array hybrid device in darkness and under *AM1.5* illumination.

A thin layer of polymer poly(3-hexylthiophene) (P3HT) was coated on the surface of Si nanoarray (see Fig. 23(a) and (b)) with the purpose of acting as hole transport layer and electron blocking layer.
Fig. 23(c) shows the principle of charge separation and transport in the graphene/P3HT/Si hybrid solar cells . The lowest unoccupied molecular orbital (*LUMO*) and the highest occupied molecular orbital (*HOMO*) for P3HT are 3.2 and 5.1 *eV*, respectively, and their alignment with respect to Si and graphene is pictured in Fig. 23(c). *e-h* pairs photogenerated in Si which diffuse at the P3HT/Si interface are separated by the built-in electric field of the heterojunction. The band alignment favors the diffusion of photogenerated holes from Si to P3HT due to the positive LUMO-$E_c$ offset and their collection in the graphene anode. Diffusion of photoelectrons in same direction is suppressed by the HOMO-$E_c$ barrier energy, but is enhanced in the opposite direction, towards the In/Ga electrode (cathode). Therefore, the P3HT layer acts as the hole transport layer and as an electrons blocking layer and helps reducing the carrier recombination at the anode. The polymer layer increases the ratio $I_{ph} / I_0$ and leads to a larger $V_{OC}$, according to eq. (59).
Through passivation of the Si surface, optimization of the thickness of polymer layer, and controlling the graphene $HNO_3$ doping level and layer number, substantial improvement of device performance was achieved, leading to a *PCE* of 9.94% and 10.34% for SiNW and SiNH arrays-based hybrid solar cells, respectively. As an example, Fig. 23(d) depicts the typical J−V curve of a 5-layer Gr/P3HT (10 *nm*)/CH3−Si nanohole array hybrid device measured under *AM1.5*. $J_{SC}$, $V_{OC}$ and *FF* of the device are $J_{SC} = 37.8\,mAcm^{-2}$, *0.48 V*, and *0.57*, respectively, yielding a *PCE* as high as 10.34%. We emphasize that $J_{SC}$ of this device is close to the best value ( $J_{SC} = 42.7\,mAcm^{-2}$ ) obtained for the conventional single-crystalline Si solar cells [214]-[216]. Moreover, its *EQE* is ~85% on the wide spectrum range from ~400 to ~1000 *nm*.
To push the *PCE* of GSJ solar cells beyond the 10% threshold for commercialization, E. Shi et al. [217] proposed adding an antireflection layer on graphene. In such way they demonstrated cells with efficiency up to about 15%, a figure competitive with the best CNT-Si and with µ-Si *p-n* cells [218] [219]. E. Shi et al. [217] deposited a thin film (50−80 *nm* thick) by spin-coating a colloidal solution containing $TiO_2$ nanoparticles (3−5 *nm*) on the entire surface of graphene (a schematic of the device is shown in the inset of Fig. 24(a)). Graphene had been synthesized by CVD on Cu and transferred onto the ethanol-cleaned surface of a n-Si substrate. The purpose was to increase light absorption by minimizing reflection from the polished Si surface, trying to enhance the limited antireflection action of graphene on Si. The choice of $TiO_2$ (*vs* other materials) was determined by its refractive index, large bandgap, and easy production of a thin, uniform coating layer. Fig. 24(b) shows the light reflection spectrum of the device before and after $TiO_2$ coating. A relatively high reflectance of nearly 40% across the wavelength range of 400−1100 *nm* is initially observed. After coating, the light reflectance in the same wavelength range is significantly reduced, with a dip to about 10% in the visible region (500−800 nm). This result is the proof that a suitable refractive index and film thickness can strongly inhibit reflected light from polished Si surface.

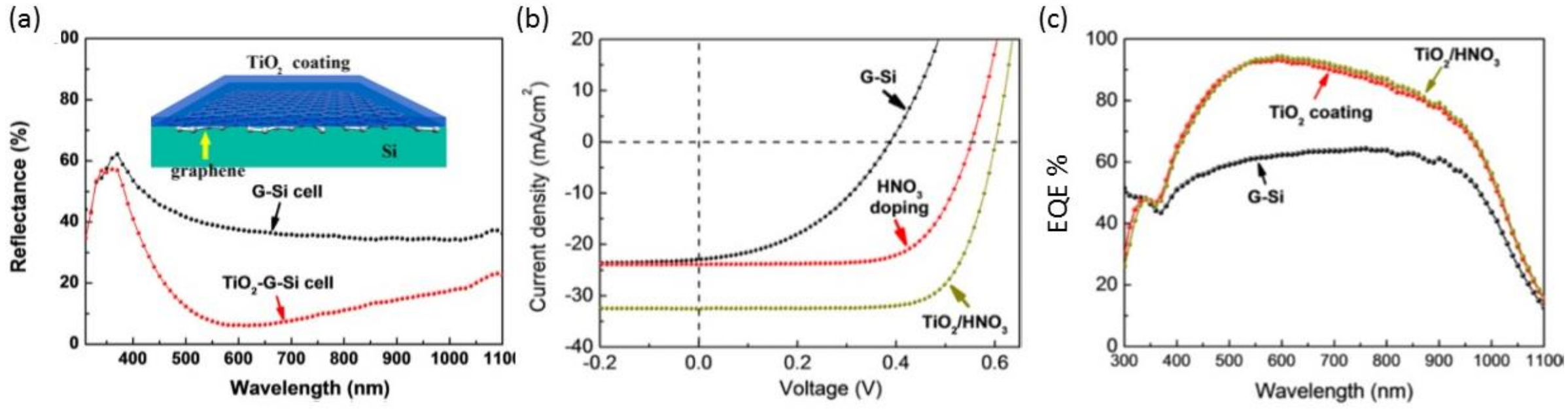

Fig. 24 - (a) Light reflection spectra of a G/Si solar cell before (black) and after (red) coating the $TiO_2$ colloid. The inset shows a layout of the G/Si device covered by a thin layer of a $TiO_2$ colloidal solution. (b) *J-V* characteristic of pristine G/Si device (black), after $HNO_3$ graphene doping and with the further addition of a $TiO_2$ anti-reflecting layer (d) *EQE vs* wavelength for pristine G/Si device and after doping/addition of $TiO_2$ layer. Figure adapted from Ref. [217]

Fig. 24(b) shows how the *J-V* characteristic of the G/Si cell, obtained immediately after etching the oxide layer, can be improved with $HNO_3$ doping and further enhanced with the $TiO_2$ layer. $HNO_3$ doping increases $V_{OC}$ and the *FF*, while the $TiO_2$ coating mainly enhances $J_{SC}$: the combined effects rise the efficiency of the cell. After doping the cell had an efficiency of ~10%, that was enhanced to ~14.5% by the addition of the coating layer. $TiO_2$ too might act as p-type doping, and further increase $V_{OC}$. The final $TiO_2$-G/Si cell had $V_{OC} = 0.60V$, $J_{SC} = 32.5 mA/cm^2$, and *FF* of 73%. Figure 24 (c) shows that an *EQE* up to 90% is achieved with $TiO_2$ for the improved light absorption in Si.
To investigate the device stability, the $TiO_2$ coated cell was stored in air without further encapsulation for 20 days and it was demonstrated that the antireflection effect, as seen in the improvement of the short-circuit current density, was stable over long time.
Y. Song et al. in Ref. [170] studied the effect of native oxide at the G/Si interface and recognized that tuning the thickness of the oxide can increase the open circuit voltage and the efficiency (from 10.0% to 12.4%). Furthermore, with the addition of a $TiO_2$ anti-reflecting layer, as in Ref. [217], they achieved a PCE of 15.6%, which is the highest reported efficiency for graphene−silicon solar cells thus far.
They fabricated devices on n-Si (Fig. 25(a)) cleaned in diluted (50:1) HF (for 30 s) to remove any native oxide. Oxide was afterwards regrown by leaving the substrates in air for varying amounts of time. In air, native oxide grows logarithmically with time and in this study was measured to be ∼5 *Å* after HF and ∼15 *Å* after 2 weeks. *I-V* characteristics were measured for different oxide thicknesses and a distinctive s-shaped kink (Fig. 25(b)) was observed when the oxide thickness was *>9 Å*. The kink, which is often observed in G/S solar cell, reduces the *FF* and the *PCE* of the cell, and was found more pronounced when the thickness of the native oxide and illumination level were increased, while chemical doping of graphene was observed to suppress it.
Following these observations, Ref. [170] proposed a model which explains the s-shaped kink as due to significant recombination current that reduces the tunneling current through the G/$SiO_2$/Si junction. As the oxide thickness augments, recombination dominates over tunneling and a clear s-shaped kink corresponding to a reduction of the junction current appears in the *I-V* characteristic.

Holes from photogenerated *e-h* in silicon, pulled towards graphene by the built-in electric field, in the presence of an interfacial oxide layer, pile up near the silicon/oxide interface. These holes either recombine with electrons either tunnel through the oxide barrier contributing to the junction current.

According to the Shockley−Read−Hall model (Ref. [46]), *e-h* recombination through interface trap states becomes important when the product of excess carriers $n_s p_s$ at the interface ($n_s$ and $p_s$ are the electron and hole densities at the interface, respectively) becomes higher than the square of the intrinsic concentration, $n_i^2$: $n_s p_s \geq n_i^2$. The recombination rate is limited by the less abundant carrier at the interface, that is electrons in this application. On the other hand the higher the concentration of holes at the oxide interface the greater is the tunneling current through the oxide. The tunneling current, which is the measured junction current, can be expressed as [220]

$$J_t = \frac{4\pi m_p^* e}{h^3 N_v}(kT)^2 p_s e^{-\sqrt{\varsigma}\delta}\left(1 - e^{-\Delta E_{Fp}/kT}\right), \qquad (108)$$

with $m_p^*$ the effective mass of holes in silicon, $p_s$ the density of holes at the interface, $\varsigma$ the average barrier height, $\delta$ the oxide thickness and $\Delta E_{Fp}$ the difference between the quasi Fermi level of holes in graphene and in silicon, under illumination. $p_s$ and $\Delta E_{Fp}$ are related by

$$p_s = p_{s0} e^{\Delta E_{Fp}/kT}, \qquad (109)$$

where $p_{s0}$ is the equilibrium hole concentration at the interface.

So, as the concentration of holes $p_s$ at the surface increases, both recombination rate (for the higher $n_s p_s$) and tunneling current (for the higher $p_s$ and $\Delta E_{Fp}$) increase. However, these two mechanisms are competing to determine the total current at the junction: for a given illumination, more recombination means less tunneling, *i.e.* reduced junction current.

With this in mind, the qualitative behavior of the *I-V* curves can easily be understood. Fig. 25(c) shows the band diagram for reverse bias. Because of band-bending, the electron concentration at the interface is lowered and this strongly limits the recombination rate. In this case the pile up of holes at the interface favors tunneling and most of the separated holes from the photogenerated *e-h* pairs contribute to the junction current. At zero or forward bias (Fig. 25(d)), the electron concentration at the interface increases, and so does the recombination rate. At the same time, the hole concentration $p_s$ diminishes since holes recombine with the many available electrons. According to eq. (108) and (109) the tunneling current is reduced. This reduction is seen as a kink and a drop of the *FF* value in the *I-V* characteristic of the cell (Fig. 25(d)).

If the oxide is thinner than 10 *Å*, the condition $n_s p_s \geq n_i^2$ is unlikely to be reached, given the high tunneling probability, and recombination remains negligible (unless $n_s$ becomes very large under significant forward bias) and no S-shaped kink is observed (Fig. 25 (e)). With a thicker oxide, however $p_s$ becomes larger. Therefore, $n_s$ does not need to be as large to satisfy the $n_s p_s \geq n_i^2$. In this case recombination becomes dominant at lower forward biases and causes the reduction of the current and the appearance of a kink (Fig. 25(f)).

Doping of graphene affects band-bending at the interface. For an increased band bending, $n_s$ decreases. Since the electrons limit the recombination rate, usually doping effectively suppresses the recombination contribution and thus the kink in the *I-V* characteristic.

From the technology of MIS (Metal-Insulator-Silicon) solar cells, it is well known that a thin layer of oxide, typically in the range 15−25 *Å*, is beneficial for efficiency. A thin oxide layer presents a tunneling barrier for electrons, which increases the effective SBH and reduces the reverse saturation current. According to eq. (59), lower $I_0$ means higher $V_0$ until the oxide layer becomes thick enough to suppress the photogenerated current.

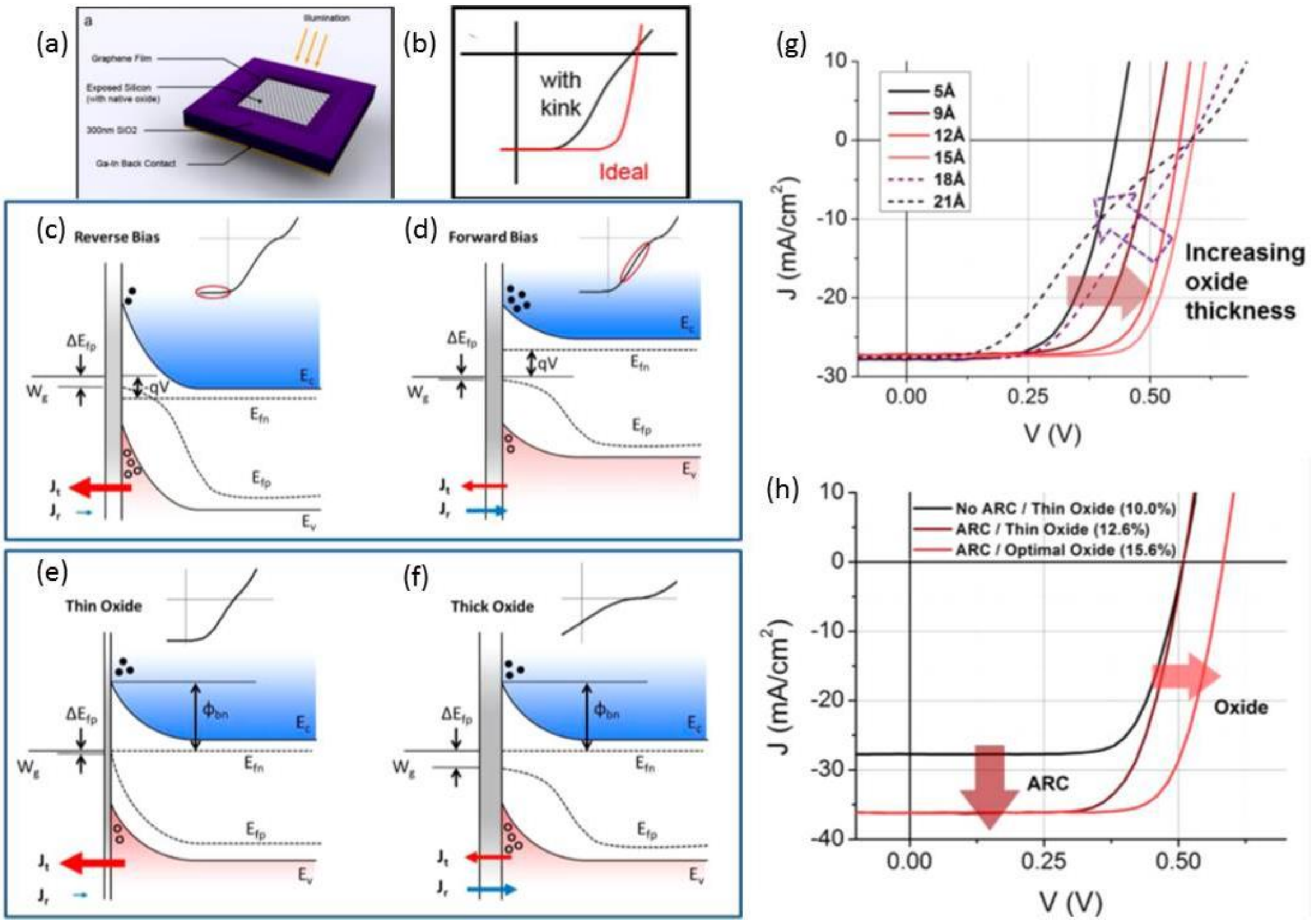

Fig. 25 - Layout of a G/$SiO_2$/n-Si solar cell (b) *I-V* characteristic with a kink commonly found in G/S solar cell devices. Energy band diagrams of a G/S junction with an interfacial oxide and under illumination, with surface electron/hole densities, in reverse (c) and forward (d) bias. Energy band diagrams and interfacial electron/hole densities for a device with thin (e) and thick (f) oxide. (g) *I-V* characteristic of the G/$SiO_2$/n-Si solar cell (after doping) for varying oxide thickness. (h) Effect of oxide interface and anti-reflecting coating on the efficiency of the cell. Figure adapted from Ref. [170]

On the other hand a too thick oxide reduces the cell current. Therefore optimal performance can be achieved only with a balanced choice of the oxide thickness. Fig. 25(g) shows the effect of the oxide thickness on the *I-V* characteristics of the G/$SiO_2$/Si cell after doping with $AuCl_3$ (sheet resistance of graphene 120 Ω/□ corresponding to a series resistance of ~9 Ω). Thickening the native oxide clearly increases $V_0$ till a plateau value (0.59 V in Fig. 25(g)); after that, a further increase of the thickness (> 15 *Å* in Fig. 25(g)), only results in the appearance of a kink, which limits the cell efficiency.

L. Lancellotti et al. [221] used the generalized equivalent circuit model of Fig. 5(e), with circuital parameters adapted to the physical properties of the graphene/semiconductor junction, to simulate devices made of graphene sheets with different numbers of layers $N_L$, coupled to n- and p-type Si or Ge. The *PCE*s of devices with different combinations of semiconductor substrate and number of layers of graphene showed values ranging from less than 1% to 11% . N- and p-type Si based devices show an opposite dependence of efficiency on $N_L$. For n-Si based devices the *PCE* increases with $N_L$, as a consequence of the augmented graphene workfunction which increases the SBH. The benefit of higher SBH compensates the transmittance reduction in this case. For p-Si based devices the efficiency decreases with increasing $N_L$ for the combined effect of SBH lowering and transmittance reduction. According to [221], for $1 \le N_L \le 4$, p-Si based devices are preferable with respect to n-Si based ones, in agreement with some experimental results (*e.g.* Ref. [132]. This is due to the fact that the barriers of G/p-Si structures are higher and the diffusion length of electrons is larger than that of holes.

For $N_L > 6$ the effect of a further increase of graphene thickness is only a transmittance reduction, and *PCE* decrease both n and p-Si.

The photovoltaic properties of graphene on p-Si substrate were investigated in Ref. [222]. Graphene was in this case obtained by borohydride ($NaBH_4$) reduction of graphene oxide. Spin-coated films of rGO of various thickness were used to study the simplest rGO/p-Si heterojunction without any doping or special configuration.

A power conversion efficiency of 0.02% (possibly extended to 1.32% for optimized devices) with $V_{OC} \approx 0.27V$, $I_{SC} \approx 0.11mA$ and $FF \approx 0.12$ fill factor (under light illumination of $1kW/m^2$) was demonstrated. A computational study was undertaken to investigate the influence of rGO thickness on *PCE* and a maximum efficiency of 6.74% was predicted using single-layer and high-quality rGO with optimized values of transmittance and sheet resistance.

Despite the low reported efficiency, rGO/Si heterojunction are still appealing devices due the simple and cheap fabrication technique and the higher transparency of r-GO [223].

*(c) Chemical sensors*

Graphene has an intrinsic high sensitivity for detecting chemical species. Its all-surface nature allows the π-conjugated system to be entirely exposed to external influences. Even the adsorption of a single molecule can affect the electronic properties of graphene, as demonstrated by Schedin et al. [224] in ultrahigh vacuum conditions.

Absorption of molecules can enable transfer of charge and modulate the graphene Fermi level. In a GSJ, molecules can affect the graphene conductivity, the SBH and have dramatic effects on the *I-V* characteristics of the junction. These effects can be used for identification purpose or to measure the concentration.

Chemical sensors are highly required for environmental and air quality control, process monitoring, safety, etc. G/S Schottky junction have been proposed as cheap and highly performant new type of sensing platforms.

Most graphene based sensors reported so far are in the form of chemiresistor or chemical field effect transistors. In these sensors adsorbed molecules change the charge carrier density and hence the conductivity in direct proportion with their number [225]-[232]. However, the detection sensitivity of these sensors, for a specific analyte, is determined by the mobility of the carriers in graphene, and is largely affected by the substrate on which graphene is synthesized or transferred [233]. Devices, based on the utilization of a gate, can achieve higher sensitivity at the cost of design complexity, power consumption and more complicate fabrication process [234]. In addition, the mass production of devices with the required reproducibility can be challenging.

H.-Y. Kim et al. [235] successfully used monolayer G/Si diodes as sensors exposing graphene to liquids and gases. These sensors showed sensitivity to liquid and gaseous electron donor (ED) and acceptor (EA) substances, such as anisole, benzene, chlorobenzene, nitrobenzene, and gaseous ammonia. The G/Si junction parameters were found to be very sensitive to the charge transfer from various adsorbates. Any change introduced by exposure was quite reversible and stable in time, two important requisites to qualify G/Si diodes as sensors.

The devices studied in [235] were fabricated with graphene grown by CVD on Cu foils, transferred on n and p-type Si substrates, with preformed Ti/Au electrodes. Both n and p substrates formed a Schottky junction with graphene well described by thermionic emission with ideality factor and barrier height $\eta \approx 1.41$ and $\Phi_B \approx 0.79\,eV$ for n-Si and $\eta \approx 1.31$ and $\Phi_B \approx 0.74\,eV$ for p-Si.

The bare graphene surface of the G/Si junctions was exposed to various liquids, with droplet volumes of 60−120 μL, covering the entire G/Si area. The analytes were exchanged by rinsing the chip with a solvent and blow-drying with nitrogen.

Fig. 26(a) shows the *J−V* characteristics of a G/n-Si diode, after applying the aromatic molecules anisole, benzene, chlorobenzene, and nitrobenzene, which have increasing electron-accepting behavior. Different curves, with different ideality factors and SBHs, are obtained, depending on the molecule used. As shown in Fig. 26(b), the ideality factor decreases with stronger EAs for the G/n-Si diode (the opposite effect is observed for the p-Si case) while the SBH increases from 0.79 to 0.80 *eV* with stronger EAs (the SBH decreases from 0.75 to 0.73 *eV* for the p-Si devices). Similarly, changes were observed in the series resistance of the diode: when exposed to liquid aromatic molecules, $R_s$ increased with EDs, and decreased with EAs, independent of the substrate type (Fig. 26(c)).
This behavior was explained by considering the doping difference between pristine and exposed (doped) graphene, as shown in the schematic band diagram of Fig. 26(d)-(f). Exposure to EDs causes extra electrons to be transferred to graphene with an up-shift of the Fermi level (*i.e.* a lowering of the SBH, Fig. 26(d)); conversely, EAs induce extra holes giving rise to an increase in the SBH because the Fermi level is downshifted (Fig. 26(f)). The liquid on graphene directly controls the SBH and the current of the G/Si junction, which can be used to evaluate the doping behavior of liquids and gases. Furthermore, the change in $R_s$ is a way to determine the concentration of EAs or EDs in a neutral solvent.

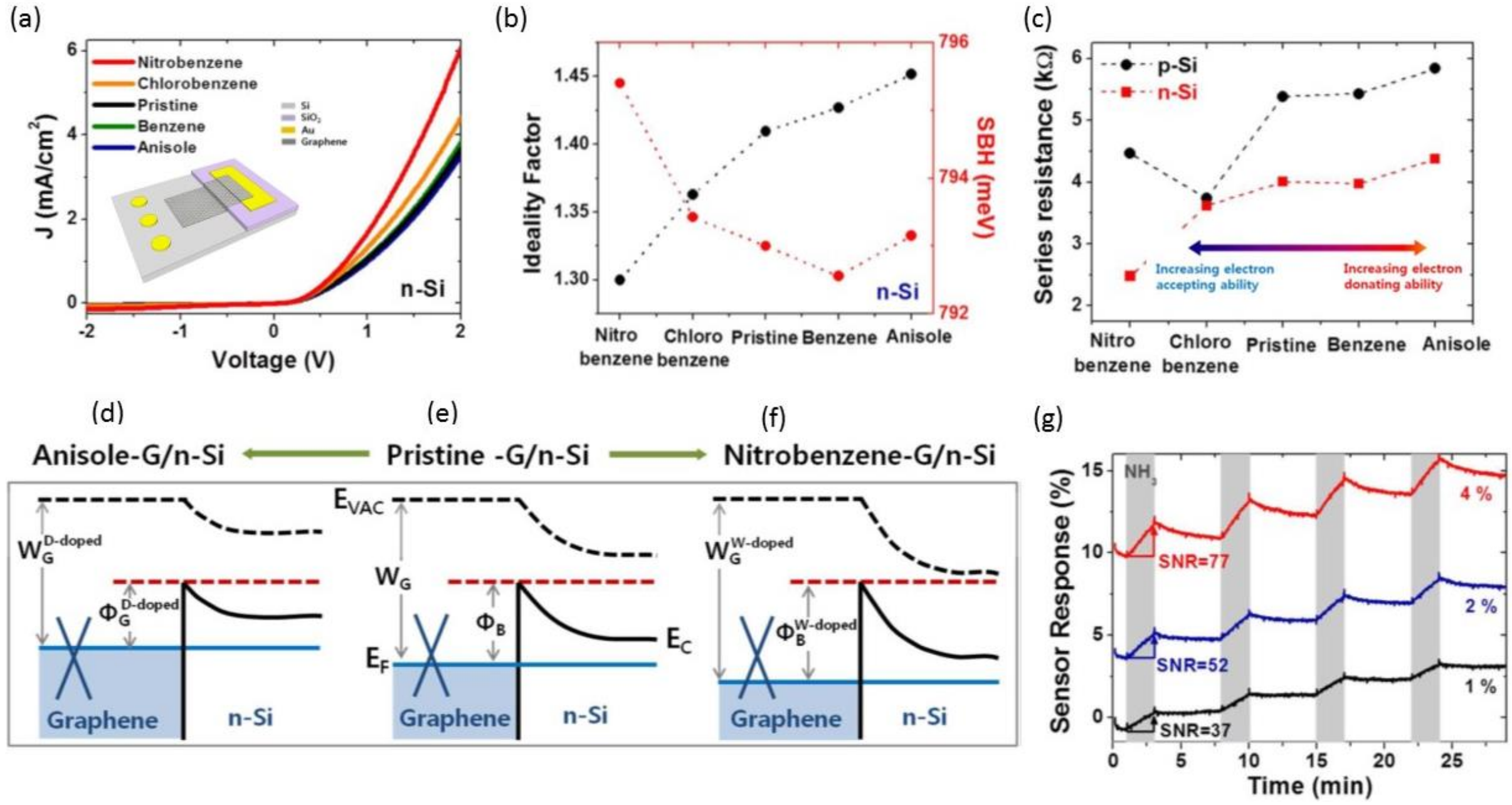

Fig. 26 - (a) $J-V$ of G/n-Si sensor when exposed to various aromatic molecules (inset shows the schematic of the device). (b) Ideality factor and SBH of G/n-Si exposed to several liquids (c) Change of $R_S$ for graphene exposed to various aromatic molecules. Schematic band diagram of the G/n-Si interface with electron donors (d), pristine state (e), and electron acceptors (f). (g) The of the G/p-Si sensor response *vs* time at 1, 2, and 4% concentration of $NH_3$. Figure adapted from Ref. [235].

H.-Y. Kim et al. [235] also demonstrated that G/Si junctions are suitable to determine the extent of charge transfer from gaseous molecules. They exposed G/p-Si devices to ammonia ($NH_3$, which is an electron donor) with concentrations from 0 to 8% in Ar (as for carbon nanotubes, one $NH_3$ molecule has been predicted to induce 0.03 electrons onto graphene [236] and showed that the current drops (since $NH_3$ increases the SBH) with increasing ammonia concentration while $R_s$ increases linearly with $NH_3$ concentration.
The percentile series resistance change:

$$S = \frac{\Delta R_S}{R_{S0}} \times 100\,\% \qquad (110)$$

($\Delta R_s = R_s - R_{s0}$ where $R_s$ and $R_{s0}$ are the series resistances with and without $NH_3$), also referred to as sensor response, is plotted *vs* time in Fig. 26(g), where the resistance changes are recorded for repeated $NH_3$ injections (gray boxes) at various concentrations. Each injection results in a resistance change with respect to gas atmosphere, with a delay mainly caused by the large volume of the sensing chamber. The steady up-shift is due to incomplete recovery between measurements since $NH_3$ desorbs slowly from graphene at room temperature. The recovery time can be accelerated by vacuum annealing or UV illumination in fully integrated sensors.
The sensors of H.-Y. Kim et al. [235] uses the variation of the forward characteristic of the G/Si junction as detection principle. Some peculiarities of the G/Si junction are not fully exploited in this approach. On the other hand, using the reverse part of the G/Si Schottky junction can have the further advance of higher and bias tunable sensitivity and low power consumption, as proposed in Ref. by [237].
The alteration of SBH due to molecular adsorption on graphene surface affects the junction current exponentially when operated in reverse bias, thus resulting in ultrahigh sensitivity. The reverse bias operation results in a drastic reduction in the operating power of the sensor (with consequent minimization of heating effects), which is highly desirable from the sensor design perspective. In addition, by operating the device in reverse bias, the work function of graphene, and hence SBH, can be controlled by the bias magnitude, leading to a wide tunability of the molecular detection sensitivity.
Fig. 27(a) shows a schematic diagram of a G/p-Si junction, which was fabricated on the same chip together with a graphene chemiresistor (a chemiresistor is a sensor based on the variation of resistance of graphene, which is inversely proportional to number of charge carriers in graphene). The *I-V* characteristic, showing the usual rectifying behavior, is shown in Fig. 27(b) (note that the positive voltage bias is applied to the Si contact). The reverse saturation current increases monotonically with increasing bias magnitude (Fig. 27(c) and (d)), due to the graphene work function changes typical of G/S junction. Devices, fabricated both on p and n-substrate were exposed for different durations to dilute $NH_3$ (electron donor) and $NO_2$ gases (electron acceptor), both in dark and illuminated ambient conditions. The effect on the reverse characteristics of a G/p-Si is shown in Fig. 27(c) and (d).
With $NO_2$ exposure, the current increases dramatically both in dark and illuminated conditions due to lowering of the SBH (Fig. 27(c), while for $NH_3$ exposure the reduction in current is rather small in dark, but is enhanced under illumination (Fig. 27(d)). Adsorption of $NH_3$ increases the SBH, thus reducing the current. However, since the current is already small, the change in current is not very large. Under illumination, the reduction in current is much more noticeable since the current is increased significantly due to excess carrier generation and barrier lowering.
The sensor response was in this work equivalently expressed as percentage change of conductivity,

$$\sigma = \frac{\Delta\sigma}{\sigma_0} \times 100\,\%, \qquad (111)$$

where $\Delta\sigma = \sigma - \sigma_0$ with $\sigma_0$ the initial sensor conductivity and $\sigma$ the conductivity after exposure to an analyte gas.
Fig. 27(e) and (f) make direct comparison of the G/p-Si sensor with conventional graphene chemiresistor fabricated using the same transferred graphene film. The responses of the two sensors for 10 min $NO_2$ exposure (shaded region) show that the chemiresistor conductivity changes by only 7.8% while that of the diode sensor changes by 104%, under the same applied bias magnitude of 4 V. The G/p-Si sensor has a response ~13 times higher than the chemiresistor. For $NH_3$ the difference in response is less dramatic, with a factor 3 enhancement.
Figure 27(e) and (f) show that the conductivity changes faster with exposure to $NO_2$ with respect to $NH_3$. This is explained by considering that the rate for charge transfer between adsorbed molecules and graphene decreases as the graphene Fermi level moves closer to the defect level introduced by the adsorbed molecules. The band diagram in Fig. 27(g) shows how $NO_2$ and $NH_3$ change the graphene-semiconductor band alignment. Initially, the graphene Fermi level, although slightly below the Dirac point, is closer to $NH_3$ than to $NO_2$, therefore the charge transfer process between $NO_2$ and graphene is much faster compared to $NH_3$. This is an important point since it suggests the way for optimized sensor design. The tunable Fermi level difference between graphene and semiconductor needs to be carefully chosen, keeping in mind the specific analyte to be detected. Ref. [237] demonstrates a sensing response down to 200 *ppb* of $NO_2$ and 10 *ppm* for $NH_3$, and anticipates that with proper optimization of the sensor, higher detection sensitivity can be achieved. Furthermore, by exposing the detector to successive cycles they show good repeatability both for electron acceptor and donor gas molecules.

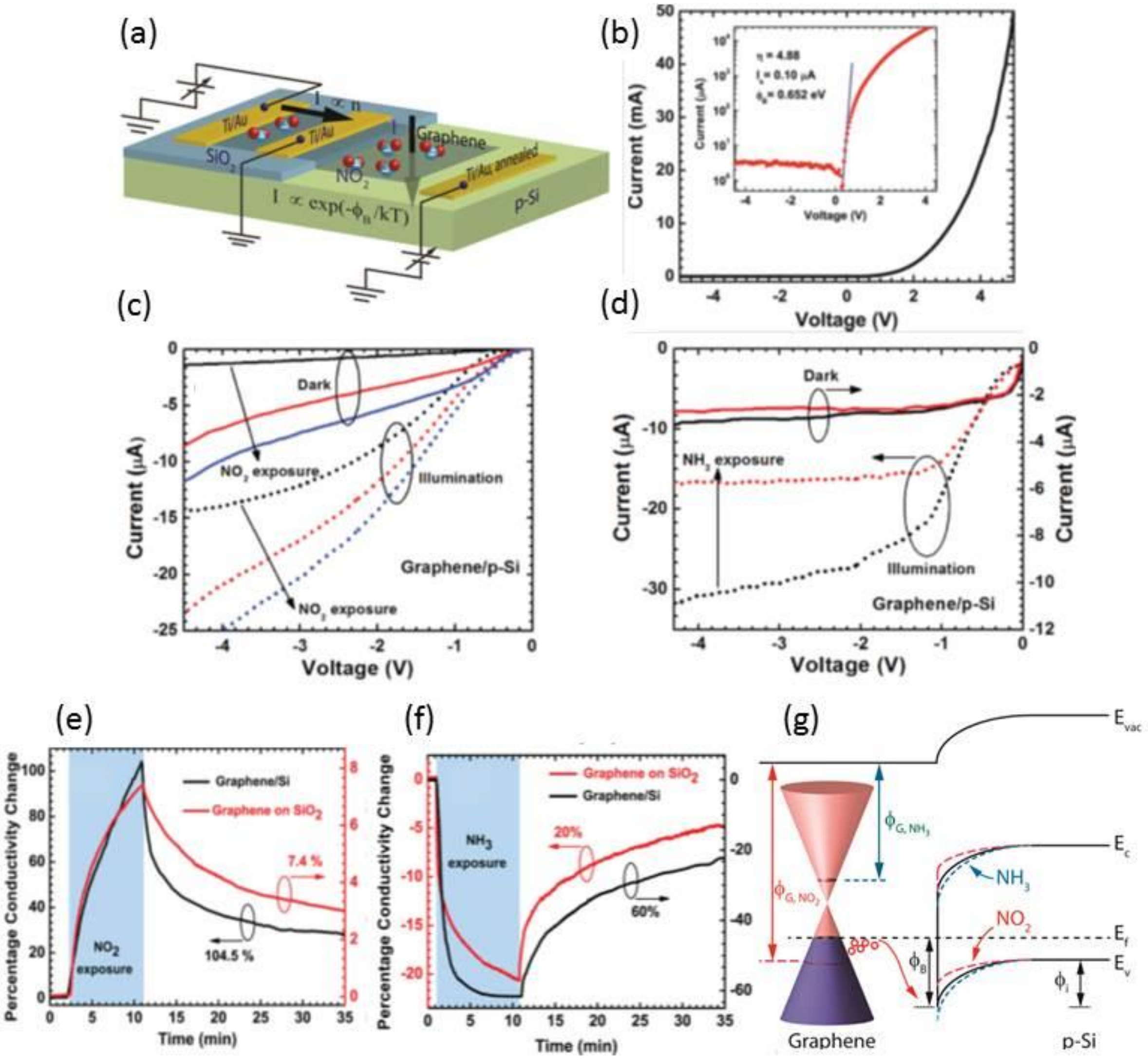

Fig. 27 - (a) Device schematic and biasing of a graphene chemiresistor and a G/Si Schottky diode sensor fabricated on the same chip. (b) *I-V* characteristics of a G/p-Si (the positive voltage bias is applied to the Si contact). (c-d) Reverse *I-V* characteristics of G/p-Si diode in dark (solid curve) and under illumination (dotted curve) for different exposure times to (c) $NO_2$ and (d) $NH_3$ . The black curves represent pre-exposure characteristics, while the red and blue curves represent characteristics after 10 *min* and 30 *min* of gas exposure. (e-f) Comparison between $NO_2$ (e) and $NH_3$ (f) responses of graphene/p-Si (black line) and graphene chemiresistor (red line) fabricated on the same chip, side by side. The exposure duration (10 *min*) and bias voltage magnitude (4 *V*) is the same in both cases. (g) Energy band diagram of G/p-Si in three different conditions, showing reduction in SBH for $NO_2$ and increase in SBH for $NH_3$ exposure, as compared to the pre-exposure condition. Figure adapted from reference [237].

The tuning the SBH and barrier width at the tiny area of contact between graphene and $SnO_2$ nanowires through the adsorption/desorption of gas molecules is at the origin of the outstanding $NO_2$ gas sensing properties of monolayer G/$SnO_2$ nanowire Schottky junction devices presented in Ref. [238]. The devices were prepared by directly growing single crystal $SnO_2$ nanowires on interdigitated Pt electrodes via thermal evaporation. A CVD graphene monolayer was subsequently transferred on top of the nanowire chip. The Schottky junction-based sensor showed sensitivity to $NO_2$ gas with the remarkable detection limits of about 0.024 *ppb* at the low operating temperature of 150 *C* and bias voltage of 1 *V* and with a response/recovery time of less than 50 *s*.

*(d) Barristors and other applications*

H. Yang et al. [239] proposed adding a top gate to the G/Si junction to control the Schottky barrier height and achieve a large modulation of the diode current (with on/off ratio up to $10^{+5}$ ). They fabricated three terminal devices, of the type sketched in Fig. 28(a), which are known as graphene barristors, *i.e.* variable barrier devices, consisting of CVD graphene on hydrogen terminated n- or p-Si. An optimized transfer process was used to yield atomically sharp interfaces (as illustrated in the inset of Fig. 28(b)) with minimum number of atomic defects or silicon dioxide formation to prevent charge trapping sites. The purpose was to avoid Fermi level pinning at the G/Si interface and make the graphene $E_F$ controllable by the gate through field effect.

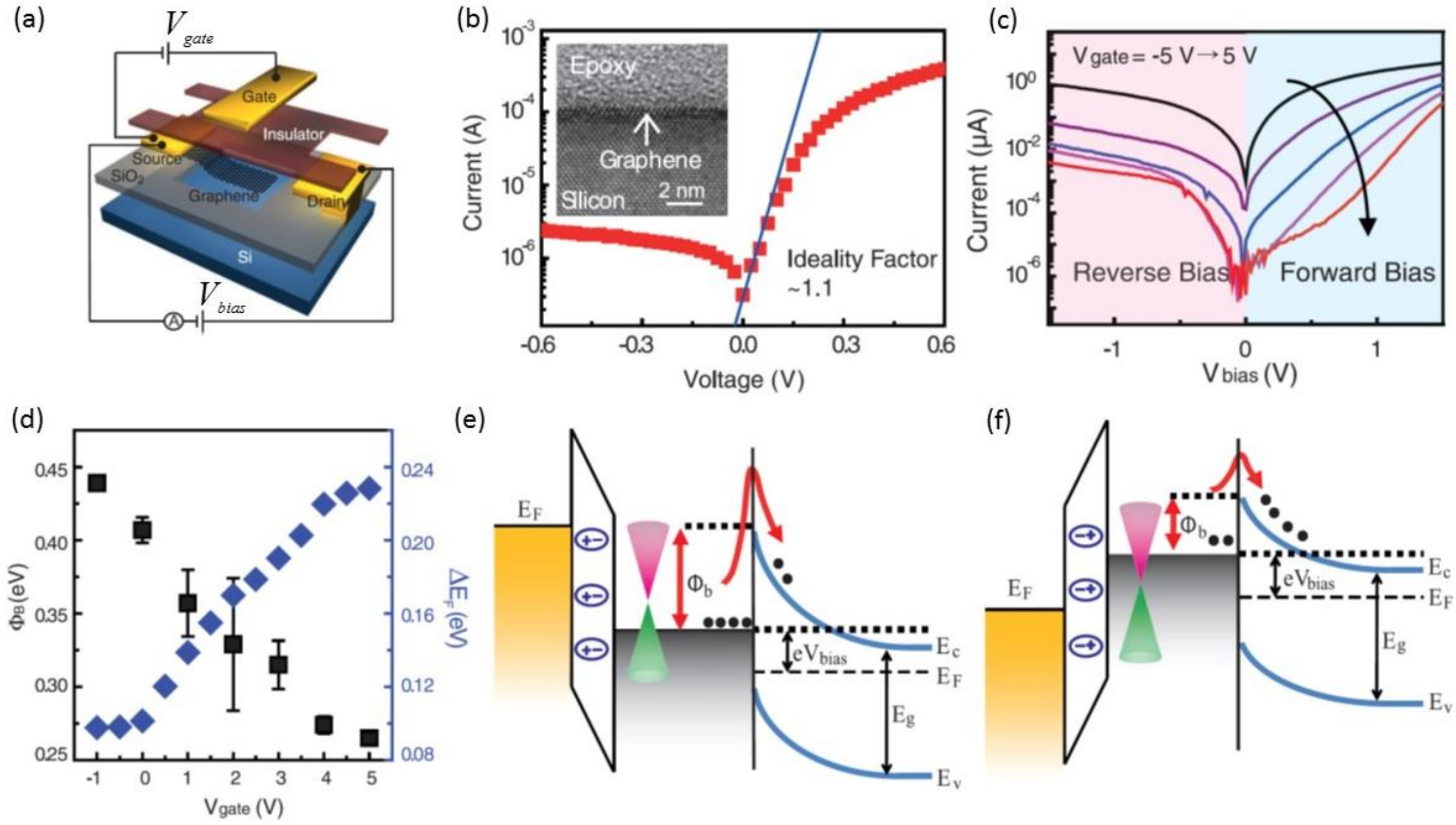

Fig. 28 - Layout of a graphene barristor with top gate. (b) *I-V* characteristic of graphene/p-Si barristor at $V_{GS} = 0V$ . (c) *I-V* characteristics of the p-type barristor for biases in the range -1.5 *V* to 1.5 *V* and gate in the range (-5, 5 )*V* by steps of 2 *V*. (d) SBH and field-effect induced Fermi level change, $\Delta E_F$ . (e-f) Band diagram of graphene/n-Si barristor (from left to right: gate-insulator-graphene-silicon) for (e) negative voltage on the gate ($V_{gate} < 0V$ and holes induced in graphene), and (f) positive voltage on the gate ($V_{gate} > 0V$ and electrons induced in graphene). Figure adapted from Ref. [239].

Fig. 28(c) shows the *I-V* characteristic of a G/p-Si barristor with zero voltage on the gate, $V_{GS} = 0V$ . The ideality factor is very close to 1 confirming the high quality of the junction. It can be noticed that, both in forward and reverse bias, the G/p-Si current is strongly modulated by the gate voltage: by stepping $V_{GS}$ from -5 V to +5 V reduces the current by a factor that can be as high as $10^{+5}$ .

The barristor works on the principle that the SBH (Fig. 28(d)) and the current are modulated by the electric field of the gate [69]. The field of the gate induces holes or electrons in graphene (Fig. 28(e) and (f)) which shift the graphene Fermi level and modify the SBH. Fig. 28(d), referred to a G/n-Si barristor, shows that increasing $V_{GS}$ decreases the SBH and increases the Fermi level variation in graphene, $\Delta E_F$ ( $E_F$ is estimated from the measurement of Hall carrier density *n* and from eq. (75)). $\Delta E_F$ and the SBH are strictly correlated, $\Delta\Phi_B = -\Delta E_F$ , confirming that the field-induced modulation of $E_F$ in the absence of Fermi-level pinning is fully responsible for the variation of the SBH of the barristor.

H. Yang et al. [239] also demonstrated inverter and half-adder logic circuits by developing a fabrication process of n- and p-type G/Si barristors on a 150 *mm* wafers with CVD transferred graphene.
We quickly mention here a few other remarkable applications of the G/S Schottky junction.
M. Liu et al. [240] proposed a graphene based optical modulator for on-chip optical communications that has the advantage of compact footprint, low operation voltage and ultrafast modulation speed across a broad range of wavelengths. They fabricated a waveguide-integrated electro-absorption modulator based on monolayer graphene, which achieved modulation of the guided light at frequencies over 1 *GHz*. Integration of graphene with an optical waveguide greatly increased the interaction length through the coupling between the evanescent waves and graphene.
T. Gu et al. [241] used G/Si junctions as optoelectronic devices for coherent four-wave mixing.
Electrophoresis was used by K. Wu and coworkers [242] to fabricate G/Si electrodes which displayed high photoresponse and high stability in aqueous solution, thus constituting solar energy materials to use in aqueous solution.

**8. Van der Waals heterostructures with 2D layered semiconductors**

Following the great success of graphene, a new class of 2D non-graphene materials, including insulators such as BN, semiconductors such as the transition metal dichalcogenides (TMDCs) $MoS_2$ and $WS_2$, or metals such as $NbSe_2$, has emerged as promising constituents of electronic and optoelectronic devices [243] [244] [245]. 2D layered non-graphene materials have become the subject of a new vibrant research field since 2010. Many 2D layered materials have a semiconductor behavior with intrinsic direct band-gap already in bulk state or undergo a crossover from an indirect-to-direct band-gap when going from bilayer to monolayer. The direct bandgap results in enhanced optical properties which makes single-layer materials advantageous for optoelectronic applications. Furthermore, differently from traditional semiconductors, 2D layered semiconducting materials have ultrathin thickness, smooth surface and high flexibility, and are suitable for flexible and wearable electronic devices. They are finally promising materials to solve the current challenges of the semiconductor industry such as the reduction of the short-channel effects, the power dissipation and the higher vertical integration degree.
TMDCs [246] [247] are the most studied 2D layered materials. Their electronic properties range from metallic to semiconducting with sizable bandgaps around 1–2 *eV* [248]. They have formula $MX_2$, where M stands for a transition metal element as Mo, W, Nb, Ni,Ta,Ti, Zr, etc. and X is a chalcogen as S, Se or Te. Each TMDC monolayer consists of three planes of atoms of the form X–M–X, with a triangular lattice of metal atoms sandwiched between two triangular lattices of chalcogen atoms, as shown in Fig. 29. There is strong covalent bonding between the atoms within each layer, while adjacent layers are weakly held together by van der Waals forces and can be easily separated. The bulk crystal forms in a variety of polytypes, which differ for stacking orders and metal atom coordination (Fig. 29 b), and can be easily exfoliated in single layers similarly to graphene from graphite.
Novel ultrathin heterojunctions obtained by stacking 2D mono-layer materials in a chosen sequence, held together by van der Waals forces, known as van der Waals heterojunctions (vdWH), have become one of latest research trends [249]. These heterostructures consist of atomically sharp interfaces and do not pose critical requirements on lattice matching between contacting layers, which is significantly different from the traditional heterostructures. Van der Waals heterostructures can be fabricated by mechanical transfer or an *in-situ* growth methods with vertical (stacked) or seamless in-plane (lateral) configuration and controllable new features . The possibility of exploiting the properties of the single components to synthesize more superior composites, achieving synergistic effects is a great opportunity offered by these structures. Although research on vdWHs is at its beginning and many unique properties as well as fundamental issues are still unclear and need to be addressed, such heterojunctions have been already demonstrated as competitive *pn* junctions [250][251] [252], vertical [253] and tunneling [254] FETs, photodetectors and LEDs [255], [256] [257] or solar cells [258].
The fast experimental progress has left theory slightly behind. Traditional bulk junction theory seems not well suitable for this emerging class of 2D heterojunctions and some new fundamental theory is next urgent topic.
Given the purpose of this review, we consider here vdWHs involving graphene/2D layered semiconductors for Schottky barrier studies and applications.

The current through a Schottky barrier into a 2D system can be derived from the thermionic theory exactly as in the 3D case and is expressed by Eq. (35) , $I = I_0\left(e^{\frac{e(V-R_S I)}{\eta kT}} - 1\right)$, with a modified expression of the reverse saturation current, which takes into account the 2D density of states, which is energy-independent. In 2D [35][156], eq. (26) of the density of states is replaced by

$$N_{2D}(E) = \frac{4\pi}{h^2} m* \qquad (112)$$

and the carrier density becomes

$$n = \int_{E_C}^{\infty} N_{2D}(E) f(E) dE = \frac{4\pi}{h^2} m^* kT e^{-\frac{E_C - E_F}{KT}} , \qquad (113)$$

which, considering that in 2D $\bar{v}_{th,x} = \sqrt{\frac{2kT}{\pi m^*}}$ and proceeding as in eq. (28)-(33), yields

$$I_0 = WA^*_{2D} T^{\frac{3}{2}} e^{-\frac{\Phi_B}{kT}} , \qquad (114)$$

where $A^*_{2D} = \frac{2e}{h^2}\sqrt{2\pi\, m^* k^3}$ is the 2D Richardson constant and W the width of the 2D junction. From eq. (35) and (213), when $V > \eta kT$ and $V > R_s I$ ,

$$\ln\left(\frac{I}{T^{3/2}}\right) = -\frac{1}{kT}\left(\Phi_B - \frac{eV}{\eta}\right) + \ln\left(WA^*_{2D}\right). \qquad (115)$$

One effect of the 2D density of states is the reduced power low $T^{\frac{3}{2}}$, which replaces $T^2$ in $I_0$ of eq. (33).
The optical and electrical properties of graphene and the exceptional light absorption of TMDC monolayers suggests the possibility to design ultra-thin photodetectors or solar cells based on just two stacked monolayers, where graphene act mainly as transparent electrode and TMDCs as good photoactive material. Such structures are able to capture a significant fraction of incident sunlight in a subnanometer thickness and reach over 1% PCE in about 1 *nm* thickness. When packed together, a power density higher than 2.5 *MW/kg* can in principle be achieved, a value that is far superior to any known energy conversion or storage device [259].

*(a) The graphene/$MoS_2$ heterojunction and its applications.*
$MoS_2$, which is one of the most popular TMDCs, has been thus far the most considered 2D layered semiconductor. Bulk $MoS_2$ has an indirect bandgap of ~1.3 *eV* which transforms in a direct bandgap of ~1.8 *eV* in single-layer form (6.5 *Å* thick). The direct bandgap favors interaction with light and opens the possibility of many optoelectronic applications. $MoS_2$ exhibits dominant n-type behavior in FET devices [243]. The electronic structure of $MoS_2$ also enables valley polarization, which is another important property for new-generation devices [260][261] [262].
There are two main methods to form the heterojunction between graphene and $MoS_2$ (or other TMDCs). One is the mechanical transfer of exfoliated graphene onto the exfoliated $MoS_2$. The other is using large-area CVD-grown graphene combined with exfoliated $MoS_2$. Both of these methods usually need PMMA to assist in the transfer, even though random exfoliating method without PMMA has been sometimes used to avoid photoresist contamination and obtain cleaner interfaces. Early work was carried out mainly with the double exfoliation method, with the limitations of micrometer-scale devices and the problem of reciprocal alignment of the graphene and $MoS_2$. The ability to reproducibly generate large-area heterostructures with a method suitable for large scale production has been highly demanded for both fundamental investigations and technological applications. A significant progress in the fabrication was the synthesis of large-area, continuous, and uniform $MoS_2$ monolayers directly on graphene by chemical vapor deposition. This was reported by McCreary et al. [263], who demonstrated uniform single-layer growth of stoichiometric $MoS_2$ for heterostructure samples on the centimeter scale with the possibility for even larger dimensions.

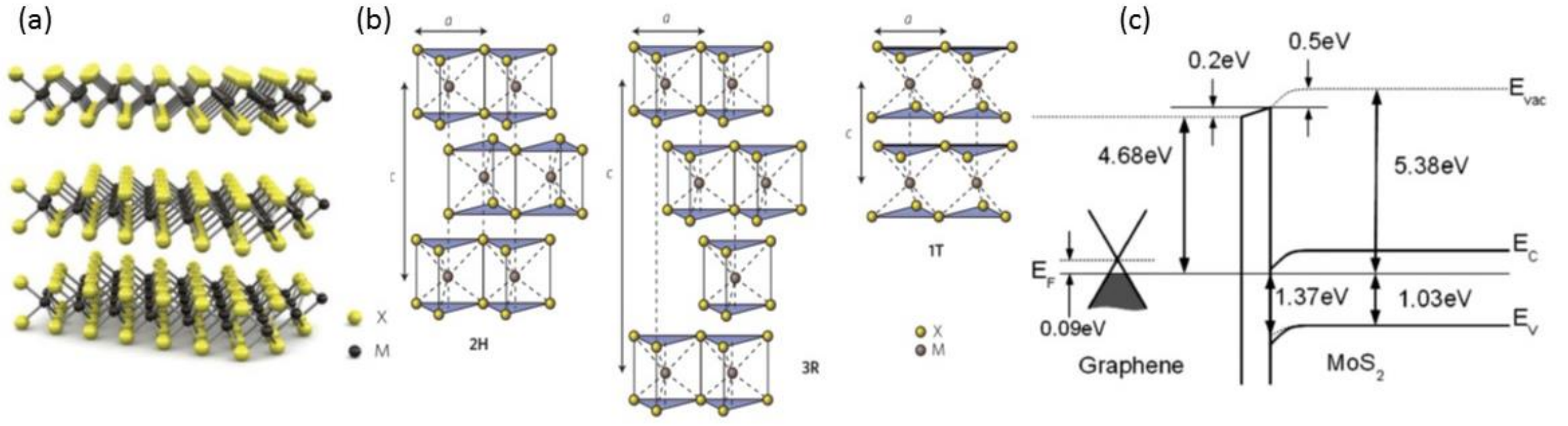

Fig. 29 – (a) Three-dimensional representation of a typical $MX_2$ monolayer, showing the metal atoms (M, in grey) between two planes of chalcogen atoms (X, in yellow). As for graphene, single layers of $MX_2$ can be extracted by using scotch-tape micromechanical cleavage. (b) Schematics of the structural polytypes: 2H (hexagonal symmetry, two layers per repeat unit, trigonal prismatic coordination), 3R (rhombohedral symmetry, three layers per repeat unit, trigonal prismatic coordination) and 1T (tetragonal symmetry, one layer per repeat unit, octahedral coordination). The lattice constants *a* are in the range 3.1 to 3.7 *Å* for different materials. (c) Energy band alignment at the graphene/$MoS_2$ interface on $MoS_2$ single crystal as derived from photoemission studies. Figure adapted from Ref. [246] and Ref. [265].

Using density functional theory, Ma et al. [264] studied the geometric and electronic structures of graphene adsorption on $MoS_2$ monolayer to facilitate the design of devices where both the finite band-gap of $MoS_2$ and the high carrier mobility of graphene are needed. They found that graphene is bound to $MoS_2$ with a binding energy of $-23\ meV$ per C atom irrespective of adsorption arrangement and that the graphene/$MoS_2$ interlayer spacing is 3.32 *Å*, which corresponds to a weak interaction. Consequently, in this hybrid structure, the linear band dispersion relation of graphene is preserved with at most the opening of a small bandgap of $2\ meV$ due to the variation of on-site energy induced by $MoS_2$. In principle, this band gap is tunable by varying the interlayer spacing, and devices combining tunable bandgap and high electron mobility become conceivable.

The energy band alignment at the G/$MoS_2$ interface on $MoS_2$ single crystal was investigated in Ref. [265] where CVD grown graphene was transferred by PMMA on a single crystal of $MoS_2$. High temperature anneal (300 C) was used to remove water from the interface and obtain ultraflat CVD grown graphene as confirmed by STM images. Charge transfer between graphene and $MoS_2$ results in p-type doping of graphene and a downward band bending in $MoS_2$, corresponding to a negative space region and low charge injection barrier. Photoemission spectroscopy was used to deduce a 0.2 $eV$ interface dipole formation (consequence of interface electron redistribution), a p-type doping of graphene corresponding to ~0.09 $eV$ shift of the Fermi-level below the Dirac point, and a negative space charge region in bulk $MoS_2$ as schematized in Fig. 30 (c). Also, evidence that interlayer van der Waals interactions can modify the band structure of 2D-layered dichalcogenides was found and a ~0.1 $eV$ $MoS_2$ band gap narrowing was reported.

Yu et al. [266] have recently developed a CMOS-compatible, fully integrated process to fabricate 2D heterojunctions of graphene with monolayer $MoS_2$ (G/$MoS_2$) in large scale, based on the selective etching of 2D materials grown by chemical vapor deposition. They demonstrated high performance discrete transistors and fully integrated logic circuits using $MoS_2$ as the transistor channel and graphene as contacts and interconnects. The tunable Fermi level in graphene allows excellent work-function matching with $MoS_2$, such that graphene contacts on $MoS_2$ can yield 10 times lower contact resistance and 10 times higher on-current and field-effect mobility than conventional metal/$MoS_2$ contacts.
The device used by Yu et al. [266] is shown in Fig. 30 (a) and consists of a back-gated FET with the $MoS_2$ channel contacted by graphene used as source and drain, on top of a $SiO_2$/Si substrate (back-gate). $MoS_2$/Ti FETs were used as control devices with exactly the same geometry.

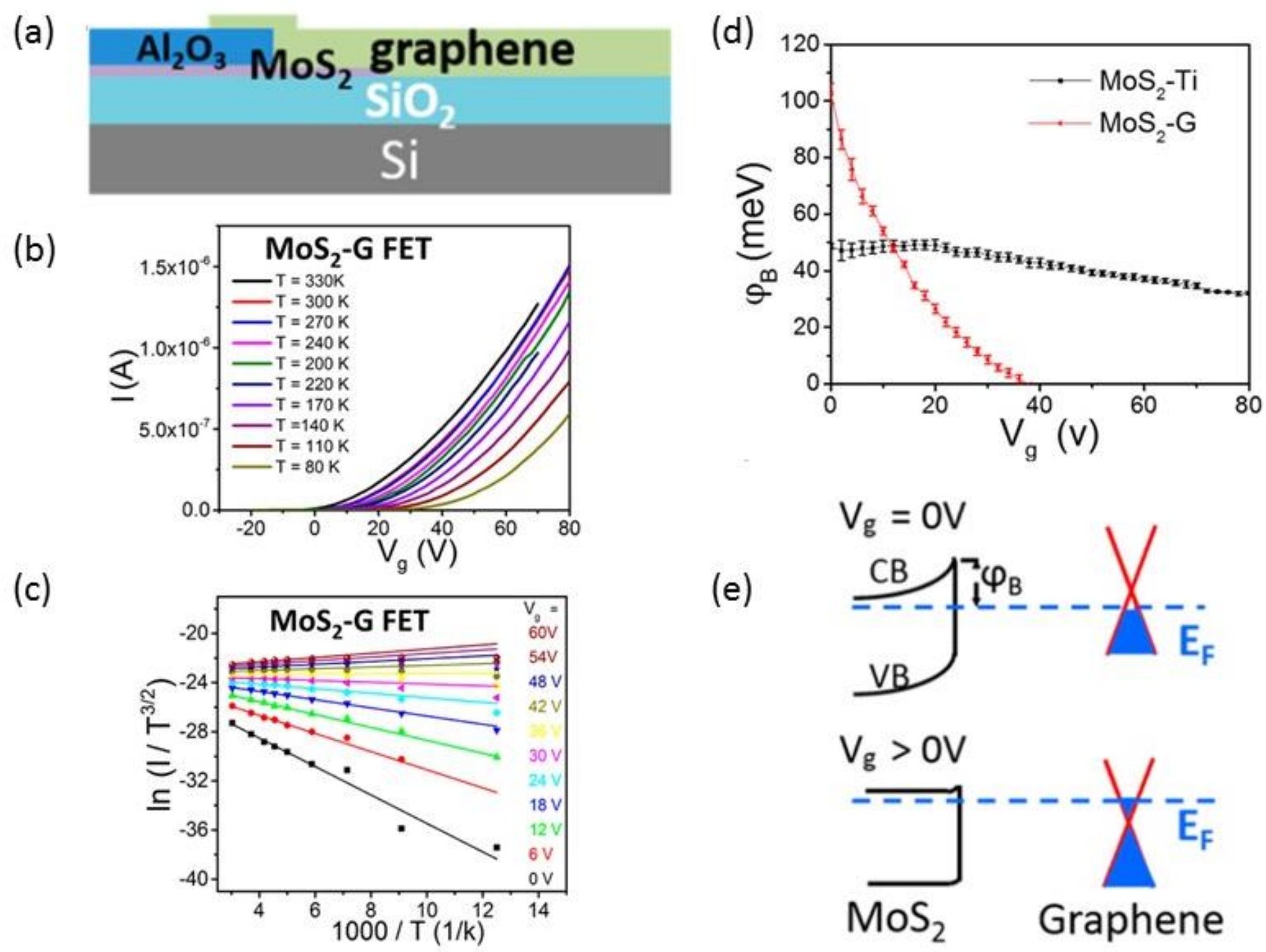

Fig. 30 – (a) Schematic of a G/$MoS_2$ back-gated FET (typical channel length and width are 12 $\mu m$ and 20 $\mu m$). Graphene is used as source/drain contact while $MoS_2$ is the FET channel. (b) Source-drain current at 0.5V source-drain bias as function of the back-gate voltage $V_g$. (c) Richardson plot $\ln\left(I_d / T^{3/2}\right)$ *vs* $1000/T$ at different $V_g$. (d) Schottky barrier height, $V_g$, as a function of the back gate bias for the G/$MoS_2$ and Ti/$MoS_2$ heterojunctions. (e) Schematic band diagram of G/$MoS_2$ heterostructure at $V_g = 0V$ and $V_g > 0V$. Figure adapted from Ref. [266].

Measurements of the current through the G/$MoS_2$ (or Ti/$MoS_2$) heterojunctions (source-drain current *I vs* source-drain voltage *V*) was performed at different temperatures and gate voltages $V_g$ as shown in Fig. 30 (b). The current smoothly decreases when the temperature is lowered, indicating the presence of a small Schottky barrier at the G/$MoS_2$ interface. The ideality factor $\eta$ and the series resistance $R_s$ of eq. (35) were extracted from the slope and the intercept of a plot of $dV / d\ln I$ as a function of $1/I$ (see eq. 43), for given temperature and $V_g$. Reported values, at room temperature and $V_g = 40\,V$, are $\eta = 13.4$ and $R_s = 212\ k\Omega$. $R_s$ which corresponds to less than $0.1\,k\Omega\cdot mm$ is ten times lower than that achieved with Ti/$MoS_2$ and represents a state-of-the-art contact resistance in $MoS_2$ device technology.
For a given $V_g$, the SBH of the heterojunction, $\Phi_B$, was extracted according to eq. (115) from the slope of the straight lines in the Richardson plot of $\ln\left(I / T^{3/2}\right)$ *vs* $1000/T$, shown in Fig. 30 (c). Fig. 30 d) shows the $V_g$-dependence of SBH for both G/$MoS_2$ and Ti/$MoS_2$ heterojunctions (notice the difference with the SB on single crystal $MoS_2$ of Fig. 29 (b), due to the different bandgap). The back-gate voltage has a minor effect on the SBH of Ti/$MoS_2$ but dramatically changes that of G/$MoS_2$. For G/$MoS_2$, $\Phi_B$ decreases from 110 to 0 $meV$ while $V_g$ goes from 0 to 35 *V*. The modulation of the G/$MoS_2$ SBH is the consequence of the change in the graphene workfunction caused by the back gate. Such a change is negligible in Ti, but as we have seen many times, becomes relevant in graphene (with 300 *nm* $SiO_2$ as back gate dielectric, a change in the value of $V_g$ by 30 *V* induces a change of around 200 *mV* in graphene work function [69]). Remarkably, when the back gate voltage is larger than 35 *V*, the SBH of the G/$MoS_2$ heterojunction approaches zero and an ohmic contact forms. The result of the SBH modulation is that graphene is capable to establish excellent contacts with $MoS_2$ and other layered semiconductors, which usually outperform those formed with traditional metals.
The tunability of the SBH by a back-gate is also exploited by Tiam et al. [267] to realize gate-controlled G-$MoS_2$ (and $MoS_2$-G-$MoS_2$) field effect Schottky barrier transistors (FESBTs), where the high mobility of graphene is combined with the high on-off ratio of $MoS_2$ transistors to overcome the low mobility and the low on/off ratio of $MoS_2$ and graphene FETs, respectively. Their back-gated FESBTs achieved a mobility around 60 $cm^2V^{-1}s^{-1}$ (typical values for similar $MoS_2$ FET are $<20$ $cm^2V^{-1}s^{-1}$) and on/off ratio $>10^{+5}$ (due to the lack of a bandgap, graphene FETs have on/off ratio < 10).

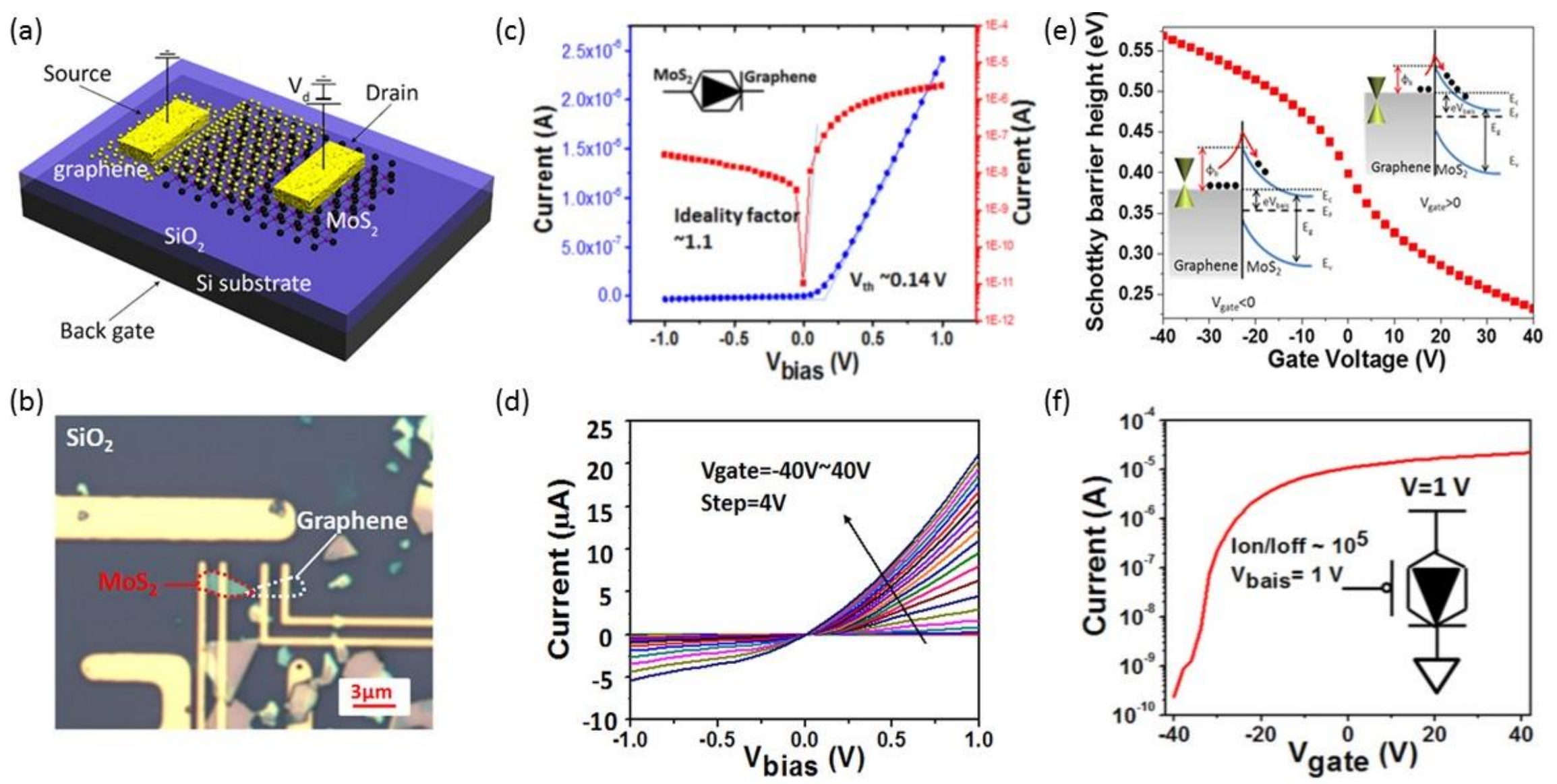

Fig. 31 – (a) Schematic of a gate controlled G/$MoS_2$ heterojunction. The Si substrate is the back-gate, while the channel includes a G/$MoS_2$ heterojunction. Source and drain, respectively connected to graphene and $MoS_2$, are contacted by evaporated Cr/Au leads. (b) Optical image of the device in (a). The $MoS_2$ and graphene are ~8 *nm* and ~3 *nm* thick, the overlapping area of graphene and $MoS_2$ is about 1 $\mu m^2$. (c) Source-drain current *vs* source drain bias ($I_d - V_d$) at $V_g = 0V$ showing rectifying behavior due to the Schottky barrier formed at the G/$MoS_2$ interface. (d) Output characteristics $I_d - V_d$ at different gate biases. (e) SBH as a function of $V_g$ at drain bias $V_d = 0V$ (ZB Schottky barrier height): a sweep of the gate voltage from -40 *V* to +40 *V* results in a variation of 0.34 $eV$ of the SBH (or of the Fermi level of graphene). (f) Transfer characteristic ($I_d$ *vs* $V_g$) of the G/$MoS_2$ FET of (a). Figure adapted from Ref. [267].

The schematic structure and an optical view of the G-$MoS_2$ heterojunction FESBT, fabricated with few-layers $MoS_2$ and graphene flakes, mechanically exfoliated onto n-Si substrates covered by 300 *nm* $SiO_2$, are shown in Fig. 31 (a) and (b). The G/$MoS_2$ layers form a Schottky junction which is the core part of the device and the reason of the rectifying characteristic shown in Fig. 31 (c). At $V_g = 0V$, the source-drain current $I_d$ shows a rapid increase at forward bias, when the souce-drain voltage $V_d$ is ramped from 0 to 1*V*, with a low threshold voltage $V_{th} = 0.14V$ (see Fig. 31 (c)). The low $V_{th}$ is desirable for low voltage applications. At $V_g = 40V$ and $V_d = 1V$, a current as large as 2400 $Acm^{-2}$ is obtained.

Similarly to the G/$MoS_2$ heterojunction of Fig. 30, the FESBT output characteristics ( $I_d - V_d$ curves ) at different gate voltages $V_g$ (Fig. 31 (d)) display a clear increase in conductance for higher $V_g$. As already pointed out, the change of the current flow is due to the moduation of the SBH by gate voltage. Such modulation, obtained by solving the Poisson equation for the carrier density in graphene and eq. (35) for the current of the G/$MoS_2$ heterojunction, is shown in Fig. 31 (e). Finally the transfer curve in Fig. 31 (f), plotted on a log scale, shows that the FESBT exhibits a $10^{+5}$ on-off ratio and is used to estimate a carrier mobility of 58.7 $cm^2V^{-1}s^{-1}$. We notice that the ability of the gate to control the SBH and thus the transistor current can be further enhanced by replacing the gate oxide with a thinner or a higher-*k* one.

Bernardi et al. [259] studied photovoltaic devices made of a monolayer graphene stacked onto a p-type $MoS_2$ monolayer (Fig. 32(a) and (b)) and concluded that such ultimate-thin heterojunctions can work as Schottky barrier solar cells with PCE of up to ~1% under *AM1.5* illumination, corresponding to at least an order of magnitude higher power density than the best existing ultrathin solar cells. Compared to the 2.3% absorbance of a single-layer graphene, $MoS_2$ monolayers possess a higher absorbance of 5-10% in the visible (Fig. 32(d)), and are able to absorb the same amount of sunlight as approximately 15 *nm* of GaAs or 50 *nm* of Si (a 300-*nm* $MoS_2$ film absorbs ~95% of the visible light). The strong absorbance of monolayer TMDC is due to dipole transitions between localized *d*-states of the transition metal atoms and to excitonic coupling of such transitions [268] [259].

The G/$MoS_2$ solar cell they considered is shown schematically in Fig. 32(b) and is interfaced to a high workfunction metal on the $MoS_2$ side and a low workfunction metal on the graphene side to enhance the built-in field. Electron–hole pairs generated in either materials are separated by the Schottky barrier formed at the interface (Fig. 32(c)).

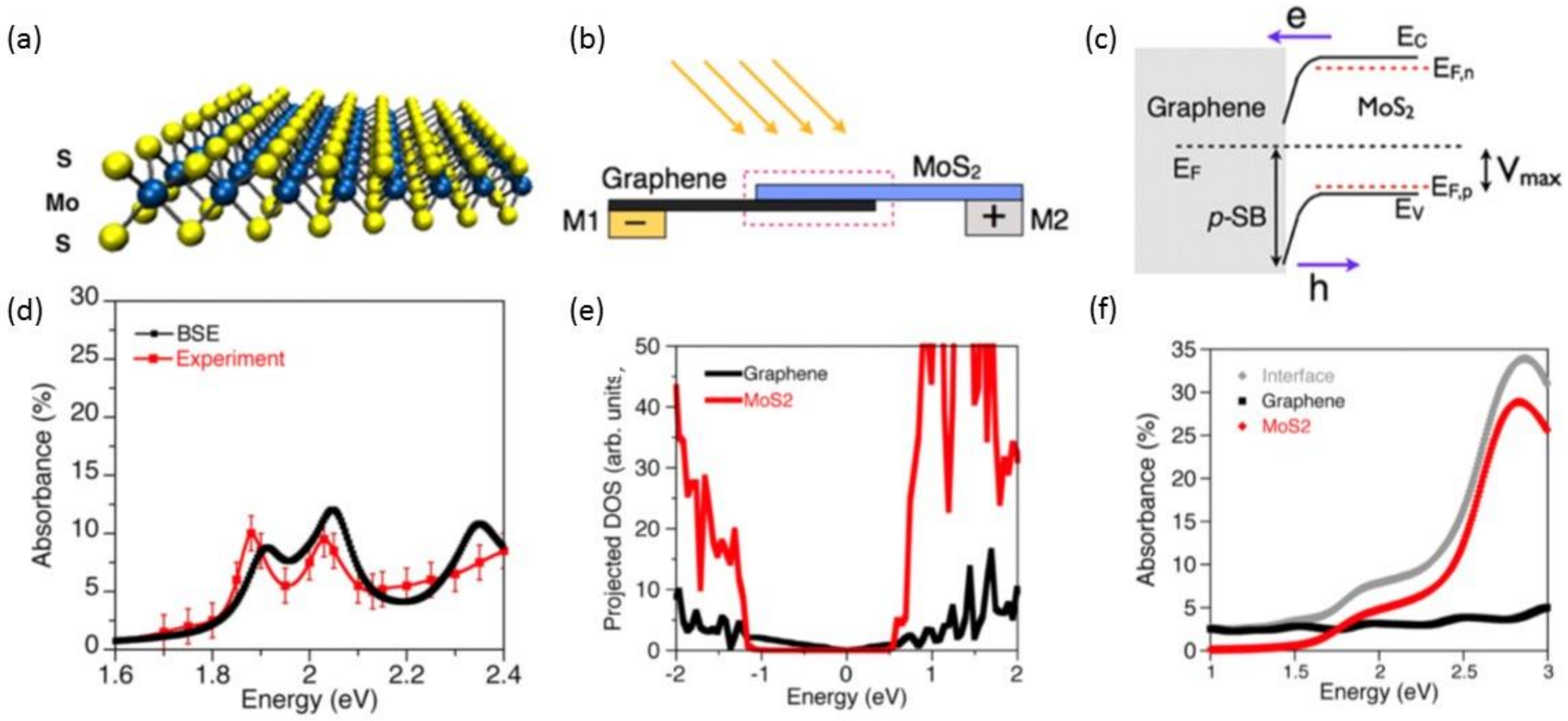

Fig. 32 – (a) Structure of monolayer $MoS_2$ with S atoms represented by yellow spheres and Mo atoms by blue spheres. (b) Schematic of the solar cell made of a G/$MoS_2$ bilayer. M1 (anode) and M2 (cathode) are low and high workfunction metals, respectively. (c) Band alignment at the G/$MoS_2$ interface as predicted using density functional theory (DFT). The valence and conduction band edges ($E_v$, $E_c$) , the Fermi level $E_F$ and the quasi-Fermi levels ($E_{F,n}$ and $E_{F,p}$) of $MoS_2$ are shown together with the direction of electron and hole diffusion. p-SB is the SBH for holes; $V_{max}$ is the maximum open-circuit voltage. Band bending in $MoS_2$ occurs in the direction parallel to the monolayer when moving away from the junction area (d) Comparison of the computed and experimental absorbance of monolayer $MoS_2$. (e) Projected DOS of the G/$MoS_2$ heterojunction around the Fermi energy (set to zero). (f) Computed absorbance of the G/$MoS_2$ and its composing monolayers, within the independent particle approximation using DFT. Figure adapted from Ref. [259].

Bernardi et al. [259] used density functional theory (DFT) combined with a lineup method to calculate a SBH of 1.2 *eV* for holes to diffuse from graphene to $MoS_2$. As sketched in Fig. 32 (c), charge separation occurs with photogenerated electrons injected from the conduction band of $MoS_2$ to graphene, while photogenerated holes are kept in the valence band of $MoS_2$ by the large SB. Under these operating conditions, the maximum achievable open circuit voltage was estimated as $V_{OC,\max} \approx 0.3\,V$ (in Fig. 32 (c) it is the difference between the SBH and the built-in potential, assuming that the quasi-Fermi hole level under illumination reaches the maximum valence band level) while for the maximum short-circuit current the value of $I_{SC,\max} \approx 4.5\,mA/cm^2$ was calculated. The proposed "lateral" geometry of the device with the electrodes spatially separated from the junction further reduces the risk of leakage currents potentially occurring when the bilayer solar cell is placed between two metallic electrodes at ~1 *nm* distance. Another predicted feature of the bilayer G/$MoS_2$ is that the electronic states of graphene and $MoS_2$ do not hybridize near the Fermi energy, as shown by the projected density of states (PDOS) of Fig. 32 (e). Furthermore, the calculations showed that the absorbance at visible photon of the G/$MoS_2$ is equal to the sum of the absorbances of isolated graphene and $MoS_2$ monolayers (Fig. 32 (f)).

An experimental realization of G/TDMC solar cells was reported in Ref. [254] with the fabrication of vertical G/TMDC/G devices (see inset of Fig. 33 (a)) using $WS_2$ or $MoS_2$ as photoactive TMDC layer. To create built-in electric fields for photocharge separation and hence achieve a large photocurrent, an appropriate positioning the Fermi level of the three layers of the structure is necessary. This can be done by simply doping the two graphene layers differently, either by electrostatic gating or chemical methods. The current-voltage characteristics of the fabricated G/TMDC/G device showed a clear dependence on illumination. At dark, the device displayed strongly nonlinear *I-V* curves (Fig. 33 (a)), right axis) with a resistance that dropped dramatically by more than three orders of magnitude under illumination giving rise to linear curves around zero bias (Fig. 33 (a)), left axis). At higher bias (> 0.2 V), the current saturates, as the number of available charge carriers in the photoactive region becomes limited.

By scanning photocurrent microscopy, where a laser spot was scanned over the sample, and the resultant photocurrent was displayed as a function of laser spot position, it was proved that photocurrent was generated only in the region where all G/TMDC/G layers overlap. The band diagrams of Fig. 33 (b) and (c) clarifies the origin of the photocurrent. In the ideal symmetric structure (Fig. 33 (b)), the electron/hole pairs generated in TMDC (by absorption of a photon with sufficient energy) have no preferred diffusion direction and, hence, no net photocurrent is measured. The *e-h* pairs are separated and a photocurrent measured if a built-in electric field (Fig. 33 (b)) across the TMDC is established, either due to a difference in the initial doping between the graphene sheets or by gating.

The photoresponsivity of such devices was estimated above 0.1 $A/W$, corresponding to an $EQE > 30$, demonstrating the potentiality of TMDCs, as well as their combinations, to create new transparent and flexible photonic and optoelectronic structures with properties that can compete or surpass current technologies.

A photodetector consisting of a vertical G/$MoS_2$ heterostructure (Fig. 34 (a)), able to reach a photoresponsivity value higher than $10^{+7}$ $A/W$ and an ultra-high gain of about $10^{+8}$ at room temperature, was reported in Ref. [269]. In that device, the metal leads only contact to the top layer graphene, which is on top of a $MoS_2$ monolayer. The CVD grown $MoS_2$ is used to absorb light and produce electron-hole pairs. The photo-excited electron-hole pairs are separated at the G/$MoS_2$ interface, where the electrons move to graphene due to the presence of a perpendicular effective electric field created by the combination of the built-in electric field, the applied electrostatic field, and charged impurities or adsorbates. With Hall-effect measurements, Zhang et al. [269] demonstrated that in dark the resistance of CVD graphene, originally p-doped in ambient by adsorbed moisture/oxygen and substrate impurities, significantly increases when graphene is in contact with $MoS_2$, suggesting that electrons move from $MoS_2$ to graphene, thus lowering the hole concentration of graphene. Moreover, Raman spectra showed a red shift and broadening of the G band of graphene on $MoS_2$ indicating that the Fermi level of graphene is raised (which corresponds to an increase in the electron concentration) with the light exposure, and that the photoelectrons generated by the Raman laser are injected into graphene.

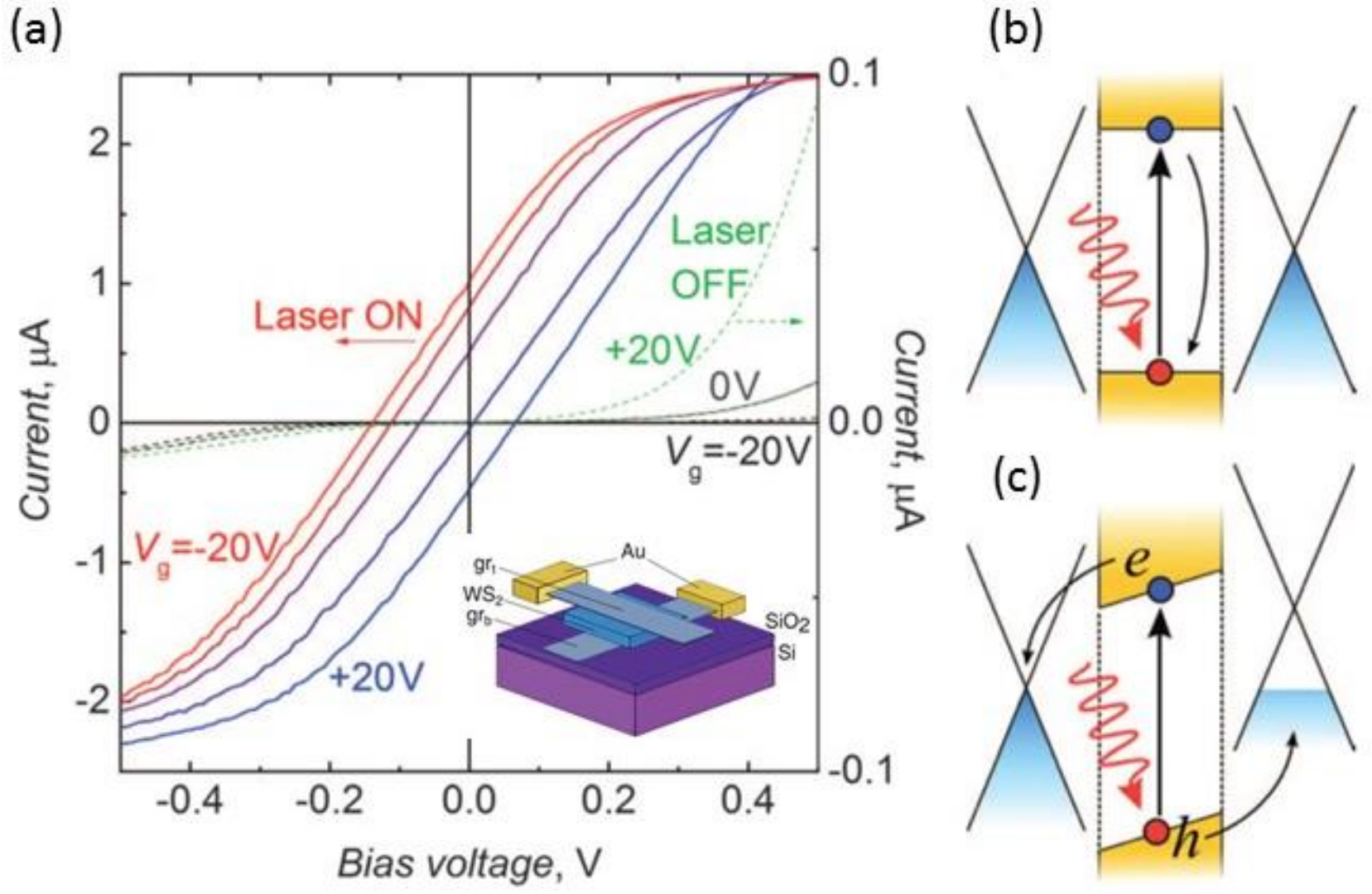

Fig. 33 - Gate-dependent *I-V* characteristics of a G/$WS_2$/G device on Si/$SiO_2$ in the dark (right axis) and under illumination (left axis) at gate voltages from –20 (red) to +20 *V* (blue) in 10 *V* steps, after graphene doping. The laser illumination energy was 2.54 *eV* and the power was 10 *mW*. Inset: schematic of the vertical G/$WS_2$/G heterostructure. Band diagrams of the G/$WS_2$/Gr heterostructure without (b) and with (c) a built-in electric field to separate the photogenerated *e-h* pairs. Figure adapted from Ref. [254]

The study of heterostructures with reversed stacking order, that is $MoS_2$ on graphene on $SiO_2$ substrate, exposed to light, consistently showed that photo-excited electrons moved from $MoS_2$ to graphene, and the photo-excited holes were trapped in the $MoS_2$ layer.

The effect of 650-*nm* light at various power densities on the transfer characteristic of the device (source-drain current *vs* gate) is displayed in Fig. 34 (b). The shape of the dark transfer curve is very similar to that obtained for pristine graphene on $SiO_2$, suggesting that the carrier transport in the G/$MoS_2$ photodetector is dominated by graphene and shows a minimum (corresponding to the charge neutrality or Dirac Point) around 10 *V*, as expected for p-doped graphene. Fig. 34 (b) shows that the drain current in the p-branch (for negative $V_g$) decreases and increases in the n-branch (for negative $V_g$), when the device is illuminated. This is consisted with the aforementioned injection of photoexcited electrons from $MoS_2$ into graphene, while the negative shift of the Dirac point indicates that the photoexcited holes are trapped in the $MoS_2$, which acts as an additional positive gate voltage for graphene.

The photoresponsivity, calculated as the ratio between the photocurrent ($I_{ph} = I_{light} - I_{dark}$) and the light power absorbed by the photodetector, is plotted in Fig. 34 (c) as a function of the light power density and reaches the remarkable value of $1.2\times10^{+7}\ A/W$ at $V_g = -10\ V$, $V_d = 1\ V$ and power density $0.01\ W/m^2$. The photogain obtained as $G = I_{ph}/(e\cdot\Delta n\cdot A)$, where $e\cdot\Delta n\cdot A$ is the number of photoexcited electron-hole pairs, $A$ is the area of the junction and $e\cdot\Delta n = C_g\cdot\Delta V_D$ is the density of the holes trapped in the $MoS_2$, resulted up to $10^{+8}$ (here $C_g$ and $\Delta V_D$ are the $SiO_2$ capacitance per unit area and $\Delta V_D$ is the light-induced variation of the Dirac point position, respectively).

Since $MoS_2$ has a prevailing n-type behavior while graphene in air is p-type, the formation of the junction is expected to result in electrons injected from $MoS_2$ into graphene, with the formation of an upward band bending at the interface corresponding to a built-in electric field from $MoS_2$ to graphene, as in Fig. 30(e) and 31(e). Hence, under illumination, the photo-generated electrons should flow into $MoS_2$. However the experimental data, in this experiment, provided evidence of the opposite electron flow. Hence, Zang et al. argued that in an atomically thin layer 2D junction, the charge concentration and polarity at the interface can be strongly affected by charge impurities on the surfaces of graphene and $MoS_2$ or by the applied gate electrostatic field.

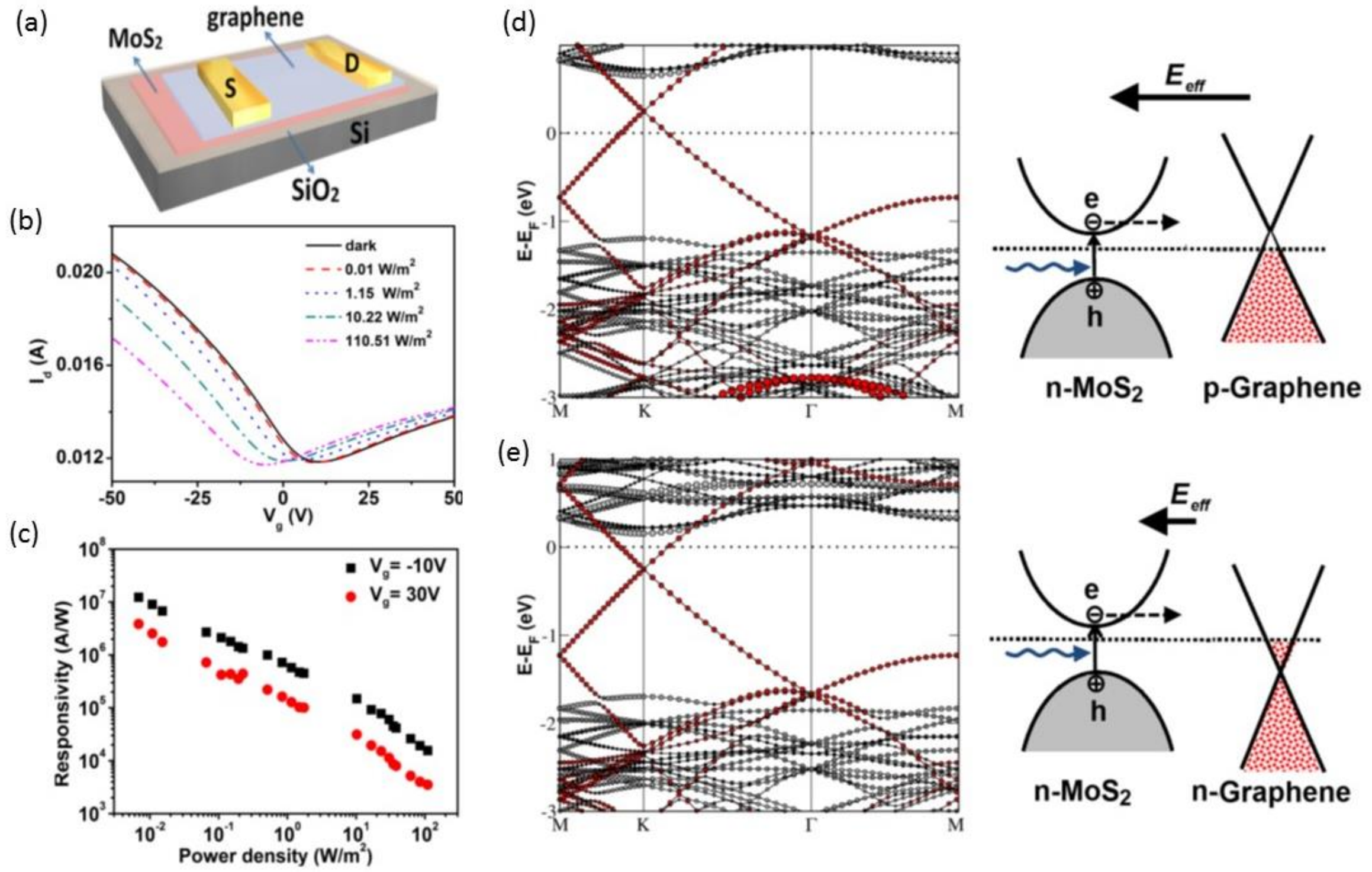

Fig. 34 – (a) Schematic of the G/MoS2 photodetector of Ref. [269] on 300 *nm* SiO2/Si and with Ti/Au contacts. (b)Transfer characteristic ( $I_d - V_g$ curve at $V_d = 0.1V$ ) and responsivity (c) of the device in (a) under exposure to light with different powers. (d) band structures for an n-doped MoS2 layer topped with (a) slightly p-doped graphene and (e) n-doped graphene corresponding to the experimental situation in air and in vacuum, respectively. The electronic states associated with graphene are represented in red while those associated with MoS2 are in gray. The photoelectron transfer process is favored by the effective electric field effective electric field which is a combination of the built-in electric field, the applied electrostatic field, and charged impurities or adsorbates at the interface. Figure adapted from Ref. [269].

The applied gate field and the external electric field from charge impurities on the surface of graphene and $MoS_2$ form an effective electric field at the G/$MoS_2$ interface which determines the flowing direction and the amount of the photoexcited electrons. In this study the direction of the effective electric field was always from graphene to $MoS_2$ both for graphene on $MoS_2$ and $MoS_2$ on graphene. The same direction holds when the device is measured in high vacuum, *i.e.* when graphene is transformed from p-type to n-type due to desorption of adsorbates and moisture. The effect of high vacuum is only a reduction of the photocurrent at the same laser power density, suggesting that the effective electric field decreases in high vacuum. With a lower effective electric field, the photogenerated electron-hole pairs are separated with more difficulty and the photocurrent is reduced. The phototocharge generation process and the photoelectron transfer to graphene is shown in the band diagrams of Fig. 34 (d) and (e) in air and vacuum, respectively. The same figures show also the band alignment as obtained from first-principles calculations based on DFT, where the electric field of the gate modulates the G/$MoS_2$ doping level and align the energy bands, thus tuning the photoresponsivity of the photodetector.
The high photogain of the G/$MoS_2$ bilayer can be explained as follows. The photoelectrons which are transferred to the graphene layer by the effective electric field circulate many times in graphene before recombining with holes in $MoS_2$, due to the high electron mobility in graphene and the long charge-trapping lifetime, leading to a very high photogain, with a mechanism similar to that illustrated in Fig. 19 and Ref. [174].
A $MoS_2$/G phototransistor, this time with $MoS_2$ on top of graphene laying on $SiO_2$, whose functioning was in contrast explained as the transfer photogenerated holes from $MoS_2$ to graphene and their many-times recirculation, is presented in Ref. [270]. A high gain corresponding to a responsivity tunable from 0 to about $10^{+4}$ $mA/W$ by the gate voltage at 2.2 $W/m^2$ light intensity irradiation was reported.

## 9. Conclusions

In this review we highlighted the interesting and new effects observed at the heterojunction between a gapless 2D system, with a limited density of states, and a 3D system, with bandgap and high density of states. The prototype of this junction was graphene on different semiconductors. Because of the low density of states, neutral or low doped graphene can exchange charge with a 3D semiconductor only varying its Fermi level position. In the absence of pinning, changes of the graphene Fermi level are strictly related to changes of the SBH formed with semiconductors. The tuning of the graphene Fermi level, and hence of the SBH, can be done electrically by means of a voltage bias or through a gate, as well as by physical doping with exposure of graphene to a dopant agent. Therefore, in a G/S junction the SBH becomes a controllable quantity.

Table 1, summarizing the zero bias SBH, shows that there is quite a large range of values of the SBH for a given semiconductor. Although some variability can be attributed to the different doping through Schottky effect (eq. (17) and (18)), most of it is caused by the preparation of graphene, the treatment of the semiconductor substrate, the transfer of graphene and the junction formation technique. Most of the G/S junctions we have discussed were obtained by transferring CVD graphene on semiconductor; however, the details of the fabrication, as the synthesis of graphene, the treatment of the semiconductor surface and the techniques of adhesion, were different in each case, and the fabrication of a G/S heterojunction is still far from standardization. A further complication is introduced by the ambiguity in the method chosen for data analysis. Although many authors report the SBH at zero bias (following the procedure outlined in section 2 (c) by eq. (40)-(42)), in many other studies the SBH is extracted from the reverse saturation current at an arbitrary reverse bias or from a fit of the forward current without extrapolation to zero bias.

From a modeling point of view, transport at the G/S junction has been mostly interpreted in the framework of the standard thermionic theory of the M/S junction. We discussed a couple of attempts of tailoring the thermionic theory to the specific G/S system, by including the low density of state of graphene, its changing Fermi level, or by using a Landauer formalism. Further research is required to fully understand the variety of transport phenomena that may occur at the interface and include them in a generalized theory of the junction. It is likely that the recent uptick of the experimental and applied investigation on the G/S interface will trigger more theoretical dedicated studies.

We have seen how the unique possibility of modulation of the SBH offered by the G/S system has paved the way for several applications. In photo-detection G/S junctions offer winning features as speed of the response, large bandwidth, easy adaptability to different levels of illumination and record current or voltage responsivity. In this context, the demonstrated high-performance and low-cost *IR* photodetector seems particularly promising. Solar cells based on G/S junctions have already reached an efficiency (~15%) competitive with M/S and microcrystalline *p-n* cells, although their large-scale manufacturability is still far-reaching. Despite of this, graphene still holds promise as a transparent, conductive electrode that can be integrated with existing photovoltaics silicon technologies to replace transparent indium-tin oxide or fluorine-doped tin oxide. As chemical sensor, the G/S junction can be superior to other devices both in terms of sensitivity and power consumption and is definitely a best option than a chemiresistor for graphene-based sensors.

We have also reviewed some properties and applications of the Schottky junction formed by graphene with 2D layered semiconductors, with special focus on G/$MoS_2$, so far the most studied device among the new emerging and promising class of van der Waals layered heterojunctions. Such 2D heterostructures constitute at the moment an important field of research with huge expectations for applications in optoelectronics and flexible electronics.

A common problem to all these applications, from an industrial standpoint, is the stability over long time. It is well-known that Schottky devices tend to drift over time as result of being very sensible to modification of the metal−silicon interface. It has still to be figured out if this is a common issue to G/S devices.

In conclusion G/S heterojunctions are promising devices for real-life applications and interesting platforms to investigate the physics at the interface between 2D-3D and 2D-2D materials. The recent growing activity on 2D graphene-like layered materials, from transition metal dichalcogenides to silicine, germanine, phosphorene, etc., and their junctions is benefitting from the experimental and the theoretical techniques developed for the G/S system.
From a practical viewpoint, we tried to show the impressive results achieved in few years of research in different application fields. These results are impressive and encouraging to envisage G/S based devices to hit the market in the next years.